\documentclass[iop,apj,numberedappendix,appendixfloats,revtex4]{emulateapj}
\pdfoutput=1 
\usepackage[T1]{fontenc}
\usepackage{apjfonts}
\usepackage{natbib}

\usepackage[pdftex,bookmarksnumbered=true,backref=true,hyperfigures=true,colorlinks=true]{hyperref} 
\shorttitle{A Census of Structure in Early-type Dwarfs}
\shortauthors{Janz et al.}
\submitted{}
\hypersetup{pdfauthor={Janz et al.},pdftitle={A Census of Structure in Early-type Dwarfs}}
\newcommand{\noprint}[1]{}

\usepackage{pdflscape}

\begin{document}
\submitted{}
\title{A near-infrared census of the multi-component stellar structure\\ of early-type dwarf galaxies in the Virgo cluster}
\thanks{Based on observations collected at the European Organisation for Astronomical Research in the Southern Hemisphere, Chile, under programme IDs 064.N-0288 and 085.B-0919. }
\author{J.~Janz\altaffilmark{1,2,*}, E.~Laurikainen\altaffilmark{3,1}, T.~Lisker\altaffilmark{2}, H.~Salo\altaffilmark{1}, R.~F.~Peletier\altaffilmark{4}, S.-M.~Niemi\altaffilmark{5}, E.~Toloba\altaffilmark{6,7}, G.~Hensler\altaffilmark{8}, J.~Falc\'on-Barroso\altaffilmark{9,10},  A.~Boselli\altaffilmark{11},  M.~den~Brok\altaffilmark{4,12}, K.~S.~A.~Hansson\altaffilmark{2}, H.~T.~Meyer\altaffilmark{2}, A. Ry\'s\altaffilmark{9,10}, S.~Paudel\altaffilmark{13,2}}%
\email{jjanz@ari.uni-heidelberg.de}

\affil{
$^1${Division of Astronomy, Department of Physics, P.O. Box 3000,  FI-90014 University of Oulu, Finland}\\
$^2${Astronomisches Rechen-Institut, Zentrum f\"ur Astronomie der Universit\"at Heidelberg, M\"onchhofstra{\ss}e 12-14, D-69120 Heidelberg, Germany}\\
$^3${Finnish Centre for Astronomy with ESO (FINCA), University of Turku, Finland}\\
$^4${Kapteyn Astronomical Institute, University of Groningen, PO Box 800, 9700 AV Groningen, the Netherlands}\\
$^5${Mullard Space Science Laboratory, University College London, Holmbury St. Mary, Dorking, Surrey RH5 6NT, United Kingdom}\\
$^6$UCO/Lick Observatory, University of California, Santa Cruz, 1156 High Street, Santa Cruz, CA 95064\\
$^7$Observatories of the Carnegie Institution of Washington, 813 Santa Barbara Street, Pasadena, CA 91101\\
$^{8}${University of Vienna, Institute of Astronomy, T\"urkenschanzstra\ss e 17, 1180 Vienna, Austria}\\
$^9${Instituto de Astrof\'isica de Canarias, V\'ia L\'actea s/n, La Laguna, Tenerife, Spain}\\
$^{10}${Departamento de Astrof\'isica, Universidad de La Laguna, E-38205 La Laguna, Tenerife, Spain}\\
$^{11}${Aix Marseille Universit\'e, CNRS, LAM (Laboratoire d'Astrophysique de Marseille) UMR 7326, 13388, Marseille, France}\\
$^{12}${Department of Physics and Astronomy, University of Utah, Salt Lake City, Utah 84112, USA}\\
$^{13}${Laboratoire AIM Paris-Saclay, CNRS/INSU, Unversit\'e Paris Diderot, CEA/IRFU/SAp, 91191 Gif-sur-Yvette Cedex, France}
}
\altaffiltext{*}{Fellow of the Gottlieb Daimler and Karl Benz Foundation.}

\begin{abstract}

{ The fraction of star-forming to quiescent dwarf galaxies varies from almost infinity in the field to zero in the centers of rich galaxy clusters. What is causing this pronounced morphology-density relation? What do quiescent dwarf galaxies look like when studied in detail, and what conclusions can be drawn about their formation mechanism? Here we study a nearly magnitude-complete sample ($-19 < M_r < -16$ mag) of 121 Virgo cluster early types with deep near-infrared images from the SMAKCED project. We fit two-dimensional models with optional inner and outer components, as well as bar and lens components (in $\sim$15\% of the galaxies), to the galaxy images. While a single S\'ersic function may approximate the overall galaxy structure, it does not entirely capture the light distribution of two-thirds of our galaxies, for which multi-component models provide a better fit. This fraction of complex galaxies shows a strong dependence on luminosity, being larger for brighter objects. We analyze the global and component-specific photometric scaling relations of early-type dwarf galaxies and discuss similarities with bright early and late types. The dwarfs' global galaxy parameters show scaling relations that are similar to those of bright disk galaxies. The inner components are mostly fitted with S\'ersic $n$ values close to 1. At a given magnitude they are systematically larger than the bulges of spirals, suggesting that they are not ordinary bulges. We argue that the multi-component structures in early-type dwarfs are mostly a phenomenon inherent to the disks, and may indeed stem from environmental processing.}

\end{abstract}


\setcounter{footnote}{0}

\section{Introduction}

Early-type dwarf galaxies populate the Universe in large numbers, their number counts
dominate over other galaxy types in clusters, and yet their formation is a widely disputed topic. 
Their name reflects their morphological resemblance to giant early-type galaxies: both are elliptical in shape and have a featureless appearance with little ongoing star formation and only small amounts of gas and dust. Therefore, they are often called  ``dwarf ellipticals" (\citealt{Binggeli:1985p849}, but see \citealt{2009ApJS..182..216K} for a different view).
They possibly follow a similar color-magnitude relation as bright ellipticals (e.g.\ \citealt{2007A&A...463..503M,2008MNRAS.386.2311S,2008A&A...486..697M}, see however recent studies for Virgo: \citealt{Ferrarese:2006p586,Janz:2009p780,chen+10}), and their sizes and locus in the Kormendy relation may or may not be explained
by a continuous variation of  the profile shape with galaxy brightness (\citealt{graham_guzman,Ferrarese:2006p586}, see also \citealt{2011arXiv1108.0997G} for a review; for an alternative view see e.g.\ \citealt{2009ApJS..182..216K}). Furthermore, the early-type dwarfs are preferably found in  high density environments of galaxy clusters and manifest the morphology-density
relation similar to giant ellipticals (\citealt{Dressler}; for dwarfs see \citealt{ferguson_binggeli}; and within the early-type dwarfs, \citealt{Lisker:2007p373}).  

Where do these galaxies come from? A priori it is, despite the similarities, not known whether the name ``dwarf elliptical'' is justified,
i.e.\ that they  are small low mass versions of giant ellipticals \citep{1984ApJ...282...85W}.
There are several indications that this is not the case (\citealt{2009ApJS..182..216K} and references therein).
Alternatively, it was proposed that  they  developed their  spheroidal non-starforming appearance 
during a transformation from a  late-type  galaxy in a cluster environment \citep{1985ApJ...295...73K}. 
Due to the morphology-density relation such a  mechanism is expected to depend highly on the environment.
Suggested processes  include ram pressure stripping \citep{Boselli,Boselli:2008jr,2010ApJ...724L.171D}, which may also heat the disk \citep{Smith11}, 
and harassment \citep{1998ApJ...495..139M,Mastropietro:2005ba,Smith10}.
While it still remains to be shown in detail that these mechanisms are
efficient enough, 
 they 
all will inevitably act in the cluster environment to some extent.
Further aspects might also play a role: the time the galaxy spent within the cluster, i.e.\ in-situ formation  \citep{Toloba:2009p3937,Lisker:2009p3975}, an infall into the cluster early on, or a more recent infall.
 Also, the relevance of mergers at these low masses  is expected to be smaller  as compared to giant ellipticals \citep{2012MNRAS.419.1324D,lisker_sam}, but nonetheless interactions take place even today \citep{MartinezDelgado:2012dv}.

 Already  \citet{binggeli_cameron} partially described the heterogeneity of this galaxy class, but it was fully appreciated only recently (\citealt{graham_guzman,Lisker:2007p373,Michielsen:2008p3941,Lisker:2009p3975,Toloba:2009p3937,Toloba:2011p3986,koleva} as a biased list of references). The diversity in different galaxy parameters  might call for a number of processes shaping the galaxies, with varying importances depending on the evolutionary history of the individual galaxy.

The SMAKCED project ($S$tellar content, $MA$ss and $K$inematics of $C$luster $E$arly-type
 $D$warfs) aims at a more complete understanding
of the origin of the early-type dwarf galaxies. We target  a complete magnitude limited
sample of early-type galaxies in the Virgo cluster in a brightness range where all
the different morphological subclasses and varieties of stellar
content and kinematical configuration coexist.
In order to reveal more details of their presumably heterogeneous evolutionary histories,
we want to gain a detailed picture of their morphology
with deep near-infrared images and, for a subsample of galaxies,  of their kinematics.
The near-infrared is particularly useful for our purpose, since it is 
sensitive to the bulk of the stellar mass.
Here we  report on our progress with deducing the 
distributions of stars with two-dimensional multi-component
decompositions for the  observed 121 galaxies, some initial results of which were presented in \citet{janz+2012}.

In Section 2 the sample selection, observations, and data reductions are described. 
Section 3 summarizes the analysis including  multi-component decompositions  and non-elliptical isophotal shapes. In Section 4 the results are presented. The analyzed  multi-component 
structures are then compared to those found in bright galaxies in Section 5.  The results are  discussed  in Section 6, and we conclude with the summary in Section~7.

\section{Sample, Observations, \& Data}
\subsection{Sample}
\begin{figure}
\begin{center}
\includegraphics[height=\linewidth, angle=-90]{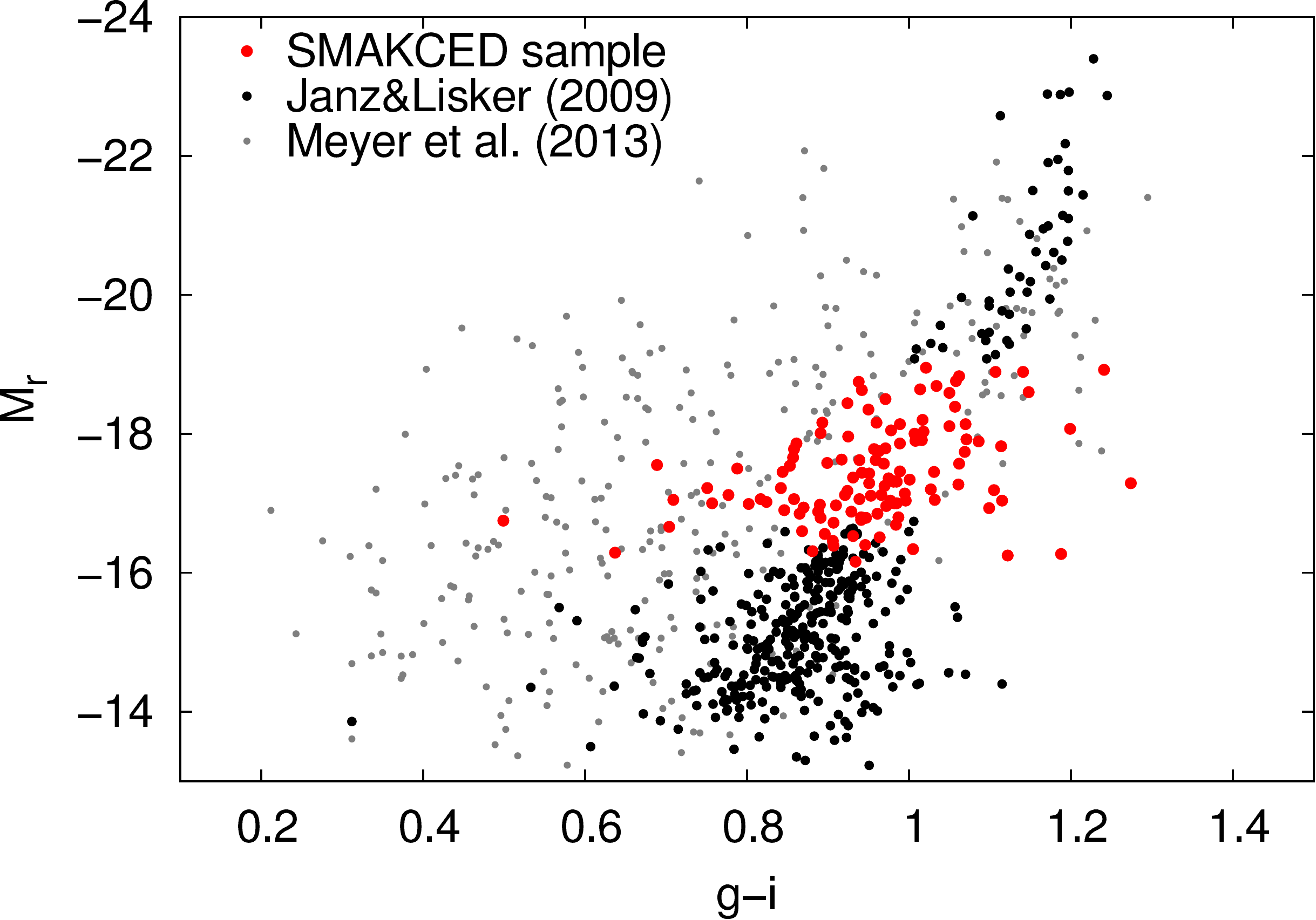}
\caption{The curved $g-i$ color-magnitude relation of early types in the Virgo cluster from \citet{Janz:2009p780}. Galaxies that we observed within SMAKCED in the near-infrared are highlighted with \textit{red} color.  For comparison the \textit{small gray} symbols in the background display the colors of late-type galaxies in the Virgo cluster \citep{meyer}.\label{fig:sample}}
\end{center}
\end{figure}

Within  SMAKCED we obtained deep near-infrared (NIR) $H$-band images for  121
 early-type galaxies in the Virgo Cluster. The sample is drawn from the Virgo cluster catalog (VCC, \citealt{Binggeli:1985p849}, with updated memberships by heliocentric velocities from the literature, \citealt{Lisker:2006p385}) and
 spans  three magnitudes within $-19\textrm{ mag} <M_r<-16\textrm{ mag}$ (Fig.\ \ref{fig:sample}, assuming a distance modulus for Virgo of $m-M=31.09\textrm{ mag}$, \citealt{Mei:2007p872}). In total there are 174 galaxies fulfilling these criteria. Our observations are complete down to $M_r=-16.76$ mag, with images for 121 of the galaxies.
The brightness range and sample size guarantee that it is possible to study galaxy parameters binned according to   the various early-type dwarf galaxy subclasses \citep{Lisker:2006p385,Lisker:2006p392,Lisker:2007p373}, galaxy brightness, local environment, or rotational support \citep{Toloba:2009p3937,Toloba:2011p3986}  with a sufficient  number of galaxies in each bin  (see Table \ref{table:bins} = Table 1).
Furthermore, some galaxies that were classified as normal (non-dwarf) early types   \citep{Binggeli:1985p849} overlap with the early-type dwarf galaxies in this brightness range (e.g.\ \citealt{Janz:2008p151}).

\begin{deluxetable}{lcccccc}
\tablecaption{Sample\label{table:bins}}
\tabletypesize{\footnotesize}
\tablecolumns{7}
\startdata
\tablehead{ \colhead{} &\colhead{dE(N)} &\colhead{dE(nN)} & \colhead{dE(di)} & \colhead{dE(bc)}  & \colhead{E} & \colhead{S0}\\
\\
 \multicolumn{7}{c}{$19\le M_r\le-18$ mag} }
$d_{M87} \le 1.5^\circ$ & 1 & 0 & 2 &0  & 1 & 1 \\
$1.5 < d_{M87} \le 4^\circ$ & 2 & 1 & 4 & 0 & 2 & 4 \\
$d_{M87} > 4^\circ$ & 3 & 0  & 2 & 0 & 2 & 2\\
   \cutinhead{$-18< M_r\le-17$ mag}
$d_{M87} \le 1.5^\circ$ & 8 & 2 & 3 & 3& 1 & 1 \\
$1.5 < d_{M87} \le 4^\circ$ & 15 & 0 & 10 & 4& 2 & 1 \\
$d_{M87} > 4^\circ$ & 3 & 0 & 4 & 3& 0 & 2\\
   \cutinhead{$-17< M_r\le-16$ mag}
$d_{M87} \le 1.5^\circ$ & 4 & 1 & 0 & 1& 2 & 1 \\
$1.5 < d_{M87} \le 4^\circ$ & 6 & 2 & 1 & 2&0  & 0 \\
$d_{M87} > 4^\circ$ & 5 & 0 & 4 & 1& 0 & 2
\enddata\tablecomments{Galaxies binned over basic parameters; $d_{M87}$ is the angular distance to M87. At the distance of the Virgo Cluster $1^\circ =0.284$ Mpc. The classification of the bright galaxies is taken from the VCC \citep{Binggeli:1985p849} and the early-type dwarf subclasses from \citep{Lisker:2006p385,Lisker:2006p392,Lisker:2007p373}.}
\end{deluxetable}

\subsection{Observations}
 The observations were done in the NIR allowing the most direct measurement of the distribution of stellar mass in galaxies (e.g.\ \citealt{BelldeJong2001}). 
  This wavelength is also less affected by dust, which is present even in some of the early-type dwarf galaxies \citep{dust}.
 We choose the $H$-band, since the observations are more efficient than in the $K$-band, and the $J$-band is less optimal for characterizing the old stellar population, i.e.\  the distribution of stellar mass, since it has a shorter wavelength.
 During the  last  two years,
 we obtained images for 97 galaxies using  SOFI at ESO/NTT, NICS at  TNG, and NOTCam at  NOT with the wide-field option (085.B-0919; AOT21/CAT\_193, AOT23/CAT\_102;  P41-217, P42-047, P42-212, P43-038, P43-206).
  The data is complemented with archival data for some bright galaxies (from ESO NTT/SOFI, 064.N-0288,\footnote{Obtained via the ESO archive, \url{http://archive.eso.org}.} P.I.\ A.\ Boselli; from NOT, Niemi). 
  
 The integration times were calculated for each galaxy individually, in order to reach a signal to noise ($S/N$)
 of 1 per pixel ($\sim$$0\farcs25\times0\farcs25$, see Table \ref{table:cams} = Table 2) at 2 half-light radii.
 The half-light radii were determined on the Sloan Digital Sky Survey (SDSS) $r$\hbox{-}band images \citep{Janz:2008p151}.
 The $H$-band surface brightness was estimated by measuring the surface brightness in the $i$-band  \citep{Janz:2009p780} and converting it  to the $H$-band with a conservative value of $i$-$H$$=1.7$ mag, based on the $i$-$H$ colors for some of the galaxies from SDSS and the archival $H$-band data. 
 Generally, we limited the total on-source integration time to three hours. Notice that, 
 the galaxies with low surface brightnesses typically have  large sizes, and therefore the effective $S/N$  is comparable to that in  galaxies with higher surface brightnesses.

 \begin{deluxetable*}{lcccccc}
\tablecaption{Telescopes and cameras\label{table:cams}}
\tabletypesize{\footnotesize}
\tablecolumns{7}
\startdata
\tablehead{ \colhead{Telescope}  &\colhead{Diameter}  &\colhead{Camera} &\colhead{Detector} & \colhead{Pixel scale} & \colhead{Field of view}  & \colhead{Read noise} }
New Technology Telescope (ESO/NTT) & 3.58m & SOFI & $1024\times1024$ HAWAII & $0\farcs288/$pix & $4\farcm92\times4\farcm92$ & $11.34\ e^-$\\
Telescopio Nazionale Galileo (TNG) & 3.58m & NICS &  $1024\times1024$ HAWAII &$0\farcs25/$pix & $4\farcm2\times4\farcm2$& $24\ e^-$\\
Nordic Optical Telescope (NOT) & 2.56m & NOTCam &  $1024\times1024$ HAWAII &$0\farcs234/$pix & $4\arcmin\times4\arcmin$& $8\ e^-$ 
\enddata
\end{deluxetable*}

 For a few bright galaxies we integrated longer than their calculated exposure times, in order to reach at least a surface brightness of 22.2 mag arcsec$^{-2}$.
Typically we reach a surface brightness  of 22.2--23.0 mag arcsec$^{-2}$, at a $S/N$ of 1 per pixel.
 In some cases the upper limit in the integration time, the weather, or
  the use of archival data resulted in an image depth shallower than desired.

 Most of the observations were carried out by dithering the galaxy to the centers of the detector's four quadrants, 
  since the detectors are sufficiently large ($\sim$$4'\times4'$). This way the sky was obtained simultaneously with the target observation. A good sampling of the bright and variable NIR sky is essential to accurately subtract it (see \S\ref{section:reduction}). A few galaxies exceeded the size of one  quadrant of the detector, or bright foreground stars hampered our dithering strategy. In those cases
 a customized 3-point dither pattern or beam-switching with separate sky observations were used.
 With each instrument and for each target we tried to optimize the exposure times for single exposures so that we stayed within the linear regime of the detector, while  minimizing the overheads. Typical values for the number of exposures per pointing (NDIT) and for the single exposure time (DIT)   were 10x6s for NOTCam and SOFI, while for NICS it was  3x25s. With NOTCam we used the ramp-sampling mode, which allows to read out the detector during the exposure: linear regression applied to the several times sampled signal  reduces the noise.
 The observations are summarized  in  Table 3.

\subsection{Reduction}
\label{section:reduction}
\subsubsection{Instrumental effects: flat fields, crosstalk}
The data reduction 
was mostly carried out using our own \textsc{iraf}\footnote{\textsc{iraf} is distributed by the National Optical Astronomy Observatory, which is operated by the Association of Universities for Research in Astronomy (AURA) under cooperative agreement with the National Science Foundation.} scripts.
Two of the detectors (SOFI and NICS) suffer from an unwanted effect called crosstalk. This effect causes ghost images of bright sources in some other quadrant of the detector and can enhance the signal in the affected rows or columns. To deal with this problem, we first remove the crosstalk from the images  using scripts provided by the observatories (\texttt{crosstalk.cl} for SOFI and a \textsc{fortran} script from the \textsc{snap} package for NICS).
The removal works well for sources that do not exceed the linearity range of the detector.
Subsequently, the images are flat fielded. For the three instruments we follow the recommended procedures and use  ``special" dome flats\footnote{The ``special" dome flats reduce the bias by a different illumination of the dome flat screen as compared to the sky by additionally considering exposures with the lamp shining only on parts of the screen, see SOFI manual.} for SOFI, differential skyflats for NOTCam, and skyflats for  NICS data.
For convenience we normalize the images to 1s exposure times. 

\subsubsection{Sky subtraction}
 It is   crucial to perform  the sky subtraction with high accuracy, since the surface brightnesses we are aiming at are on the order of 10 mag arcsec$^{-2}$ fainter than the sky.
Typically we average the sky over the frames taken in a time window of  10-15 minutes around the observation. The length of the time interval is adjusted to the variability of the sky. 
Any sources (including the target) need to be carefully masked out down to low surface brightnesses. It is unfeasible to create appropriate masks on the single exposures.
Therefore, we iterate the reduction process and create the masks on preliminary coadded images with \textsc{SExtractor} \citep{sextractor}. 

For averaging the masked sky frames, they need to be scaled to a common flux level. Since  there can be  gradients in the sky over the field of view, it is important to compensate for the masked out regions  in the scaling. We find appropriate values for the masked pixels by filling them with fits of fourth order polynomials to the unmasked pixels in each quadrant of the detector. Note that the reconstructed pixels are  used only to find the appropriate scaling,  not for  the averaged sky frame that is then subtracted from the image. 
The rows and columns of the detector can have a slightly different output.
We average over rows and columns, and subtract the result from the image to obtain the final flat foreground subtracted single exposure.  When averaging, we  again use  the masks and compensate for the masked pixels.

\begin{figure}
\begin{center}
\includegraphics[height=\linewidth, angle=-90]{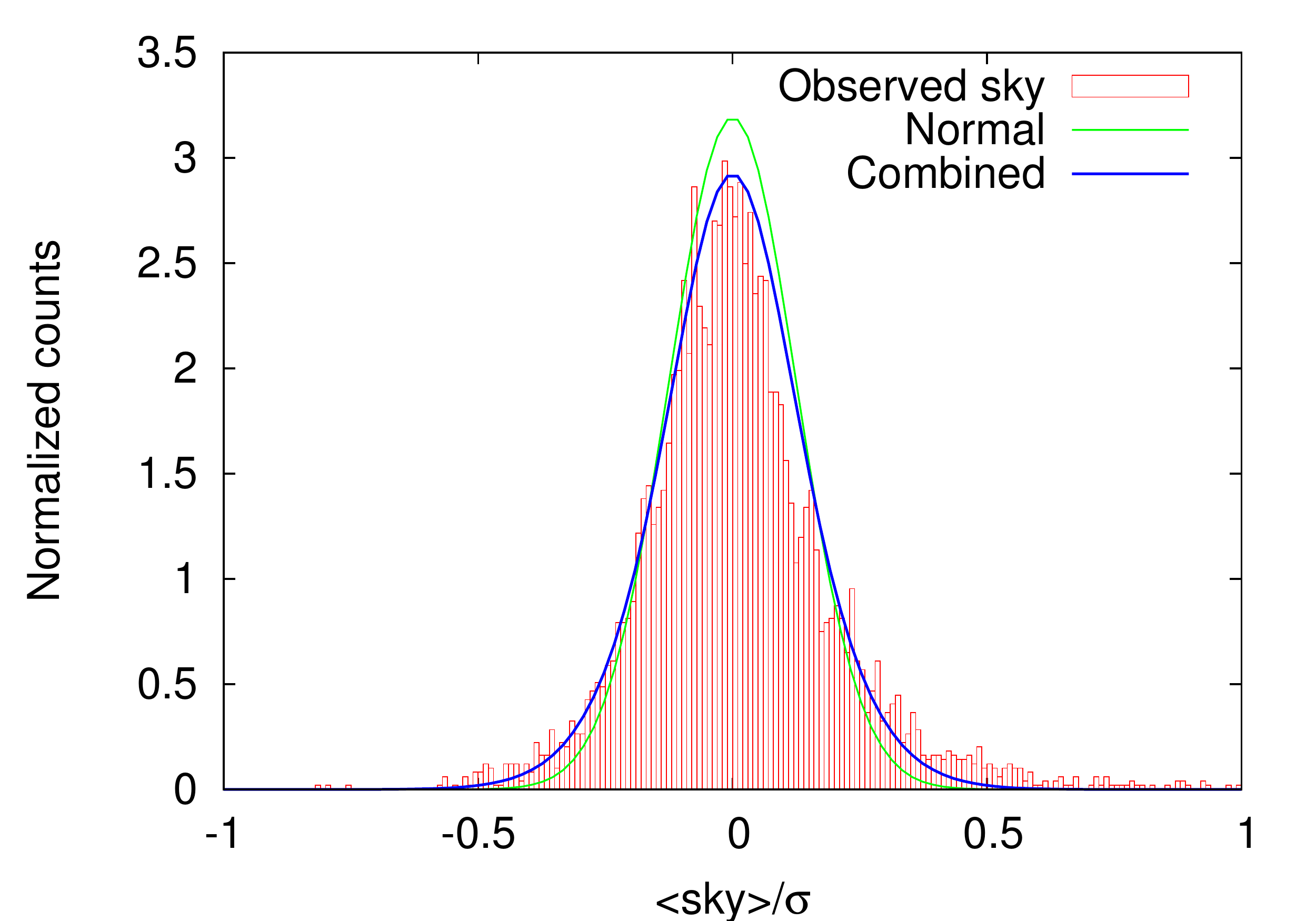}
\caption{Sky values measured in small boxes of 21x21 pixels across the images in units of the noise in the image. The Gaussian has a variance of $\sigma= 0.125$, the combined curve adds three Gaussians accounting for the overall pattern in the $\sigma$-image (see text). \label{fig:sky}}
\end{center}
\end{figure}

In order to assess the quality of the sky subtraction, we measure the sky levels in small boxes of 21x21 pixels distributed across the final coadded images. When scaled to the average noise in the individual coadded image, 
 a width of $1/\sqrt{21*21}=0.05$ is  expected for the distribution if the noise is Gaussian. The expected width 
is slightly larger if pixels in some boxes are not used due to masked out objects. 
Moreover, the noise follows overall a pattern due to the used dithering: the central part  of the galaxy is imaged every time ($\sigma_{\textrm{center}}$),  parts at the edges in half of the exposures ($\sigma_{\textrm{edge}}=\sqrt{2}\sigma_{\textrm{center}}$) and in the corners in every fourth exposure ($\sigma_{\textrm{corner}}={2}\sigma_{\textrm{center}}$),  for the standard pattern. 
Fig.~\ref{fig:sky} shows the measurements in units of the average noise in the individual coadded image. For comparison, we plot a normal distribution ($\sigma=0.125$) and a combination of three normal distributions accounting for the overall noise pattern. We conclude that the increased width in comparison to the expectation is due to correlated noise and variations of the background on larger scales. 
These variations are considered in the evaluation of the fitted models in Section \S\ref{section:fitting} and the profile figures there.

\subsubsection{Illumination correction}
During the observing run in March 2012 we observed one standard star in two nights (with NICS and NOTCam)
on a 5 by 5 grid of different positions on the detectors. 
The measured fluxes of the standard star vary at a $\sim$10\% level depending on the position on the detector
(the reason for that is not fully understood, see \citealt{hodgkin}).
The variations are systematic so that they form a smooth surface. We fit  second order polynomials to the
surfaces using \textsc{iraf}/surfit and apply the normalized fits as an illumination correction to the science frames.
The illumination correction might  vary with time. Therefore, 
we compare the fitted surface to the brightness variations across the
detector using stars in the science frames of two photometric nights in 2010 (Appendix \ref{sec:zp}). From that comparison we conclude 
that the possible error from a time-dependence of the illumination correction is smaller than the error that
would be made without the correction. 
Indeed, when applying the illumination correction, the scatter of the zeropoints (determined by comparison of sources in the field-of-view
to 2MASS and UKIDSS, see \S\ref{sec:calib}) is reduced.

\subsubsection{Distortion correction, alignment, and coaddition}
\label{sec:distortion}
NICS at TNG and NOTCam at NOT suffer from  optical distortions large enough to hamper the alignment of the dithered images for coaddition: the galaxy and the stars have  different distances depending on the positions on the detector. In the corners the pixels are shifted up to about 10 or 13 pixels for NICS and NOTCam, respectively, relative to their position on an undistorted image. 
Therefore, we use the  \textsc{iraf}/geotran task and distortion maps to transform the images to geometrically flat images. 
For NICS  the distortion map was extracted from the \textsc{snap}
pipeline,  and for NOTCam the distortions were provided   by Magnus G\aa lfalk. 
The NTT images have been found to be astrometrically flat to a 0.1\% level \citep{SOFI}.

The distortion corrected single exposures are  aligned using the \textsc{iraf}/xregister task.
The signal of the galaxy is measured in a small galaxy aperture and the noise
on all non-masked pixels. The single exposures are coadded using these values
as scales and weights, in order to optimize the $S/N$ of the final image.
During the coaddition, we mask out the detectors' bad pixels.
In some cases a perfect alignment of the images is impossible due to an imperfect modeling of the field distortions.
Then we optimize the alignment for the region of the image where the galaxy is located. 
Since stars in other parts will not be perfectly aligned and since the  seeing conditions can vary
during the integration time, it is important not to do a clipping in the averaging process.
In that case some of the flux of the imperfect stars would be lost and the calibration
of the images  would be biased. Finally, the averages in lines and columns, masking all sources,
are subtracted from the coadded image.

\subsection{$\sigma$-images and PSF}
The many single exposures are used to estimate the uncertainties of the flux in each pixel.
For all pixels in the final coadded image we  calculate  the standard deviations. These noise maps we  call subsequently $\sigma$-images. They will be used   later 
when fitting models to the galaxy's light distribution (\S\ref{section:fitting}).\footnote{For the archival data from the NTT we approximate the $\sigma$-images with a constant value of the average noise in the images.}
Finally, the full width at half maximum (FWHM) of the point spread function (PSF) is determined as the average of the FHWM values of the point sources, measured on the coadded images with \textsc{SExtractor} (see Table 3). 

\subsection{Flux Calibration}
\label{sec:calib}
The images are calibrated using point sources in the field of view and comparing their fluxes
to the magnitudes given in the 2MASS point source \citep{2mass} and UKDISS  catalogues \citep{ukidss}.\footnote{Like 2MASS and UKIDSS, we use the Vega system for our NIR magnitudes. The UKIDSS and 2MASS filter systems are slightly different. We convert the UKIDSS catalog values to the 2MASS system using \citet{Hewett:2006hy}.}  
This has the advantage of obtaining reliable photometry also for the data taken under non-photometric conditions.
We require  the sources to be sufficiently point-like and the photometric errors to be small. Also too bright sources are rejected
in order to be sufficiently in 
the linear regime of the detector (for details see Appendix \ref{sec:zp}). Typically there are 15 comparison objects in the field of view of the coadded images. The median error of the zeropoints, estimated by the scatter and number of stars used for the determination in each frame, is about 2\% (see also Appendix \ref{sec:zp}).
Both, the scatter of the zeropoints, determined by individual sources in one field-of-view,  and the scatter of the final, averaged zeropoints used for the galaxies are reduced when applying the illumination correction.

Finally, the images are corrected for galactic extinction according to \citet{schlegel}.

\section{Methodology}
{ In this paper we look at non-parametric photometric parameters of the galaxies, at the profiles of the disky/boxy characteristics of the isophotes, and carry out a two-dimensional multicomponent structural analysis. Our approach for the decompositions is different from the traditional one-dimensional method for identifying such structures, which considers the azimuthally averaged profiles as well as an isophotal analysis based on ellipse fitting.
All the analyses are described in the following paragraphs, and the choice of the two-dimensional decomposition approach is reasoned in Section \S\ref{section:fitting}.}
\subsection{Non-parametric photometry}
\label{sec:nonpar}
First,  we want to derive basic non-parametric photometry parameters for all of the galaxies in the sample.
We follow a similar procedure as in \citet{Janz:2008p151}: we determine an aperture, in which we will measure the flux, in terms of the Petrosian radius.
The Petrosian radius $r_P$ \citep{1976ApJ...209L...1P} is the radius at which the intensity drops to a fifth of the mean intensity within this radius,\footnote{Other authors use different values (e.g.\ \citealt{conselice+2002,conselice+2003}).} 
\begin{equation}{{I(r_P)}\over{<I>_{r_P}} }= 0.2\, .\end{equation}
 Instead of circles we use ellipses so that we obtain a ``Petrosian semi major axis (SMA)"  (see, e.g. \citealt{2004AJ....128..163L}). The total flux of the galaxy is then measured within the aperture of two Petrosian SMA.  Immediately, also a value for the half-light SMA can be obtained.
 The axis ratio and position angle of the aperture are  first determined by the image moments within the circular aperture \citep{Abraham}, and after one iteration of the whole process by \textsc{iraf}/ellipse at an SMA twice the half-light SMA.

  Depending on the profile shape a certain fraction of the flux falls outside the aperture. We correct our total magnitudes and half-light apertures for the missed flux according to
\citet{2005AJ....130.1535G}. However,  
 typically the correction is small, since the  early-type dwarf profiles are mostly not much more concentrated than an exponential profile.
Our method was shown to be consistent with other curve-of growth methods (\citealt{chen+10} for early types in Virgo with SDSS data).
In the following we denote the SMA of the half-light aperture by $r_e$.  

Furthermore, we want to quantify the galaxy concentrations in a model-independent way. 
We follow \citet{conselice2003} and measure the concentration within 1.5 Petrosian SMA as 
\begin{equation}C=5\log{r_{80}/r_{20}}\end{equation} \citep{bershady}, with the radii containing 20\% and 80\% of the light. 

\subsection{Ellipse fitting}
We run \textsc{iraf}/ellipse to measure light profiles, profiles of ellipticity and position angle, as well as 
parameters for non-elliptical isophotal shape.
For that purpose we fix the center of the galaxy, while  leaving the position angle and ellipticity as  free fitting parameters. 
In the case of the isophotal shapes we fit the isophotes in steps of 1.1 pixels in order to get  detailed profiles, whereas
all the other fits are done with a bigger step size, resulting in less noisy measurements. 

We also perform ellipse fitting for the two-dimensional decomposition models in Section \S\ref{section:fitting}. 
These measurements { are} done using the same apertures for  the observations and the models in order to be directly comparable.
We choose to measure  the observations, the models,  and the individual components along the isophotes obtained by fits to the models. 

\subsection{Isophotal shape}
\label{section:boxiredux}
\begin{figure}
\begin{center}
\includegraphics[height=\linewidth, angle=-90]{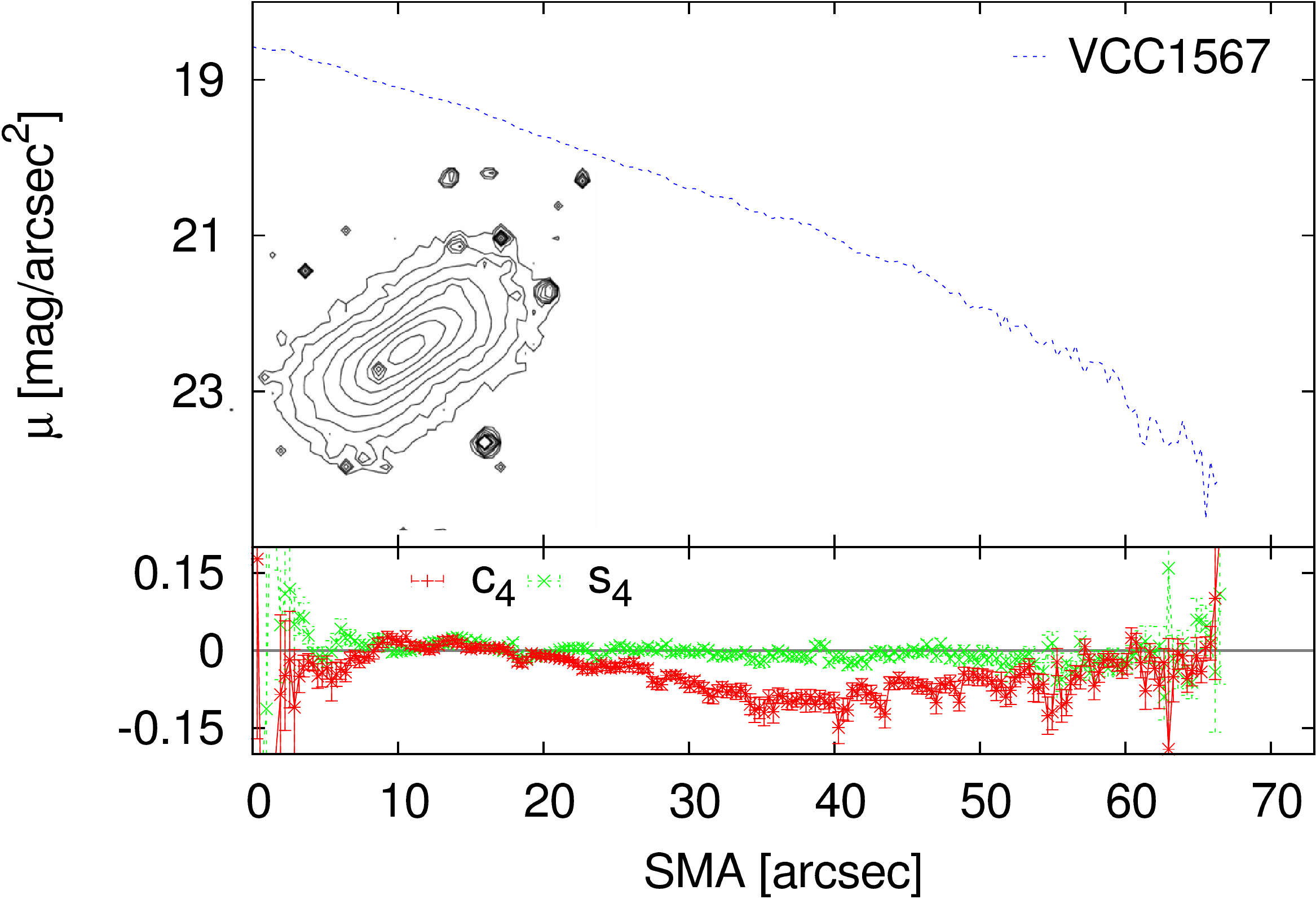}\\
\includegraphics[height=\linewidth, angle=-90]{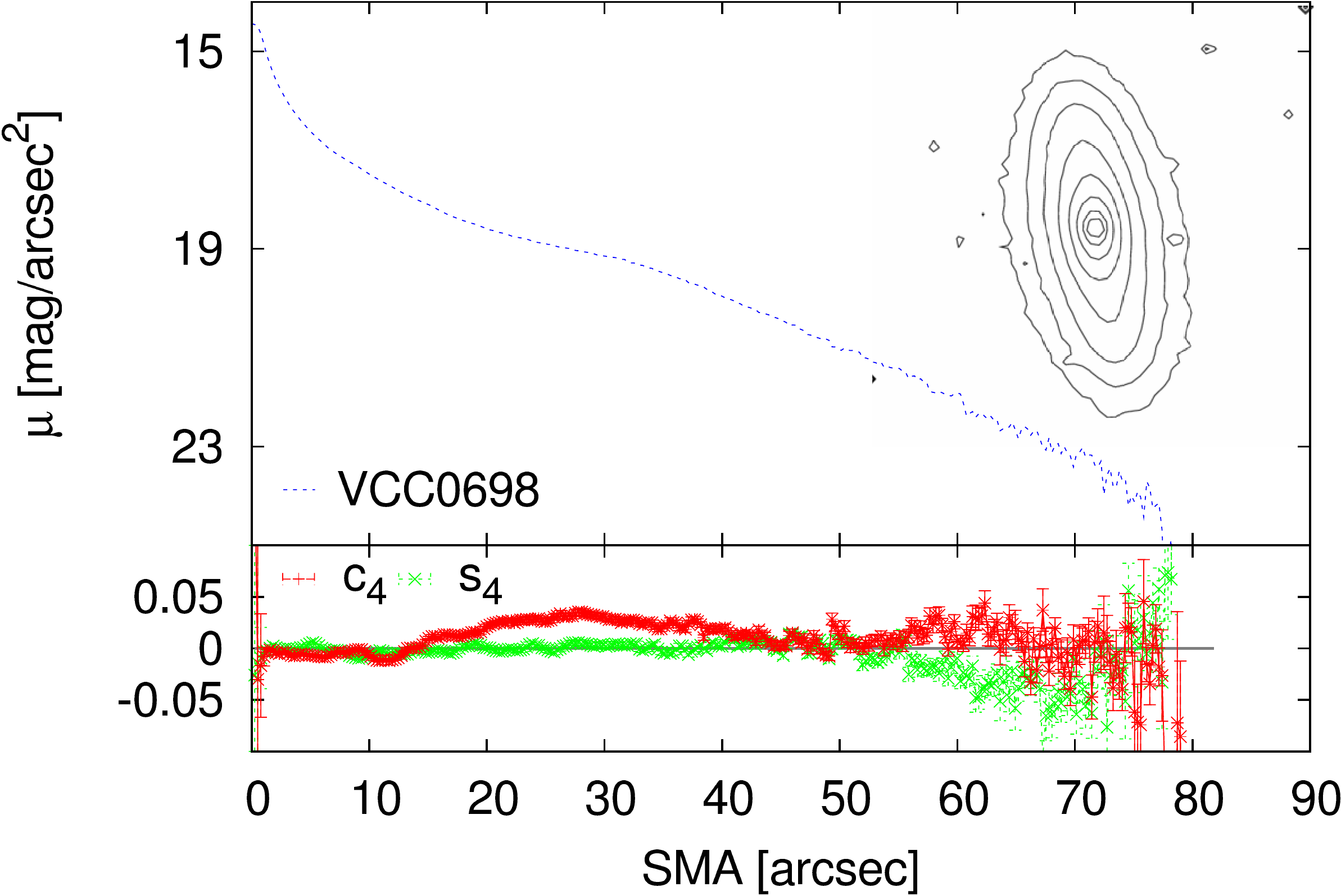}
\caption{Illustration of boxi- and diskiness. One example of a boxy galaxy (VCC1567, \textit{upper panel}) and a disky galaxy (VCC0698, \textit{lower panel}) is shown. For both the light profiles \textit{(blue)} are plotted together with the profiles of $c_4$ \textit{(red)} and $s_4$ \textit{(green)}. The insets show the isophotes, the outermost isophote correspond to $a\approx44$ and $a\approx51$ arcsecs, respectively. 
The two profiles are very consistent with those obtained in a similar manner with SDSS data.
The median values in the inner part ($1\farcs25<a<r_e$), outer part ($r_e<a<2r_e$),  and for the whole galaxy are $\bar{c_4}= -0.0158, -0.0794, -0.0454$ and $\bar{c_4}=-0.0039, 0.0270, 0.0129$, respectively, for VCC1567 and VCC0698.  VCC1567 has three features with integrated values  $\tilde{c_4} / r_e=-0.028,0.003,-0.088$, VCC0698 has two $\tilde{c_4} / r_e=-0.004,0.035$. \label{fig:boxy}}
\end{center}
\end{figure}

\textsc{iraf}/ellipse fits deviations of the isophotes from an elliptical shape  using a Fourier series:
\begin{equation}I(\theta)=I_0+ \sum_{n=1}^\infty ( C_n \cos n \theta + S_n \sin n\theta )\,.\end{equation}
Since  the nomenclature in \textsc{iraf} and the literature is not uniform, we adapt the notation of \citet{Toloba:2011p3986} with $C_n$ and $S_n$ for the cosine and sine terms (in \textsc{iraf}: $B_n$ and $A_n$, respectively).
The coefficients are normalized by the length of the SMA and the gradient of the profile, $c_n=C_n/ ( a \cdot dI/dr )$ and $s_n=S_n/ ( a \cdot dI/dr )$,  
to quantify the  distance of the isophote to the ellipse relative to its size. The most interesting coefficient is $c_4$ \citep{Carter:1978wg}, which 
is the parameter describing the boxiness ($c_4<0$) or diskiness ($c_4>0$) of the isophote (Fig.\ \ref{fig:boxy}). In the following  the parameter is referred to as boxiness.

The sine coefficient of the same order describes deviations of the isophote, not along and between the major and minor axis, but rotated by $22.5^\circ$.
For a pure boxy or disky shape this coefficient should remain close to $s_4\approx0$. The isophote can also have a more complex deviation from an elliptical shape, in which case the measurement can be less confidently attributed to a boxy or disky shape. For most of the galaxies  the profiles  are noisier than the examples in Fig.\ \ref{fig:boxy}: in some cases the values  scatter for large parts of the profile around 0, consistent with the errors associated with the measurements. 

Moreover, some galaxies  have both boxy and disky sections.
Therefore, boxiness values are assigned to the  galaxies in two ways: \textit{(a)}  the median value $\bar{c_4}$  for the inner ($1\farcs25  <a<r_e$) and  outer ($r_e<a<2r_e$)  part of the profile, and \textit{(b)} integrated values $\tilde{c_4}$ over boxy or disky features in the ${c_4}$ profiles:
\begin{equation}
\tilde{c_4} = \int_{\textrm{\footnotesize feature}} c_4(r) dr\,.
\end{equation}
These integrals have the dimension of an angular size and are normalized to the angular size of the galaxy ($r_e$) in the later figures.
The features are found by searching for sections in the profile where the sign of $c_4$ does not change for at least three consecutive measurements, $|c_4|$ is larger than its error as given by \textsc{iraf}/ellipse, and  larger than $|s_4|$.
We calculate the formal error of $\tilde{c_4}$ as the square root of the quadratic sum of the single measurement, multiplied by the bin width of the numerical integration. 
This way  multiple  disky or boxy values can be assigned to different regions in a single galaxy.

\subsection{Two-dimensional Decompositions}
\label{section:fitting}
{ 
Traditionally one-dimensional profile decompositions are used in structure analysis both
for bright and dwarf galaxies.  Although one-dimensional azimuthally averaged profiles themselves are
not very sensitive for detecting non-axisymmetric structures in galaxies
(\S\ref{sec:procedure}, \citealt{Byun:1995et,deJong:2004dw,Laurikainen:2005gm,Noordermeer:2007kp,2009ApJ...696..411W}),
 combined with inspection of the radial profiles of the ellipticities and position angles, they have been successfully used to
identify multiple components, including those of the elliptical galaxies.

However, in this study we have chosen the multi-component two-dimensional 
decomposition approach.  That enables functional fitting, not only of bulges and disks, 
but also of the non-axisymmetric structures like bars, using the complete
information of the two-dimensional light distributions.  Above all, this allows us to derive
quantitative parameters for the different structure components, in a
similar manner as has been obtained for the bulges and disks for
bright S0s and spirals.  A parametric comparison of the structures of
the sample galaxies with those obtained for the bright galaxies is one
of the goals of this paper.

 It has been shown for bright galaxies by
\citet{2006AJ....132.2634L} that the flux of the non-axisymmetric
structure can be erroneously mixed with the flux of the bulge if not
fitted separately. This is the case both with the 1D and 2D
decomposition approaches. For the dwarf galaxies this problem was
pointed out by  \citet{Mcdonald:2011p4445}, who speculated that omitting
the non-axisymmetric structures some light might erroneously be
attributed to some other, mostly the inner component. 
They estimate the effect on the parameters of the other components to be 10\% to 20\%.
Since in dwarf galaxies the multiple components are expected to be much more subtle
than in the bright galaxies, they might be even more difficult to see
in the azimuthally averaged one-dimensional profiles.  As an example of the
strength of the two-dimensional approach VCC0940 is shown in Figs.\ \ref{fig:VCC0940_special} \& \ref{fig:fitting}.
 Although the bar is hardly visible in the azimuthally averaged profile
it is clearly manifested in the two-dimensional profile, in which all the pixels of
the image are shown (Fig.\ \ref{fig:VCC0940_special}). In the two-dimensional decomposition
approach these pixels can be functionally fitted.
}

Our two-dimensional multicomponent decomposition approach was  introduced in \citet{janz+2012}, but will be described in some detail here.
The fitting is carried out with \textsc{galfit} 3.0 \citep{Peng:2010p4252}.
 \textsc{galfit} minimizes the $\chi^2$  to find the optimal solution for a given model.
 It uses an algorithm that tries to keep the probability of  falling into a local minimum of $\chi^2$ in the parameter space small.  For the  simple models  the starting values  are chosen based on the photometric parameters of the whole galaxies, while for the two-component models they are based on reasonable assumptions and improved if needed.
  It is assumed that the ellipticity and position angle of each individual component are constant with radius.
 In the case of further features such as bars  (see below) the component is first identified in the profile or in the residual image and  starting values are then chosen accordingly.
  For calculating the $\chi^2$s we provide \textsc{galfit} with the $\sigma$-images (\S\ref{section:reduction}).
  We use masks for fore- and background sources and  assume a 
 Gaussian PSF 
 with the measured FWHM.
 During the fitting process the model is convolved with the PSF image. 
 
 The decompositions of highly inclined galaxies are not considered reliable, unless the galaxy is seen exactly edge-on \citep{Laurikainen:2010p4602,2012ApJS..198....2K}. 
 Thus we limit this analysis to galaxies with axis ratios larger than the corresponding value of a flat disk inclined by $65^\circ$, $b/a>\cos 65^\circ$, which excludes 14 galaxies from this analysis.
 Furthermore, for 8 galaxies we did not find reliable decompositions and therefore they are excluded as well.\footnote{Note that,  that the excluded galaxies are not likely to be simple galaxies, so that their inclusion would lower the fraction of one-component galaxies. VCC0575 has possibly even more than two components; VCC0698 is probably highly inclined; VCC1178 and VCC1745 likely have lenses, which are not well-fitted with Ferrers functions; VCC1196 probably has a bar and some residual structures that might remind of shells; VCC1567 is probably seen edge-on and is extremely boxy (see Fig.\ \ref{fig:boxy}, the values come even close to the galaxy studied by \citet{leda}); VCC1297 has a small angular size, the profile looks overall like a S\'ersic profile, but there is a strong structure in the residual image, \citet{Ferrarese:2006p586} report an increasing ellipticity to the galaxy center that may be related to the double nucleus identified by \citet{lauer}. }

\subsubsection{The model functions}
The following functions were used for the profiles: 
\begin{itemize}
\item $\delta$-function for point-like sources.
\item  Exponential function:
\begin{equation}I(r)=I_0\exp\left(- {{r}\over{h}}\right)\,,\end{equation}
with the central intensity $I_0$, the scale length $h$, and $r_e=1.678 h$.
\item S\'ersic function \citep{1963BAAA....6...41S}:
\begin{equation}I(r)=I_e \exp\left( -b_n \left[ \left(r\over{r_e}\right)^{1/{n}} - 1 \right] \right)\,,\end{equation}
with the intensity  $I_e$ at $r_e$,  the S\'ersic index $n$ for the steepness of the profile ($n=1$  and $n=4$ corresponding to an exponential and de Vaucouleurs profile, respectively), and the $b_n$ parameter that depends on $n$. The dependency can be expressed by gamma functions \citep{ciotti} and  approximated by $b_n=1.992n-0.3271$ for $0.5<n<10$ \citep{capaccioli}. The purpose of $b_n$ is to ensure that $I_e$ is the intensity at the half-light radius.
\item  Modified Ferrers function \citep{binney_tremaine}:
\begin{equation}I(r)=I_0 \left[1-(r/r_{\textrm{out}})^{2-\beta}\right]^\alpha\,.\end{equation}
This function features a nearly flat core, with a central slope of $\beta$, and a truncation at $r_\textrm{out}$ with the sharpness $\alpha$. Beyond $r>r_\textrm{out}$ the profile is not defined and the function has a value of 0.
The flat core and sharp cutoff makes it especially useful for fitting bars.
\end{itemize}
Except for the $\delta$-function, the position angle and axis ratio are two additional fitting parameters. 
Each component is constrained  to have elliptical isophotes and constant position angle and ellipticity.
In the following we will use the term simple model for a model that uses a single S\'ersic function or a combination of a S\'ersic function and a $\delta$-function for the nucleus. 

\subsubsection{Fitting procedure}
\label{sec:procedure}
First the following models are fitted to the galaxies, listed in the order of increasing complexity: 
\begin{itemize} 
\item a one-component S\'ersic model,  
\item a combination of a S\'ersic model and a point-source,  and
\item  a two-component  model with the inner component fitted by a S\'ersic function and the outer component with an exponential function.
\end{itemize}
{ Although typically the outer component is fitted with an exponential, this condition was relaxed in cases, in which the outer profile is obviously  not exponential. In such cases a S\'ersic function is used with the S\'ersic parameter $n$ as a free fitting parameter. }
 The position angle and the axis ratio of the outer component in the latter model are fixed to an average value of the outermost isophotes.

These basic  models are then compared  and visually inspected using \textsc{galfidl}.\footnote{H.\ Salo, \url{http://cc.oulu.fi/~hsalo/galfidl.html}.}
Most important for our evaluation were the residual images after subtracting the model, the profiles of the two-dimensional light distributions of the observation, the whole model, and the models for the different components. By 
two-dimensional we mean that the surface brightnesses of all pixels are shown as a function of their angular distance to the galaxy center \citep{Laurikainen:2005gm,AlexeiGadotti:2008hz}.
 \begin{figure}
\begin{center}
\includegraphics[height=\linewidth, angle=90]{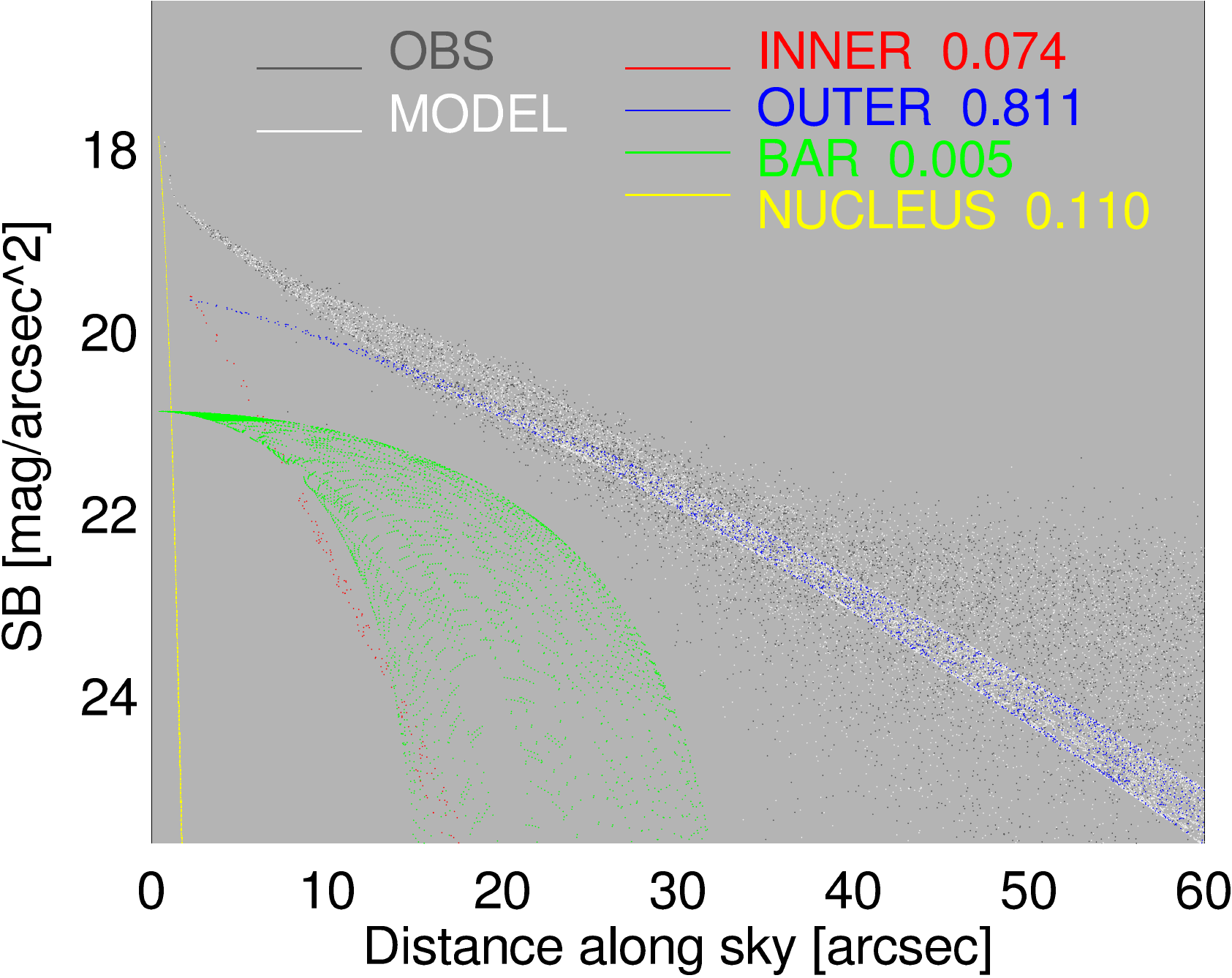} \\
\includegraphics[height=\linewidth,angle=-90]{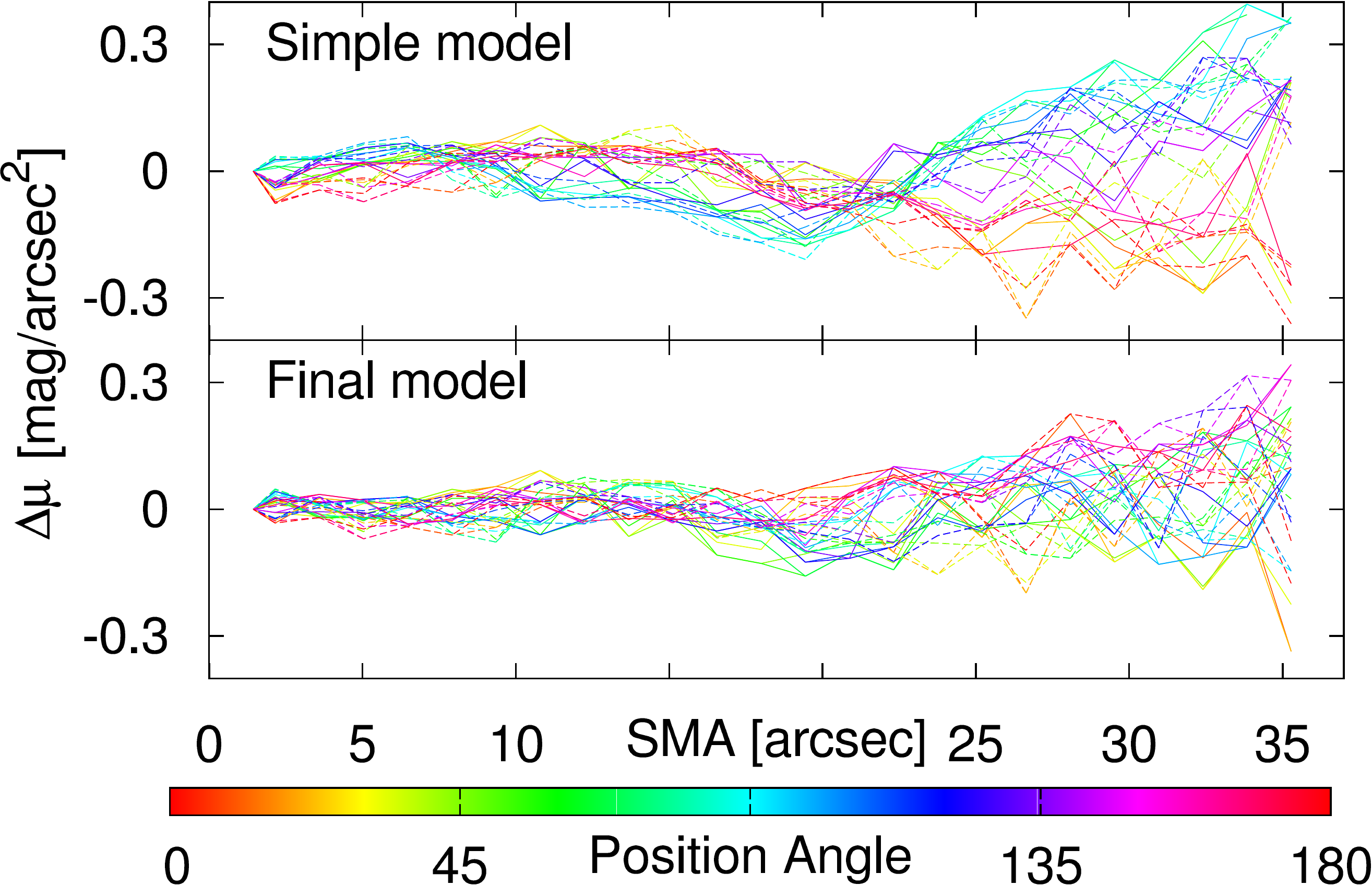} 
\caption{\textit{Upper panel:} Profile of VCC0940 and its final model displaying all pixels (an image of the galaxy is found in Fig.\ \ref{fig:fitting}). The total model (white) is plotted twice: without noise and including noise according to the image. The model without noise is not a line due to the apparent flattening of the light distribution. It is visible as an area filled with white points. At distances larger than $30\arcsec$  the model is dominated by the outer component and thus blue points showing the disk component are plotted on top of the complete model. The model including the noise has a broader distribution at larger radii, due to an approximately constant noise, displayed on the logarithmic magnitude scale. These white points spread around the dark points of the observed pixels.
Colors indicate different components; the number after the label indicates the fractional contribution to the total flux.
 \textit{Middle and lower panel:} residual profiles  along different position angles (displayed with different colors, see colorbar above the panels, solid and dashed lines show the profiles in sectors rotated by $180^{\circ}$) for the simple and final model; plotted is the difference of surface brightnesses $\Delta\mu$.\label{fig:VCC0940_special}}
\end{center}
\end{figure}
The upper panel of Fig.\ \ref{fig:VCC0940_special} illustrates the information  contained in such profiles. 
For a perfectly round component all the pixels at a certain distance from the center will have the same surface brightness, while for increasing apparent flattening of the  component the range of the surface brightness at this distance will also increase, thus making the wedge-shaped profile. 
In Fig.\ \ref{fig:fitting}, displaying the same galaxy, { but showing the azimuthally averaged one-dimensional profiles instead,} the bar is hard to detect, both in the image and in the azimuthally averaged profile. The deviation of the model from the observation in the azimuthally averaged profile is less than 10\% (cf. \citealt{2005AJ....130..475A}). However, in the residual image the bar is clearly visible. Moreover, at about 10-20 arcsecs the bar thickens the pixel distribution in the two-dimensional profile  because of the ellipticity maximum (Fig.\ \ref{fig:VCC0940_special}, upper panel;  generally, the distribution becomes thicker at larger radii due to the nearly constant noise and the logarithmic scale). 
{ Notice that in Fig.~\ref{fig:fitting} the observations are measured in the same apertures as the models, in order to allow a direct comparison of the profiles and not to mix it with effects of differences in the isophotes. The observed profile measured along the simple model isophotes is equivalent with determining the profile with fixed ellipticity and position angle. Along the final model isophotes the profile comes very close to a profile determined along the isophotes of the observation.}

That the overall shape of the profile is not well represented by a simple model is manifested  as a clear structure in the residual  profile (Fig.\ \ref{fig:VCC0940_special}, middle panel).  
In addition to the overall behavior, the residual profiles in  sectors at  different position angles 
deviate systematically  from each other. This behavior is due to the non-axisymmetric bar. 
  All line up nicely in the final model (lower panel), where the thickened pixel distribution in the bar region (upper panel) is accounted for.
Additional guidance for the model evaluation are  the profiles of the orientation parameters (Fig.\ \ref{fig:fitting}).  
 In less clear cases we flag the decomposition as ``uncertain'' and in doubtful cases we opted for the simpler model.

In those cases where none of the previous models is satisfactory, we try  one of the following changes: 
\begin{itemize}
\item vary the starting values (since \textsc{galfit} can end up in a local minimum of $\chi^2$ in the parameter space when degeneracies occur with the increasing number of components). Or,  in a few cases,  
\item use  an outer component with S\'ersic $n < 1$ (instead of an exponential), or 
\item use different centers for the different components.  
\end{itemize}
A typical profile type turns out to be the one with a ``down-bending'' part, that is, a component with a flat inner profile and a steep outer decline causing an edge, similar to the profile of a bar, but with a round shape.
Since the axis ratio is close to unity this can be seen as a nearly round extended excess of light in the residual image, with an insufficient modeling of the central flux.
This kind of component we call ``lens'' (see \S\ref{sec:groups},  and also \citealt{binggeli_cameron})  in analogy to lenses frequently detected in bright early-type disk galaxies, which have similar profiles \citep{Kormendy1979,2013MNRAS.430.3489L}. We model it  with a Ferrers function{, in a similar manner as done for the bright galaxies in our comparison samples.}
Typically these components replaced the inner component, and only in a  few cases we  add it to the two-component model or use it in combination with an inner component.
Therefore, the galaxies with lenses may also be thought of as special cases of two component galaxies.

Finally,  if a bar is revealed in the model subtracted images, we fit the bar with an additional Ferrers function.\footnote{In two galaxies with a lens we find a small bar. These are not fitted, since we want to avoid to have overly complex models. }
The complete set of models with their residuals, light profiles, and profiles of the orientation parameters is shown for one galaxy, VCC0940, in Fig.\ \ref{fig:fitting}.\footnote{In a few cases, also outer components fitted by S\'ersic functions with $n>1$ were used, see Table 5.} 
In principle a small error in the sky subtraction can alter the profile at low surface brightnesses enough to change the shape of the profile, i.e.\ mimic a complex structure. We estimate the potential influence of  large scale variations by  the standard deviations of mean sky values, measured in small boxes  on the image.  
When evaluating the decompositions, 
we include new components only if the deviations of the profile shape exceed the uncertainties due to large scale background variations (the uncertainty is indicated as blue shaded areas in the profiles of Fig.~\ref{fig:fitting}).

\begin{figure*}
\begin{center}
\includegraphics[height=7cm, angle=0]{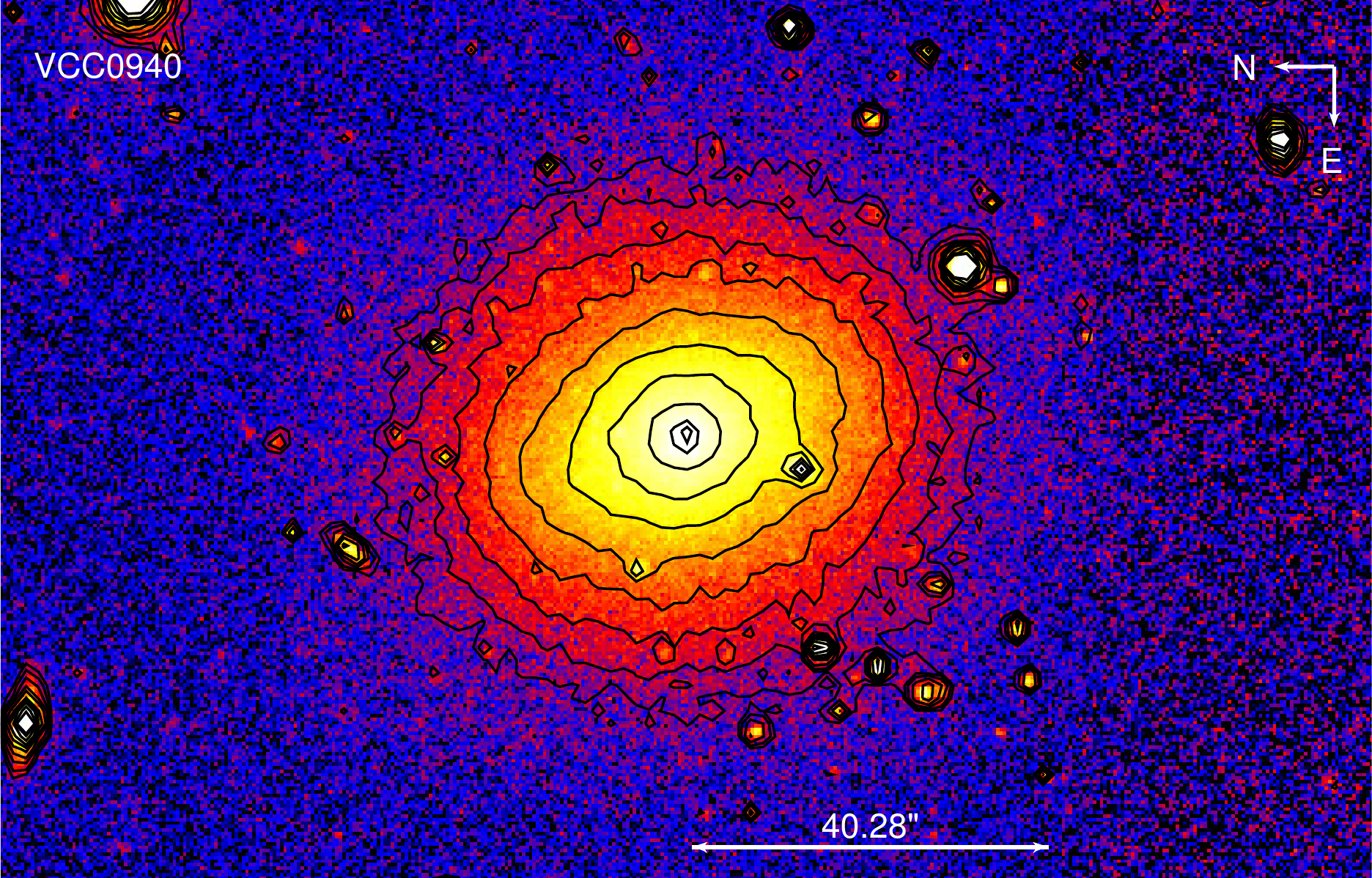} \\
\raisebox{3cm}{\includegraphics[height=3cm, angle=0]{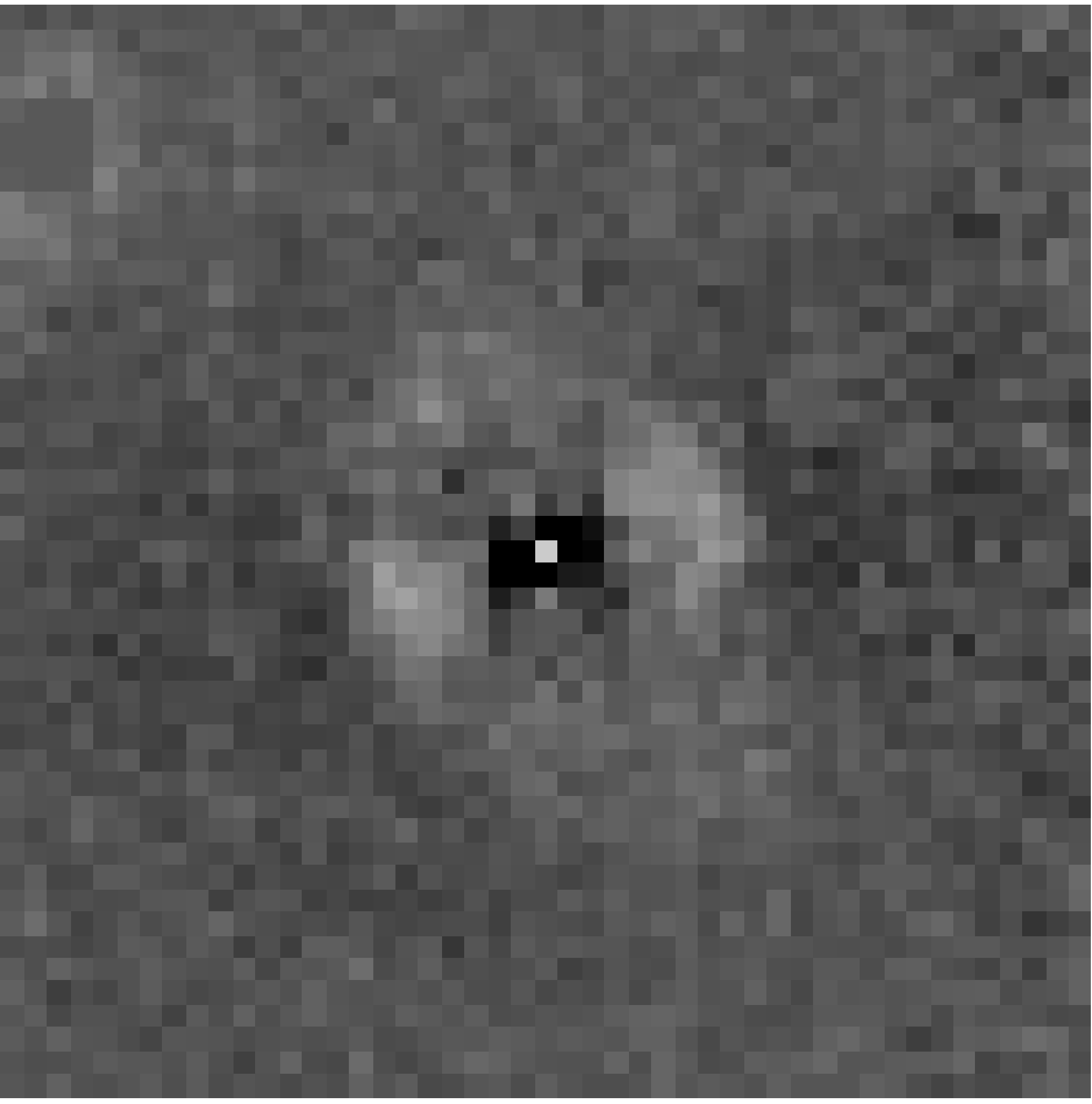}}
\includegraphics[height=10.2cm, angle=-90,origin=br]{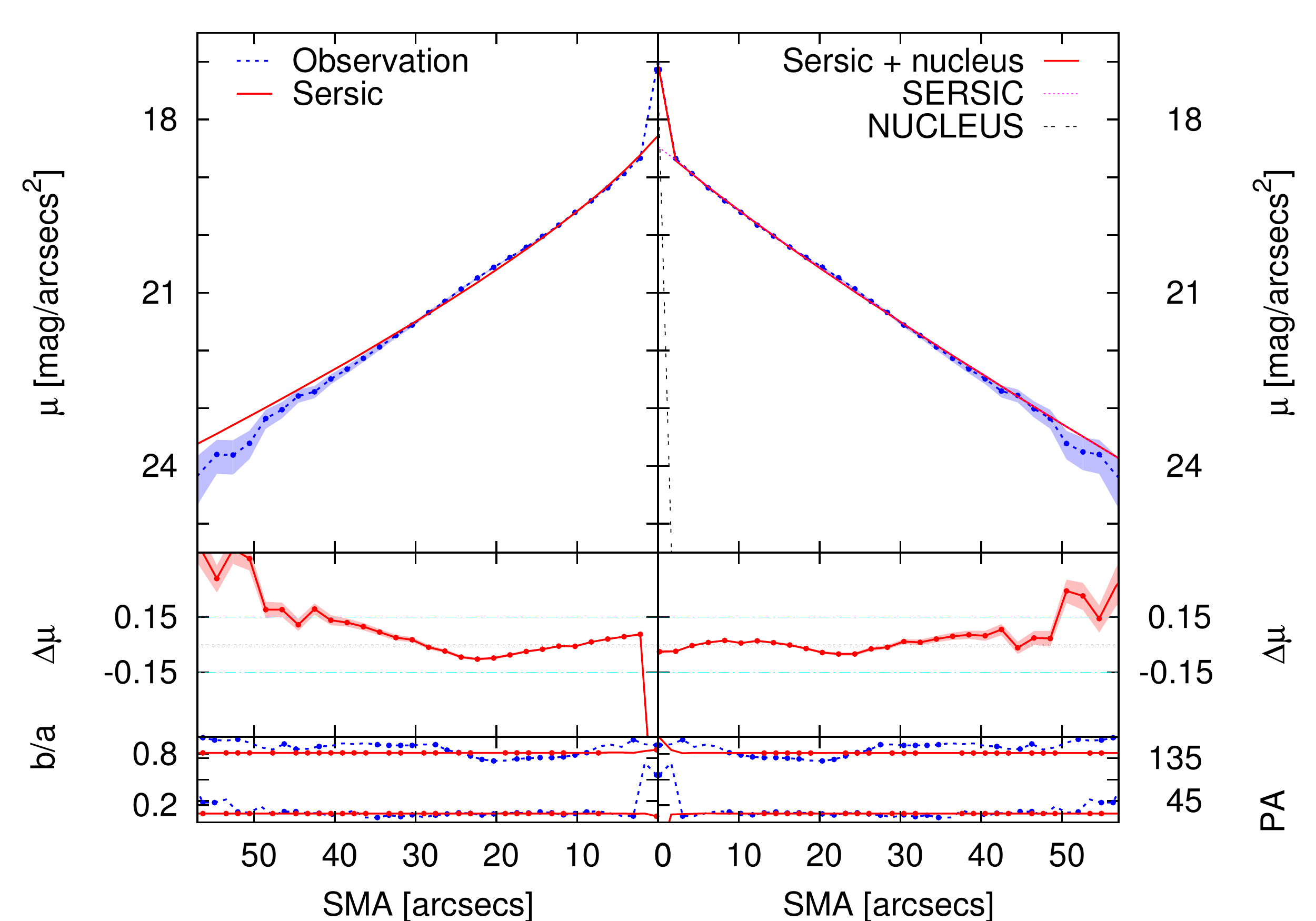}
\raisebox{3cm}{\includegraphics[height=3cm, angle=0]{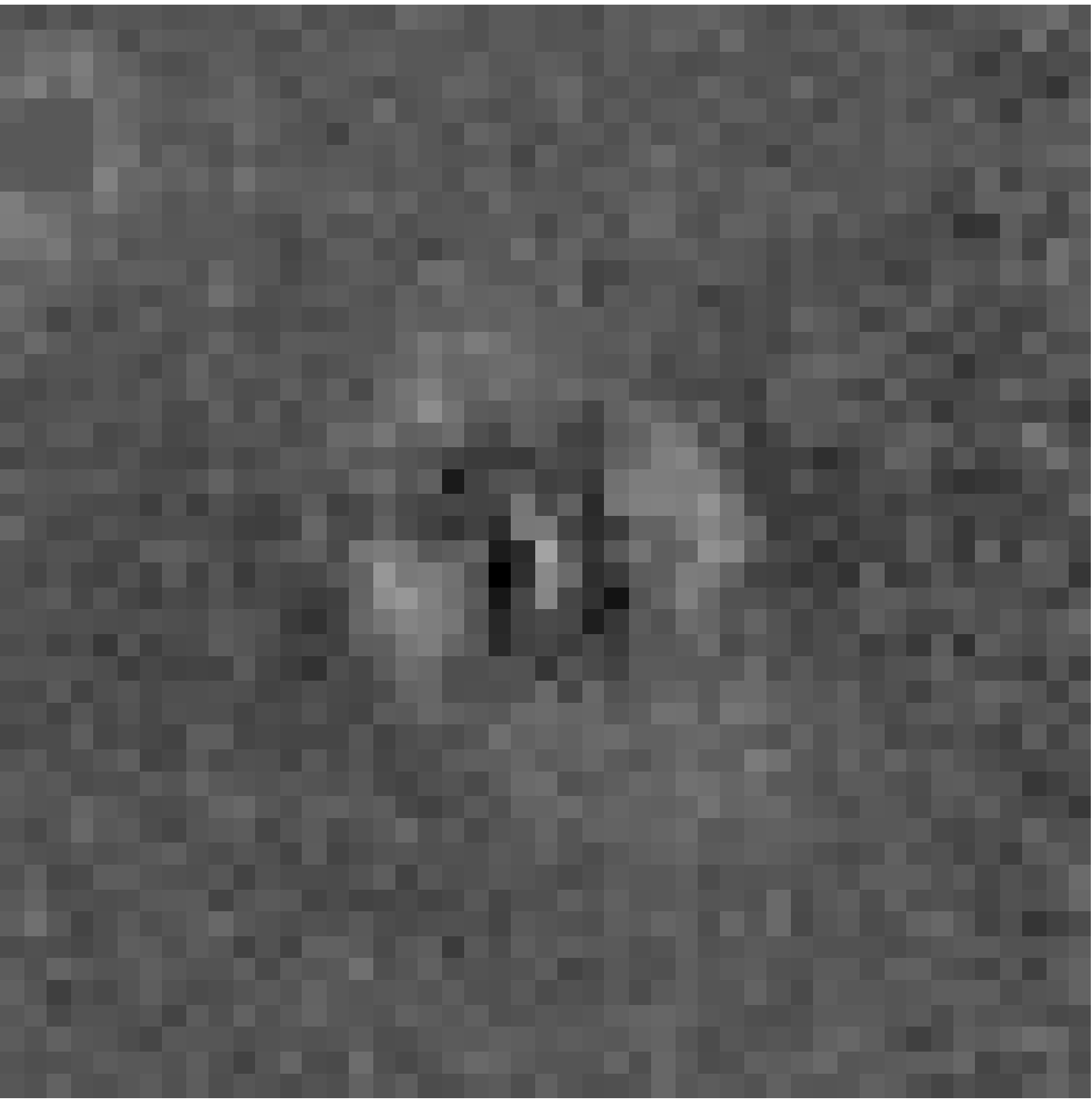}} \\
\raisebox{3cm}{\includegraphics[height=3cm, angle=0]{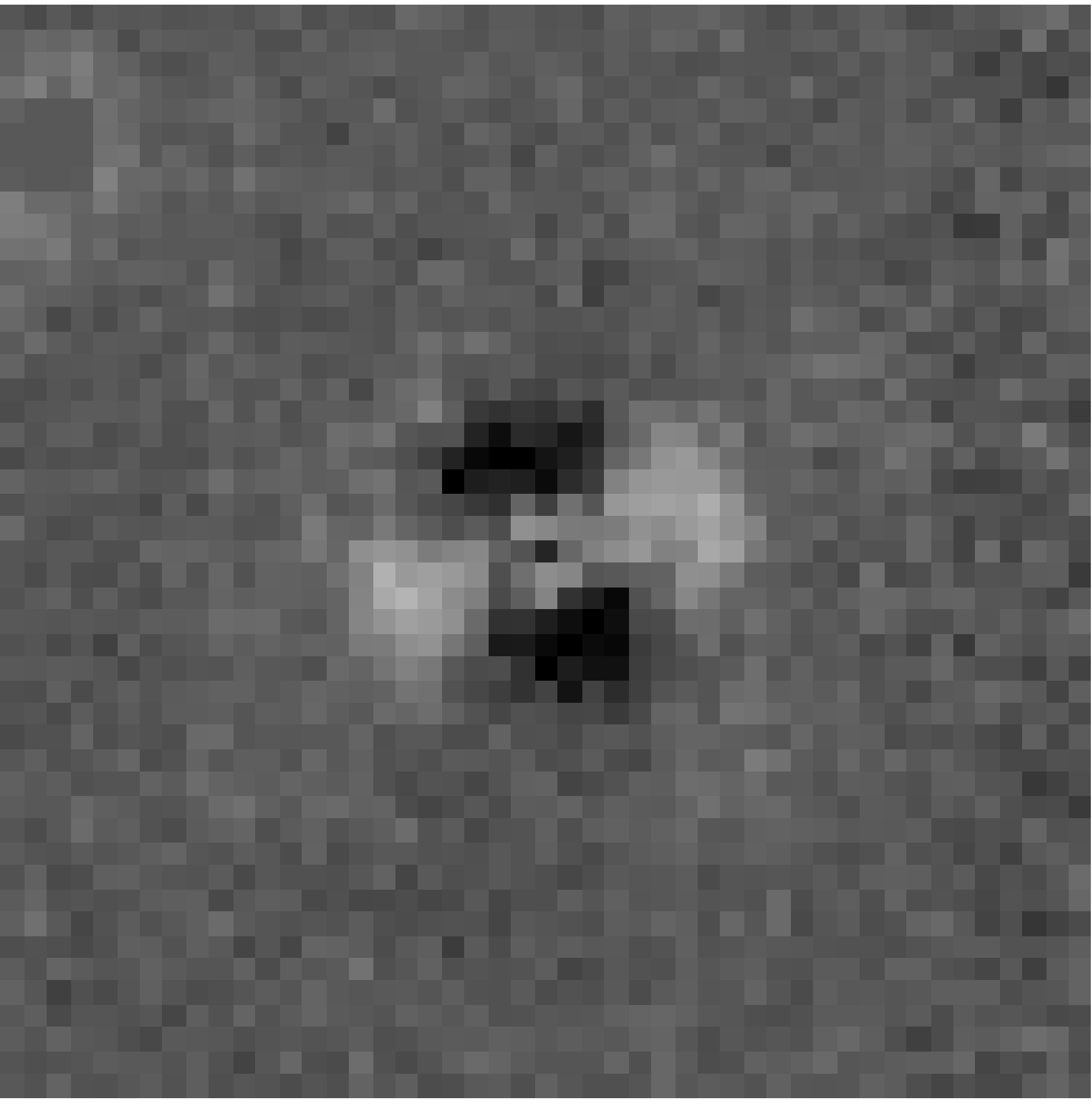}}
\includegraphics[height=10.2cm, angle=-90,origin=br]{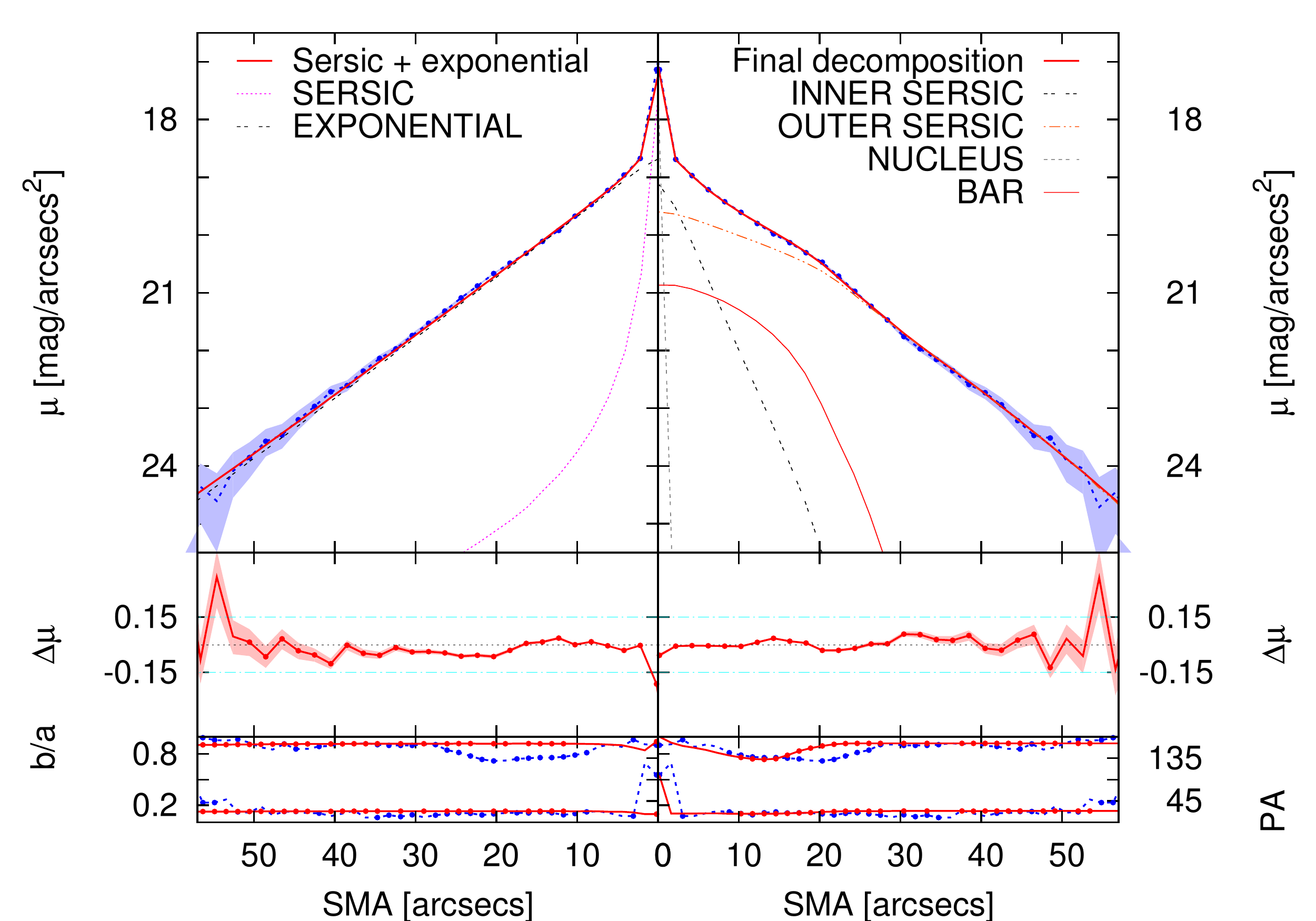}
\raisebox{3cm}{\includegraphics[height=3cm, angle=0]{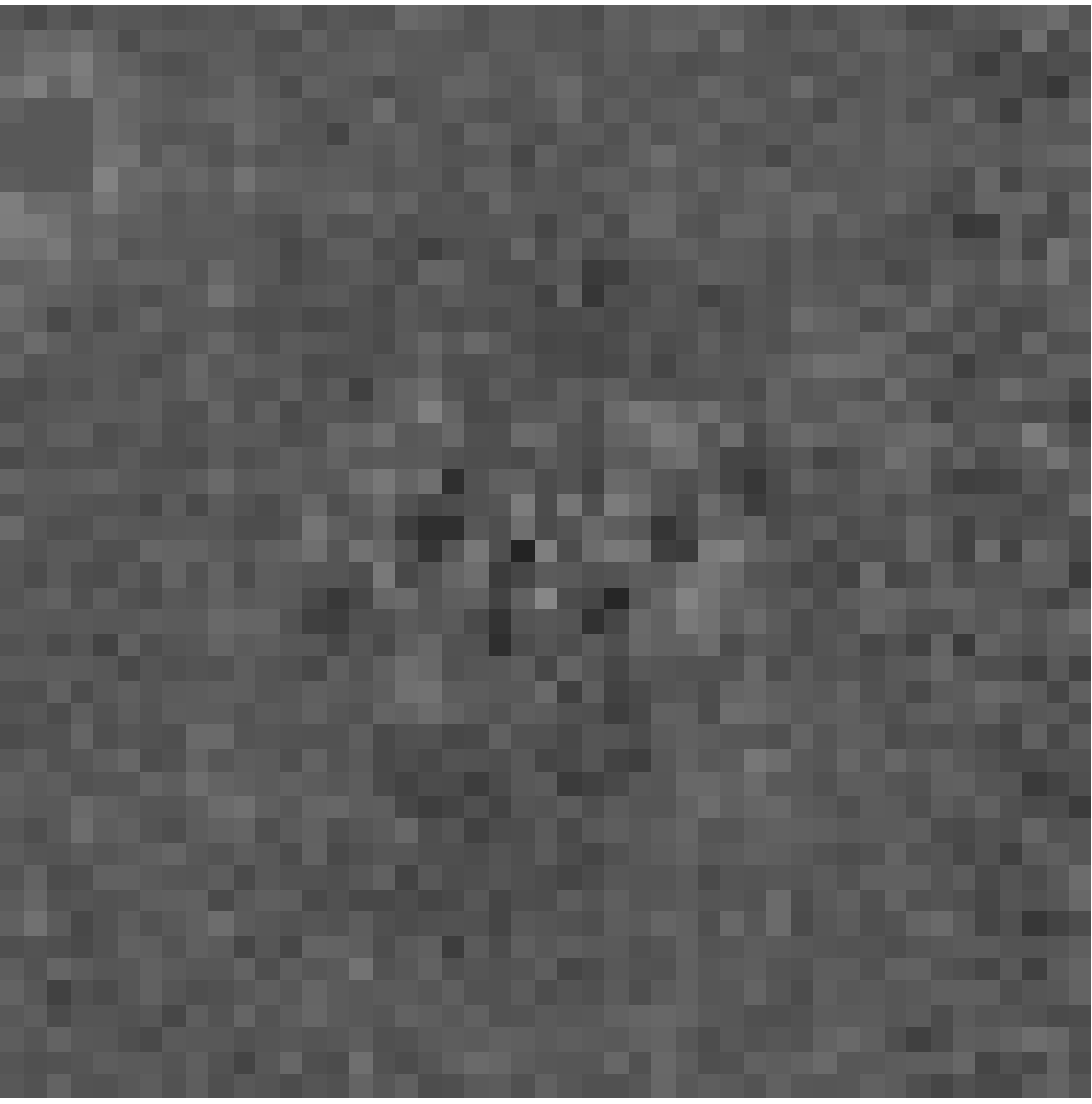}}
\caption{Top panel: image of VCC0940. The other panels illustrate the visual model evaluation in the fitting procedure: residuals,  azimuthally averaged light profiles, and 
orientation parameter profiles.  The simple models (one S\'ersic component and S\'ersic+nucleus) are shown in the middle panels. The bottom row displays the two-component model and the final model with a bar. 
 The observations are measured in the same apertures as the models, using the model isophotes, which can lead to slight variations in the observed profile displays between the panels.
The residual images display $\pm3\sigma$ on a linear scale, and are re-binned for the small scale display. The width of the residual stamps is about half an arcminute. In the bottom part of the profile panels the axis ratio $b/a$ \textit{(upper lines)} and the position angle (PA, \textit{lower lines}) are shown. The shaded regions  and error bars in the profiles and residual profiles illustrate the estimate of large scale background variations and the surface brightness uncertainties given by \textsc{iraf}/ellipse, respectively. 
 A similar presentation of models can be accessed for all galaxies in our sample as online-only figures and via \url{http://www.smakced.net/data.html}.
 \label{fig:fitting}}
\end{center}
\end{figure*}

Furthermore, the detection of  components, especially in the outer parts, obviously depends on the image depth. The integration time for an individual galaxy was chosen to reach a certain $S/N$ per pixel at  $2 r_e$, so that the galaxies with lower surface brightnesses were observed for longer times. Therefore,  we do not expect to have missed components of similar relative flux and extent at the faint end of the sample. The decomposition for a single galaxy might be different if yet deeper images were available. For example, in VCC0170  the outer part, with different orientation, is not visible in shallower images (see illustration in Fig.\ 1 of \citealt{janz+2012}). For VCC0490 we did a two component decomposition, with the outer component having a S\'ersic $n < 1$. However, on a very deep optical (white filter) WFI image \citep{Lisker+AN} this part looks more like a bump in the profile on top of an even more extended component.

\subsubsection{Quantification of fit quality}
\label{EVIRFF}
\begin{figure*}
\begin{center}
\includegraphics[width=0.32\linewidth, angle=-90]{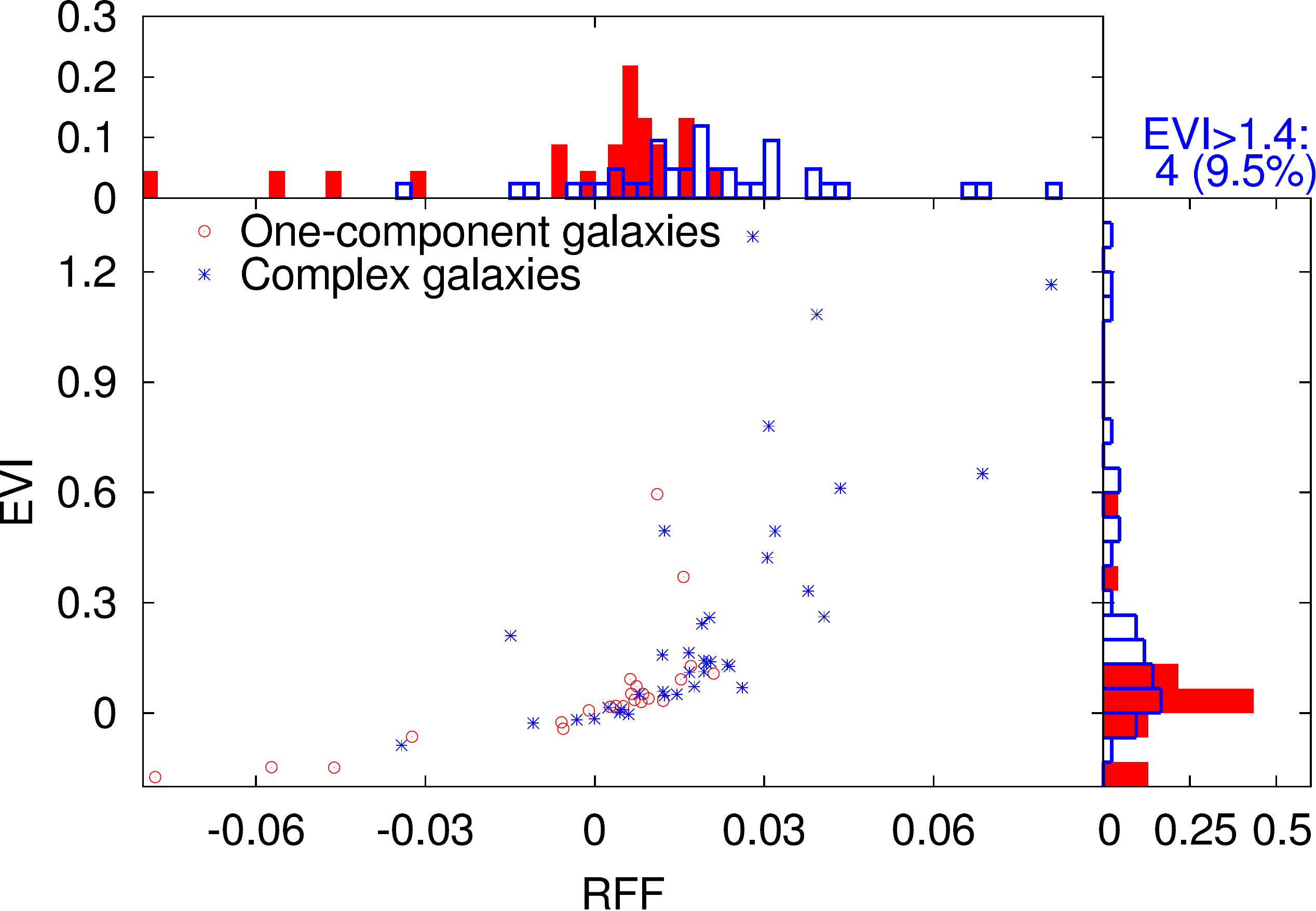} \ \ \ \ \ \ \ 
\includegraphics[width=0.32\linewidth, angle=-90]{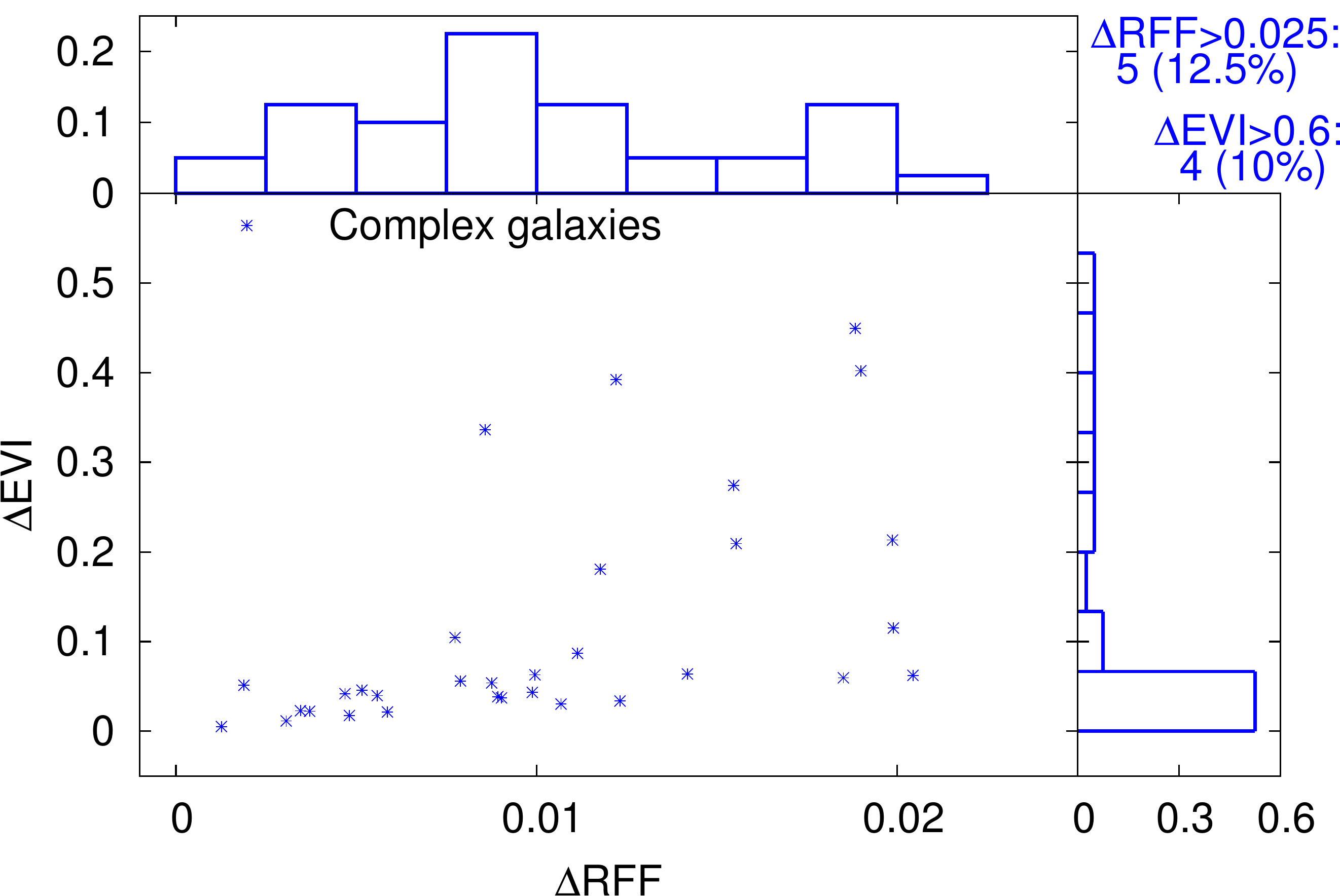}
\caption{RFF and EVI values for all galaxies when using a simple model \textit{(left panel; red -- one-component galaxies, blue -- complex galaxies)}, and the improvement from the simple to the final model for two-component galaxies,  galaxies with lens, and barred galaxies \textit{(right panel)}. For both indices is $\Delta$INDEX$=$INDEX$_\textrm{simple}-$INDEX$_\textrm{final}$. The small panels show histograms of distributions on  the corresponding axis.\label{fig:hoyos}}
\end{center}
\end{figure*}
 Although  the model selection in the decompositions is done in a visual manner, 
we seek to quantify the quality of the fits, and to quantitatively justify that for many galaxies 
in our sample a non-simple model is indeed needed.
We follow the approach of \citet{Hoyos:2011p4620} , who used the following two indices. 
The residual flux fraction (RFF) (introduced by \citealt{blakeslee+2006}) is defined as
\begin{equation}\textrm{RFF}={{\sum \left( \left|\textit{Res\,}\right|-0.8*\sigma \right)}\over{\sum \textit{Flux}}}\,,\end{equation}
where  $\left|\textit{Res\,}\right|$ is the absolute value of the residual flux, $\sigma$ is the $\sigma$-image, and \textit{Flux} is the observed flux. The summation is carried out over a chosen aperture. The excess variance index (EVI) is 
\begin{equation}\textrm{EVI} = {{1}\over{3}}\times\left( {{\sigma^2_R}\over{\left<\sigma^2\right>}}-1\right)\,,\end{equation} 
where $\sigma_R^2$ is the mean square of the residual and $\left<\sigma^2\right>$ the mean value of the $\sigma$-image squared, both again measured in a chosen aperture. The factor $1/3$ was chosen according to their data based on a visual examination.
The expectation values of both indices for a perfect model are zero.  
As \citeauthor{Hoyos:2011p4620} point out, RFF is close to zero for an extended low surface brightness galaxy even for cases where  in the model the inner part shows a significant deviation  from the observation. 
Therefore, they  introduced the EVI index, which is sensitive to the variations in the bright inner part of the galaxy.  
Fig.\ \ref{fig:hoyos} compares RFF and EVI for our simple models, using apertures  containing the outer isophotes of the galaxies. 
More than 85\% of the objects with RFF$>0.013$ are complex galaxies, increasing to 95\% with the additional criterion of EVI$>0.13$. The one-component galaxies have on average smaller values of RFF and EVI, and generally at least one of the indices is rather small. In the right panel of Fig.\ \ref{fig:hoyos} the improvements { (which are less dependent on the noise estimate)} in the final models are shown: $>95$\% of complex galaxies  show an improvement, i.e.\ a decrease, of both RFF and EVI from the simple to the final model.

In addition we also evaluated the behavior of the reduced $\chi^2$ and of the tessellated excess residuals, as detailed in Appendix \ref{section:areas}. Both quantifications support our visual choices for the necessity of  complex models. { However, we want to stress at this point that we consciously chose to make the decision about the final decomposition in a non-automated way. }

\clearpage
\section{Results}
\subsection{Structural types}
\label{sec:groups}
\begin{figure}
\begin{center}
\includegraphics[height=7.1cm,angle=-90]{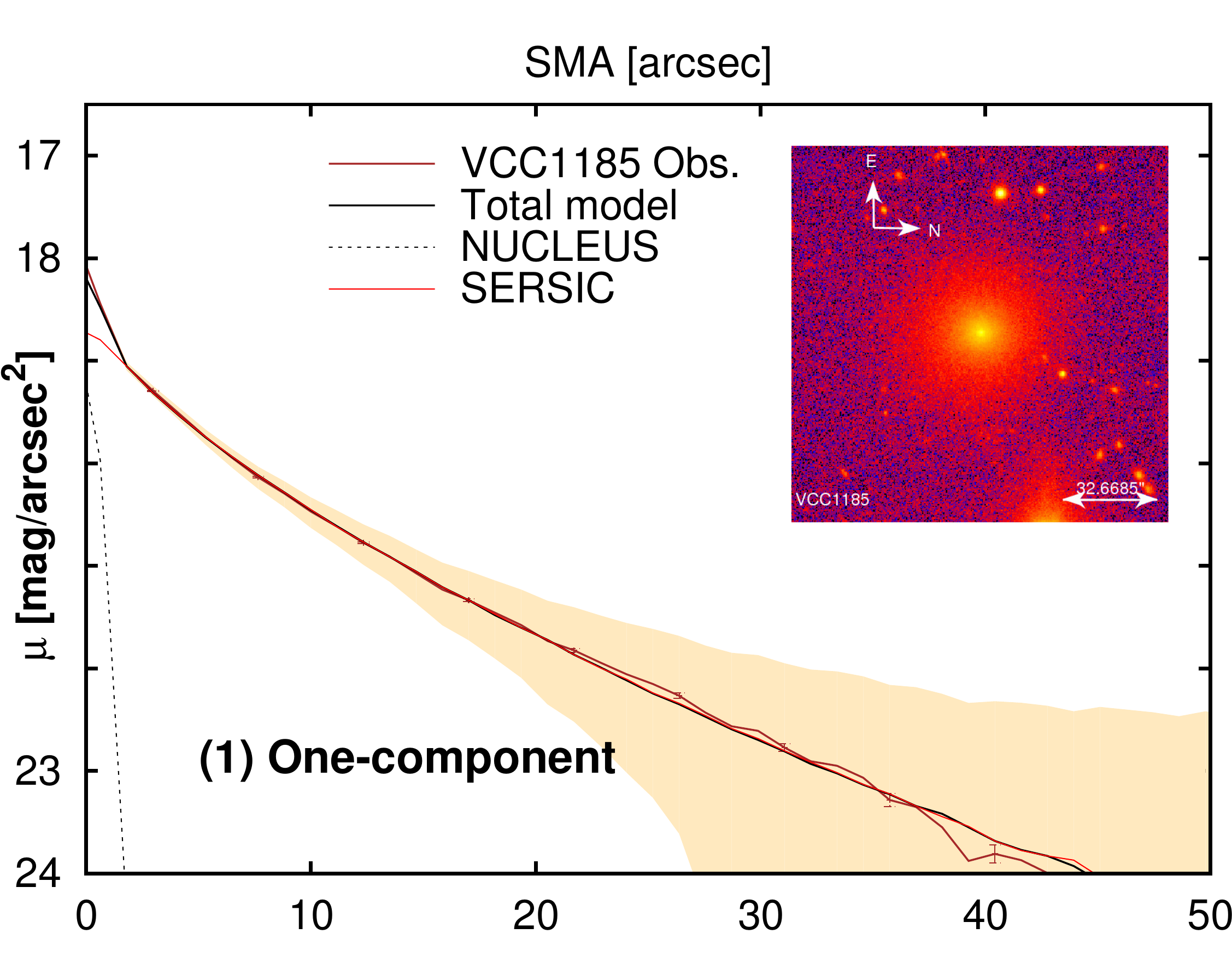} \\
\includegraphics[height=7.1cm,angle=-90]{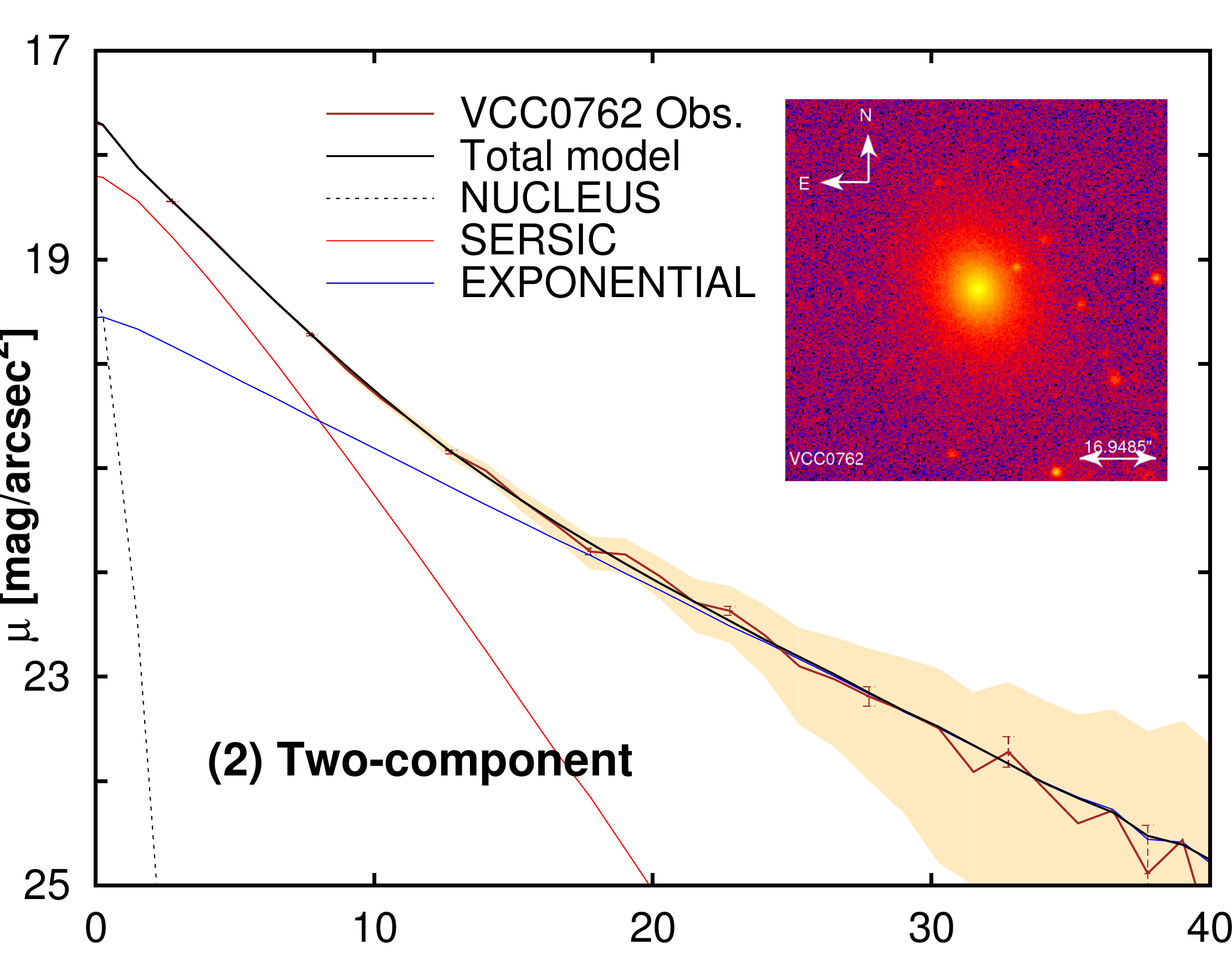} \\
\includegraphics[height=7.1cm,angle=-90]{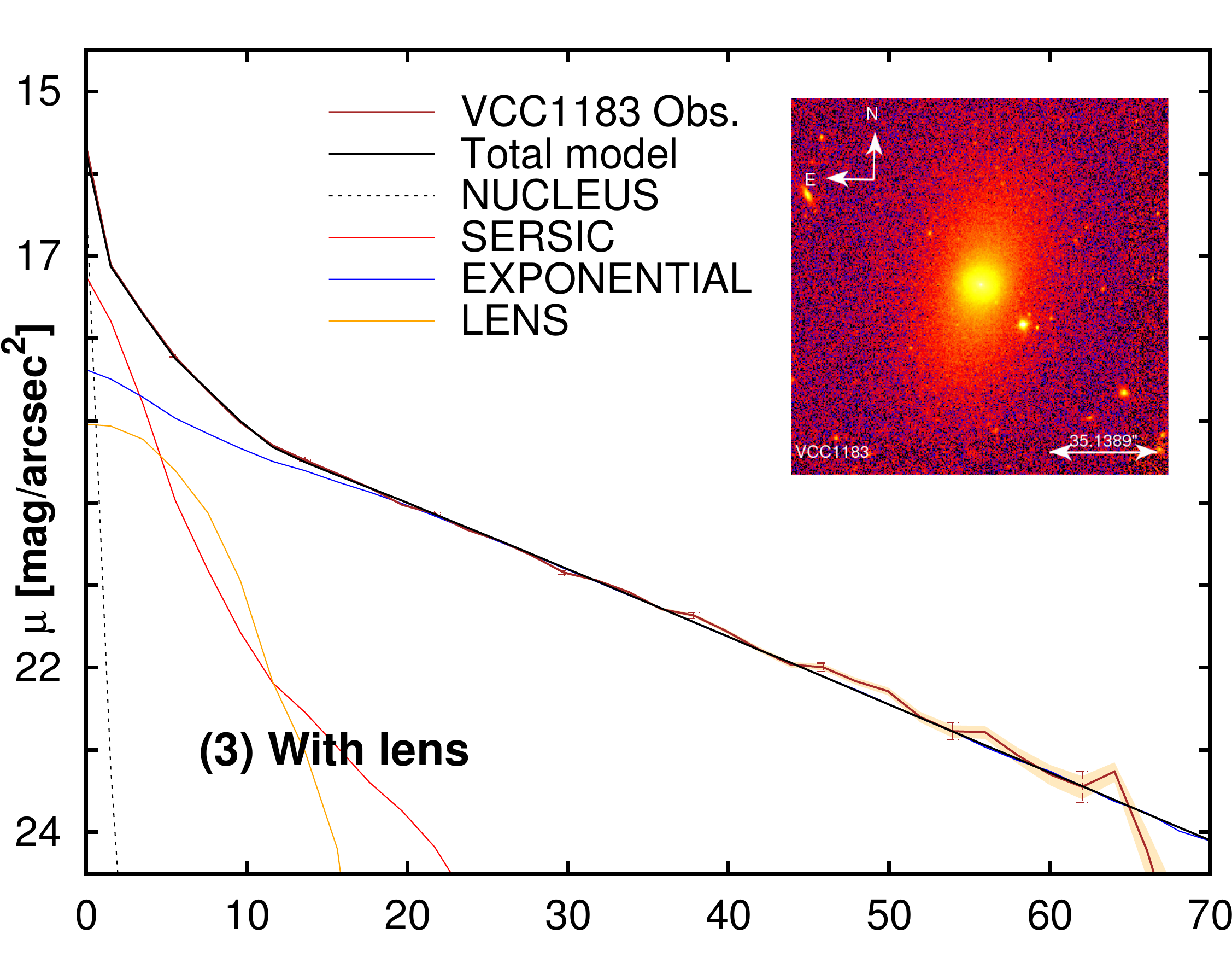} \\
\includegraphics[height=7.1cm,angle=-90]{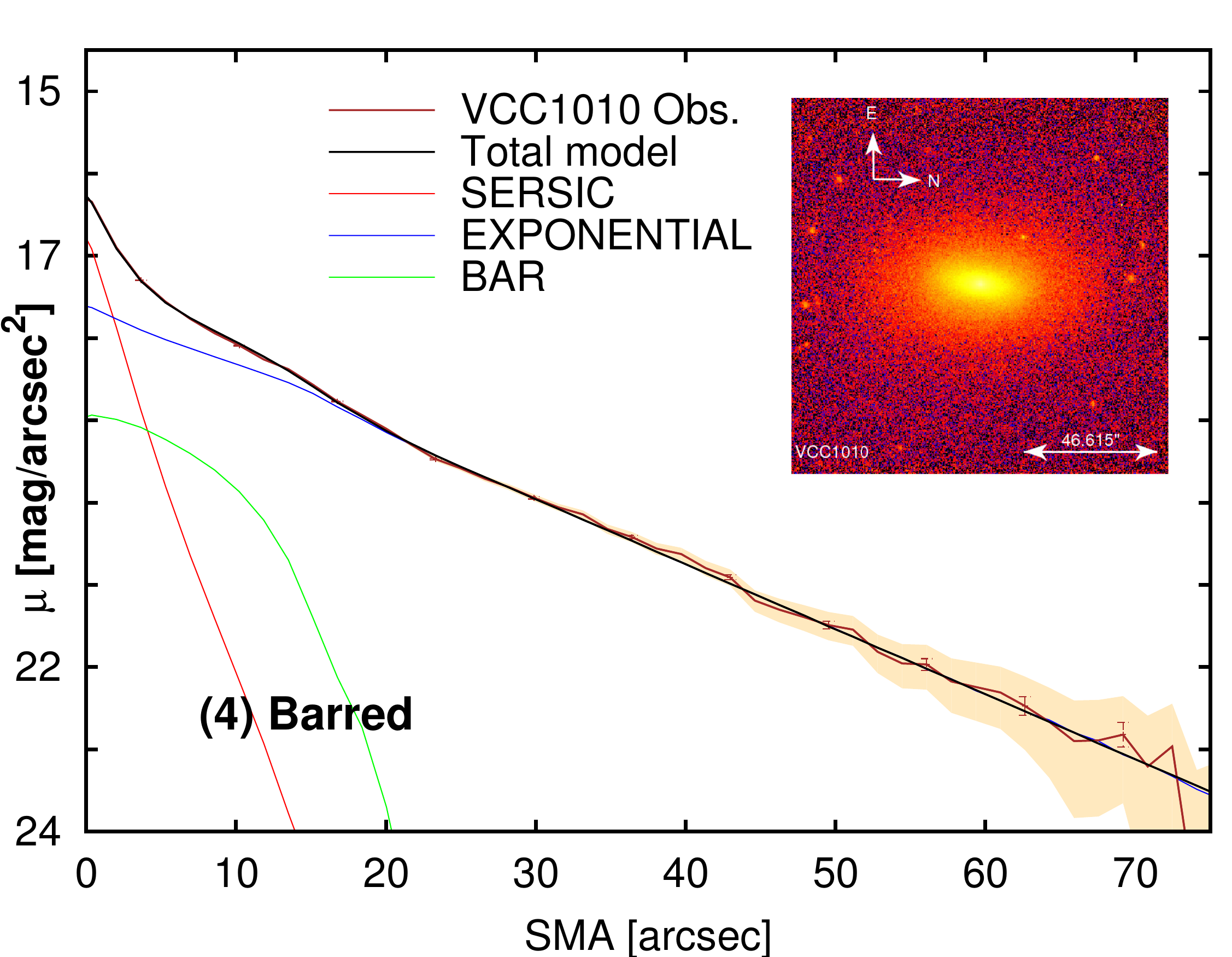} 
\caption{Representative galaxies for the structural groups (\S\ref{sec:groups}), from top to bottom: VCC1185 -- single component, VCC0762 -- two  components, VCC1183 -- with lens,  and VCC1010 -- barred. The shaded areas display the maximal systematic error caused by large-scale background variations and the error bars display the surface brightness uncertainties as measured by \textsc{iraf}/ellipse.\label{fig:groups}}
\end{center}
\end{figure}

Following  \citet{janz+2012}  we group the galaxies into four  structural types, based on the decompositions:
(1) one-component galaxies that satisfy a simple model { (with or without a nucleus, first two models in \S\ref{sec:procedure} and Fig.~\ref{fig:fitting}),} (2) two-component galaxies
that have an inner and outer component (with or without an additional nucleus), 
(3) galaxies with a lens component, and (4) barred galaxies (Fig.\ \ref{fig:groups}; more examples in \citealt{janz+2012}, their Fig.\ 2).

The concept of barred galaxies is clear: these galaxies have two or in some cases  three  components (not counting the nucleus), one of them being a bar. 
  The  galaxies with a lens can be considered as a special group of two-component galaxies.
  In the surface brightness profiles the lens-like component is fairly flat and has a rather steep
  truncation (Fig. \ref{fig:groups}, panel 3). While this profile shape is  quite similar to the profile of a bar (Fig. \ref{fig:groups}, lower panels), the lens is not strongly elongated. 
  As its shape clearly differs from
 that of an inner component with $n>1$, galaxies with such a characteristic bump in the profile are considered as a separate group.
  
    Most importantly, not all galaxies follow a simple model. In fact, a large fraction falls into
one of the three  complex groups.\footnote{ With a different type of analysis, several of our two-component galaxies may be modeled as a single component with \emph{varying} position angle and ellipticity, which is the case in the ACS VCS (Advanced Camera for Surveys Virgo Cluster Survey; \citealt{Ferrarese:2006p586}; L.\ Ferrarese, priv.\ comm.). This does, however, not touch the fact that they show a \emph{complex} light distribution.}

\begin{deluxetable*}{l|rr|rrr|rrr|rrr|r}
\centering
\tablecaption{Frequencies of structural types\label{table:groups}}
\tabletypesize{\footnotesize}
\tablenum{4}
\tablecolumns{13}
\tablehead{
\colhead{Group} & \multicolumn{2}{c}{1} &  \multicolumn{3}{c}{2} &  \multicolumn{3}{c}{3} & \multicolumn{3}{c}{4} & \colhead{Total}\\
 \colhead{} &  \multicolumn{2}{c}{One-component} &  \multicolumn{3}{c}{Two-component} &  \multicolumn{3}{c}{With lens} & \multicolumn{3}{c}{Barred} & \colhead{analyzed}}
\startdata
All & 32 & $32\%\pm4\%$ &  39 & (8) & $39\%\pm4\%$ &  14 & (5) & $14\%\pm3\%$ &   14 & (5) & $14\%\pm3\%$ & 99 \\
\tableline
$-19\le M_r\le-18$  & 1 & $6\%\pm5\%$ &  8 & (0) & $47\%\pm12\%$ &  4 & (3) & $24\%\pm10\%$ &   4 & (0) & $24\%\pm10\%$ & 17 \\
$-18<M_r\le-17$ & 16 & $30\%\pm6\%$ &  23 & (5) & $43\%\pm6\%$ &  6 & (0) & $11\%\pm4\%$ &   8 & (5) & $15\%\pm4\%$ & 53 \\
$-17<M_r\le-16$ & 15 & $52\%\pm9\%$ &  8 & (3) & $28\%\pm8\%$ &  4 & (2) & $14\%\pm6\%$ &   2 & (0) & $7\%\pm4\%$ & 29 \\
\tableline
dE(all) & 31 & $37\%\pm5\%$ &  33 & (7) & $39\%\pm5\%$ &  9 & (3) & $11\%\pm3\%$ &   11 & (5) & $13\%\pm3\%$ & 84 \\
\tableline
dE(N) & 20 & $49\%\pm7\%$ &  11 & (2) & $27\%\pm6\%$ &  7 & (2) & $17\%\pm5\%$ &   3 & (2) & $7\%\pm3\%$ & 41 \\
dE(nN) & 3 & $38\%\pm17\%$ &  5 & (1) & $63\%\pm17\%$ &  0 & (0) & $0\%$ &   0 & (0) & $0\%$ &  8 \\
dE(di) & 4 & $15\%\pm7\%$ &  11 & (2) & $42\%\pm9\%$ &  2 & (1) & $8\%\pm5\%$ &   9 & (3) & $35\%\pm9\%$ & 26 \\
dE(bc) & 4 & $33\%\pm13\%$ &  8 & (2) & $67\%\pm13\%$ &  0 & (0) & $0\%$ &   0 & (0) & $0\%$ & 12 \\
E\&S0 & 1 & $7\%\pm6\%$ &  6 & (1) & $40\%\pm12\%$ &  5 & (2) & $33\%\pm12\%$ &   3 & (0) & $20\%\pm10\%$ & 15 \\
\tableline
$D_{M87}<1.5^\circ$ & 6 & $22\%\pm7\%$ &  9 & (3) & $33\%\pm9\%$ &  5 & (2) & $19\%\pm7\%$ &   7 & (3) & $26\%\pm8\%$ & 27 \\
$1.5^\circ\le D_{M87}<4^\circ$ & 16 & $36\%\pm7\%$ &  18 & (4) & $41\%\pm7\%$ &  6 & (2) & $14\%\pm5\%$ &   4 & (1) & $9\%\pm4\%$ & 44 \\
$D_{M87}\ge4^\circ$ & 10 & $36\%\pm9\%$ &  12 & (1) & $43\%\pm9\%$ &  3 & (1) & $11\%\pm5\%$ &   3 & (1) & $11\%\pm5\%$ & 28 
\enddata
\tablecomments{We list numbers and fractions of galaxies in each group  binned over the brightness,  $dE$ subclass (dE(N) -- nucleated, dE(nN) -- no nucleus, dE(bc) -- blue core, dE(di) -- disk features; \citealt{Lisker:2006p385,Lisker:2006p392,Lisker:2007p373}), and  angular distance to M87  ($1^\circ = 0.284$ Mpc). The total sum in the subclass binning is larger than the total number of galaxies, since $dE$s can belong to  multiple subclasses.
The less certain cases in each group, shown in parenthesis, are included in the numbers and fractions. The uncertainties of the percentages are calculated according to the Poisson statistics. Additionally two small bars were visually identified but not fitted and are therefore not counted in this Table.}
\end{deluxetable*}

\begin{figure*}
\begin{center}
\hskip 2mm\includegraphics[height=0.31\linewidth, angle=-90]{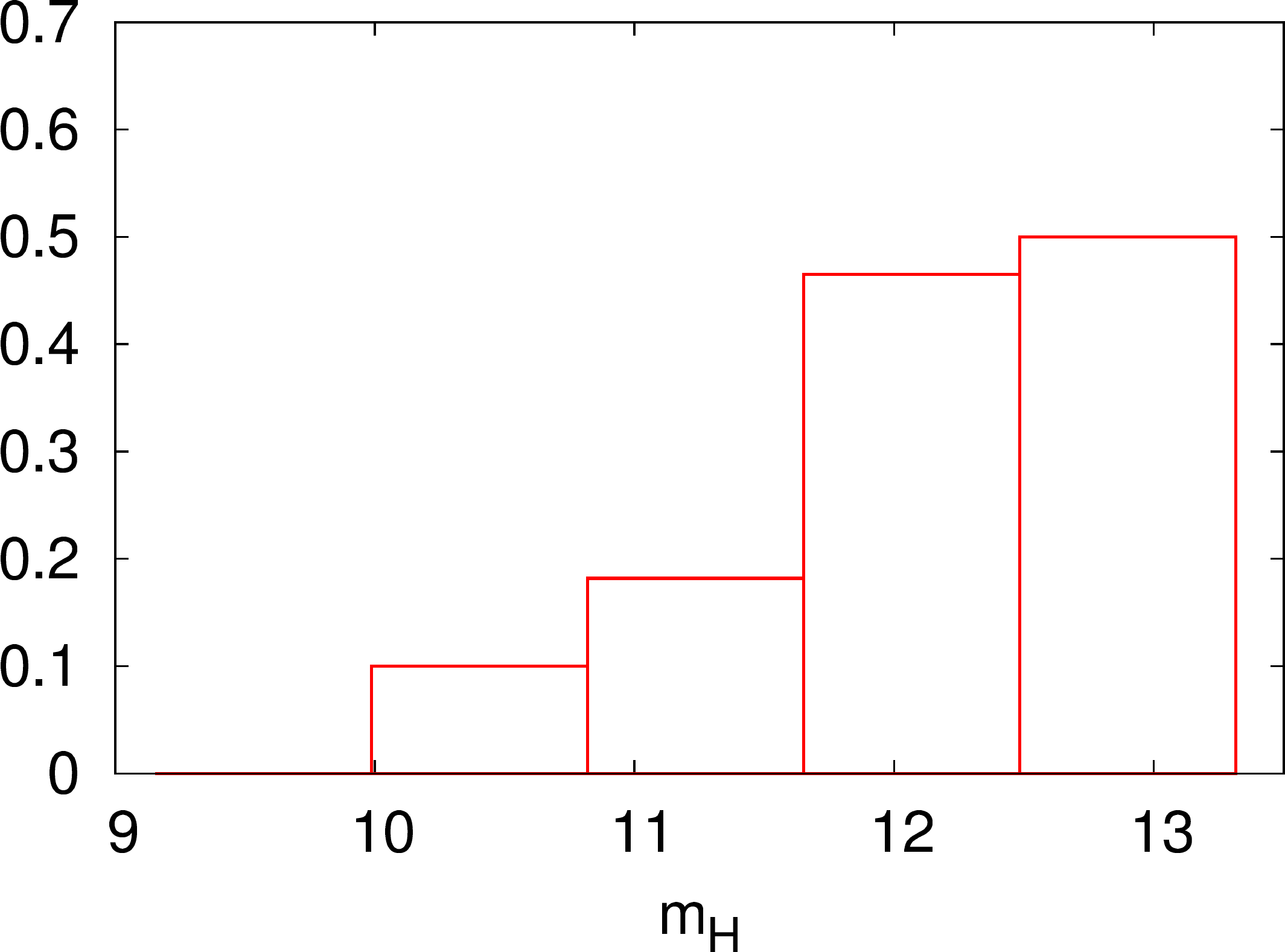}\hskip 5mm\includegraphics[height=0.31\linewidth, angle=-90]{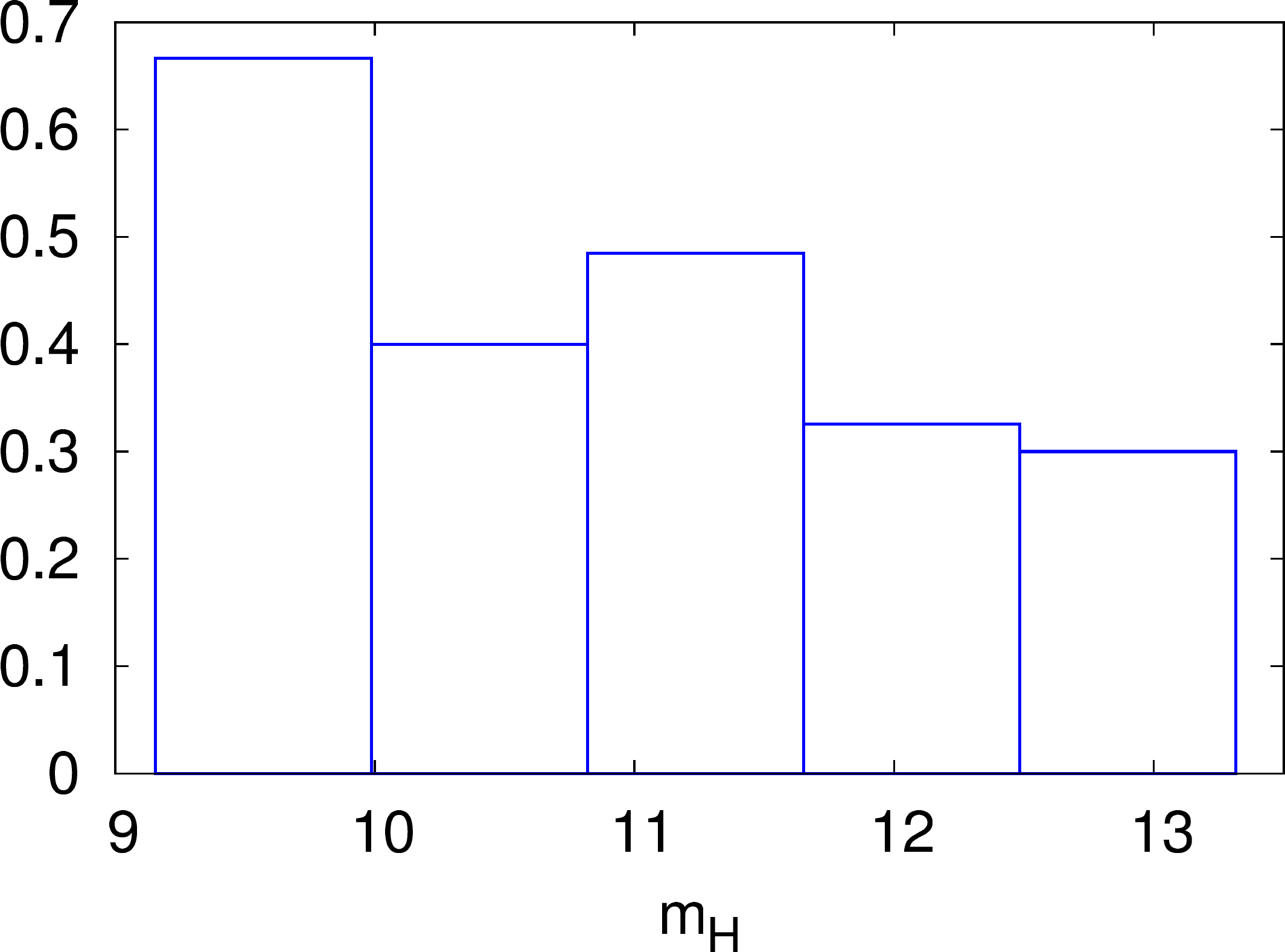}\hskip 5mm\includegraphics[height=0.31\linewidth, angle=-90]{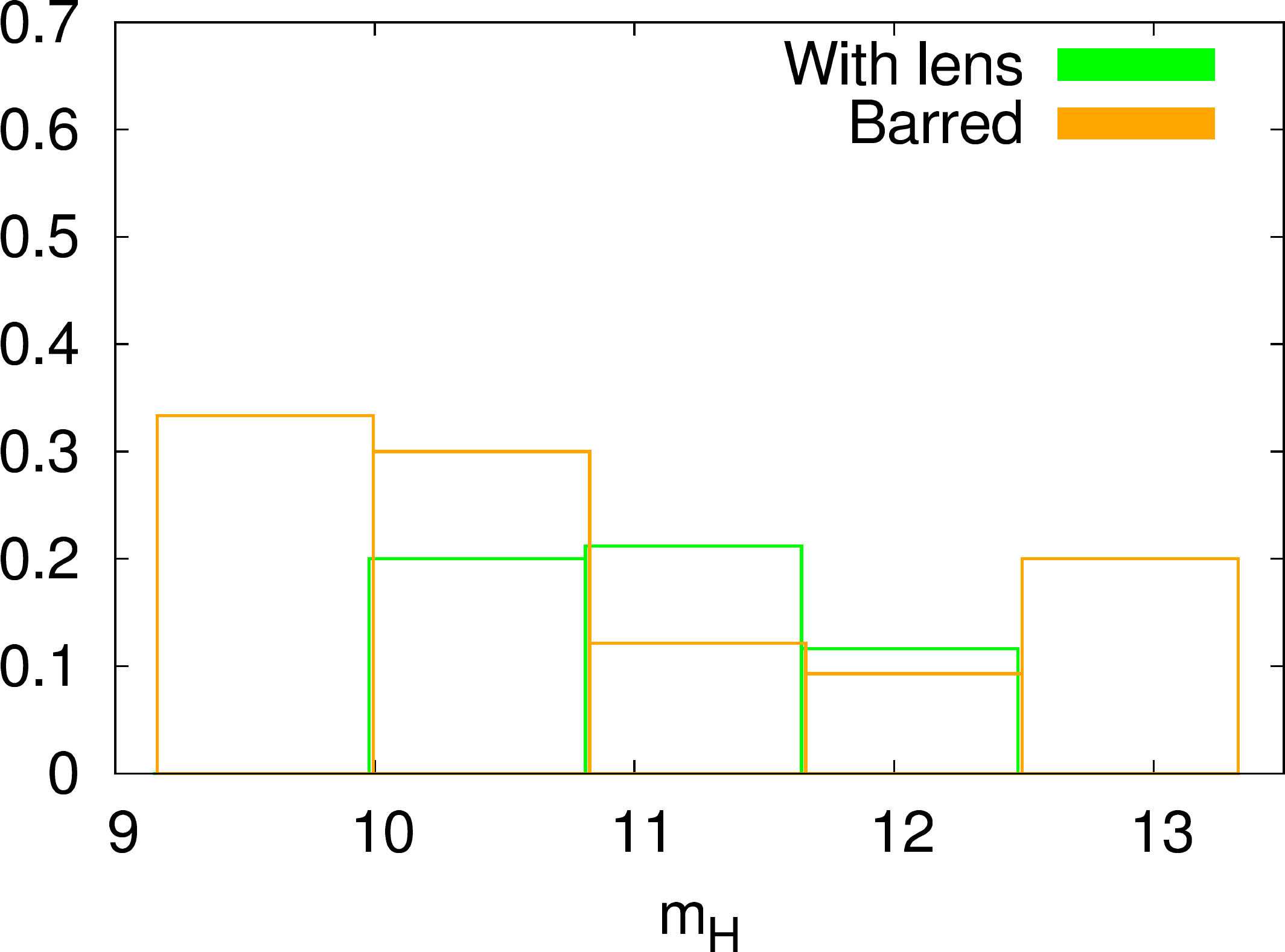}\\
\includegraphics[width=0.95\linewidth, angle=0]{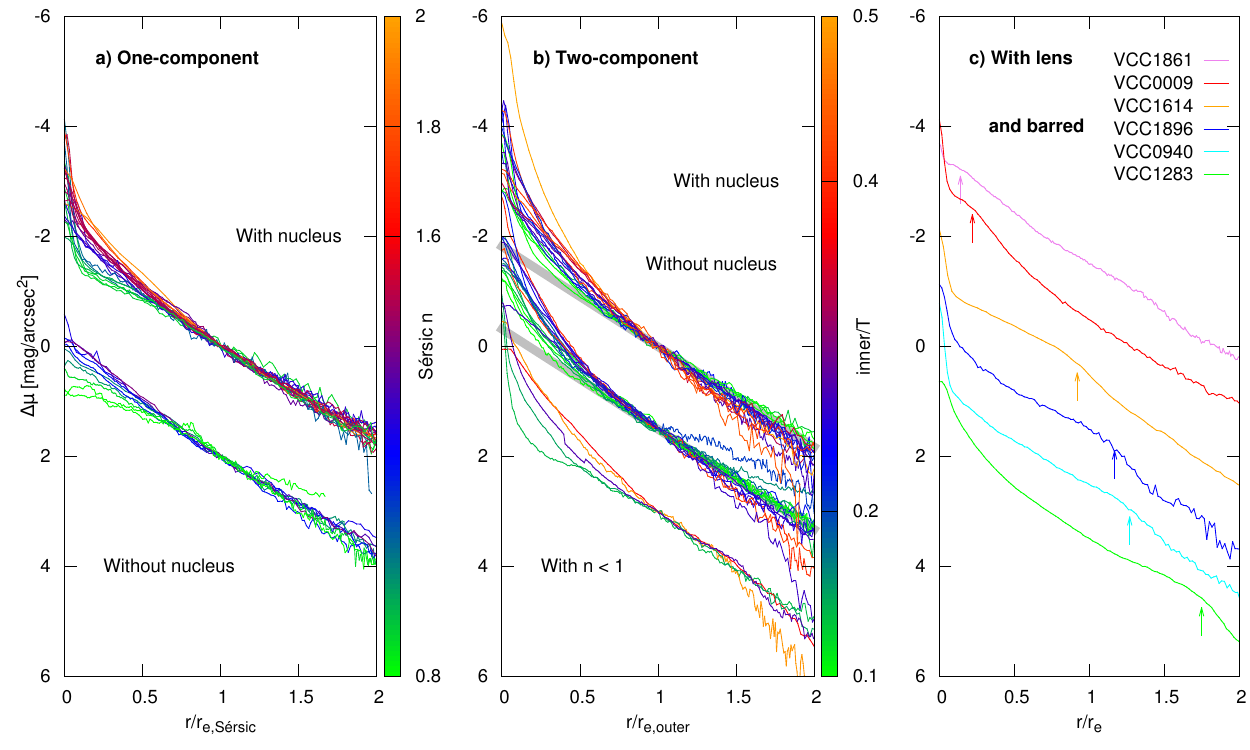}\ \ \ \ \ \ 
\caption{ The top panels  show from left to right the fractions of the one-component galaxies, two-component galaxies, and galaxies with lenses and bars in bins of galaxy brightness of 0.5 mag.
The bottom panels display the azimuthally averaged profiles of  the galaxies in the one-component and two-components groups, and a few examples of  the groups of galaxies with lenses and barred galaxies. The radii are normalized to the effective radius of the S\'ersic function (\textit{a}), the outer component (\textit{b}), and the whole galaxy's half-light radius (\textit{c}). The y-axis is normalized to the surface brightness at that radius, and for different galaxy groups or galaxies arbitrary offsets are applied. In  panels \textit{a} \& \textit{b} the S\'ersic index ($n$) and the ratio of flux in the inner component to that of the whole galaxy ($\textit{inner}/T$), respectively, are coded in color. In panel \textit{b} the profiles are grouped from top to bottom: galaxies with nucleus, without nucleus, and galaxies with an outer S\'ersic component with $n < 1$.
The gray lines in the middle panel display an exponential profile. In the right panel examples for profiles of galaxies with lenses and bars are shown, from top to bottom the first three galaxies have lenses, while the last three have bars. 
The arrows indicate the radius where the lens or bar contributes most significantly to the light.\label{fig:profiles}}
\end{center}
\end{figure*}

We briefly summarize 
the findings already presented in \citet{janz+2012} and provide an updated Table of the frequencies of structural types, with observations for 20 new galaxies (Table \ref{table:groups} = Table 4). \citeauthor{janz+2012} found the fraction of one-component galaxies to be close to a quarter. The fraction is a strong function of galaxy brightness and increased with the observations of fainter galaxies.  This  is not expected to be a simple selection effect, since the fainter galaxies had longer integration times by design of the observations. 
With the updated sample a  Kolmogorov-Smirnov (KS) test gives a statistically significant difference of the brightness distributions of simple and complex galaxies ($p<1\%$). The updated fraction of a third for the complete sample is still remarkably low. Even if we exclude galaxies classified as E or S0 as well as count less certain two-component galaxies as one-component galaxies the fraction is only 44 \%.
The fraction 
is highest among the early-type dwarf galaxy subclasses   where no disk feature or central star formation have been detected \citep{Lisker:2006p392,Lisker:2006p385}.  
The bar fraction  drops towards fainter galaxies. The updated fraction is slightly smaller than in  \citet{janz+2012} due to the increased sampling of fainter galaxies. With two small bars in galaxies with lenses, which were not fitted and therefore do not appear in the Table, the fraction is 16\%. If only certain bars  and  galaxies with a dE or dS0 classification are considered, the bar fraction is 10\%.
 The same is true for the fraction of galaxies with lenses (14\% and 7\%, respectively).

\subsection{Profile shapes}
The overall profile shapes of the galaxies in our sample are not much steeper than exponentials.
They follow the known relation of S\'ersic $n$ and galaxy brightness (e.g.\ \citealt{1993MNRAS.265.1013C,1994A&A...286L..39C}). 
However, we  show with our two-dimensional multi-component decomposition analysis that this view is oversimplified. 
Instead of simple S\'ersic profiles becoming steeper with increasing galaxy brightness, the trend of the profile shape with galaxy brightness can be, at least  partly, attributed to the more frequent multi-component structures in the brighter galaxies (Fig.\ \ref{fig:profiles}, for more quantitative analysis see Appendix \ref{sec:sersic}). 

In Fig.\ \ref{fig:profiles} we show all the profiles together, scaled to a common radius and surface brightness. 
{ While the choice of a fixed exponential shape for most outer components (see \S\ref{sec:procedure}) could in principle influence the resulting parameter space, the figure shows that the assumption is justified for the vast majority of galaxies (see also e.g.~\citealt{binggeli_cameron}).}
We  point out that also the inner component in the complex models is close to an exponential profile (most of the inner components have $n<1.2$). 
Bars and lenses have similar extents. 

The profiles of the one-component  or complex galaxies do not show systematic differences with the presence of a nucleus. 
We note that in  Fig.\  \ref{fig:profiles}  the nucleus is actually hard to see, since the plotted radial range of the profiles is much larger than the PSF.
In some of the profiles of the one-component galaxies it is visible as a peak at a few central pixels.
 However, for the complex galaxies the profiles are scaled to even larger radii.  For those, the inner excess of light is due to the inner component and not due to the nucleus.

\subsection{Morphological parameters} 

\begin{figure*}
\begin{center}
\includegraphics[height=12cm, angle=-90]{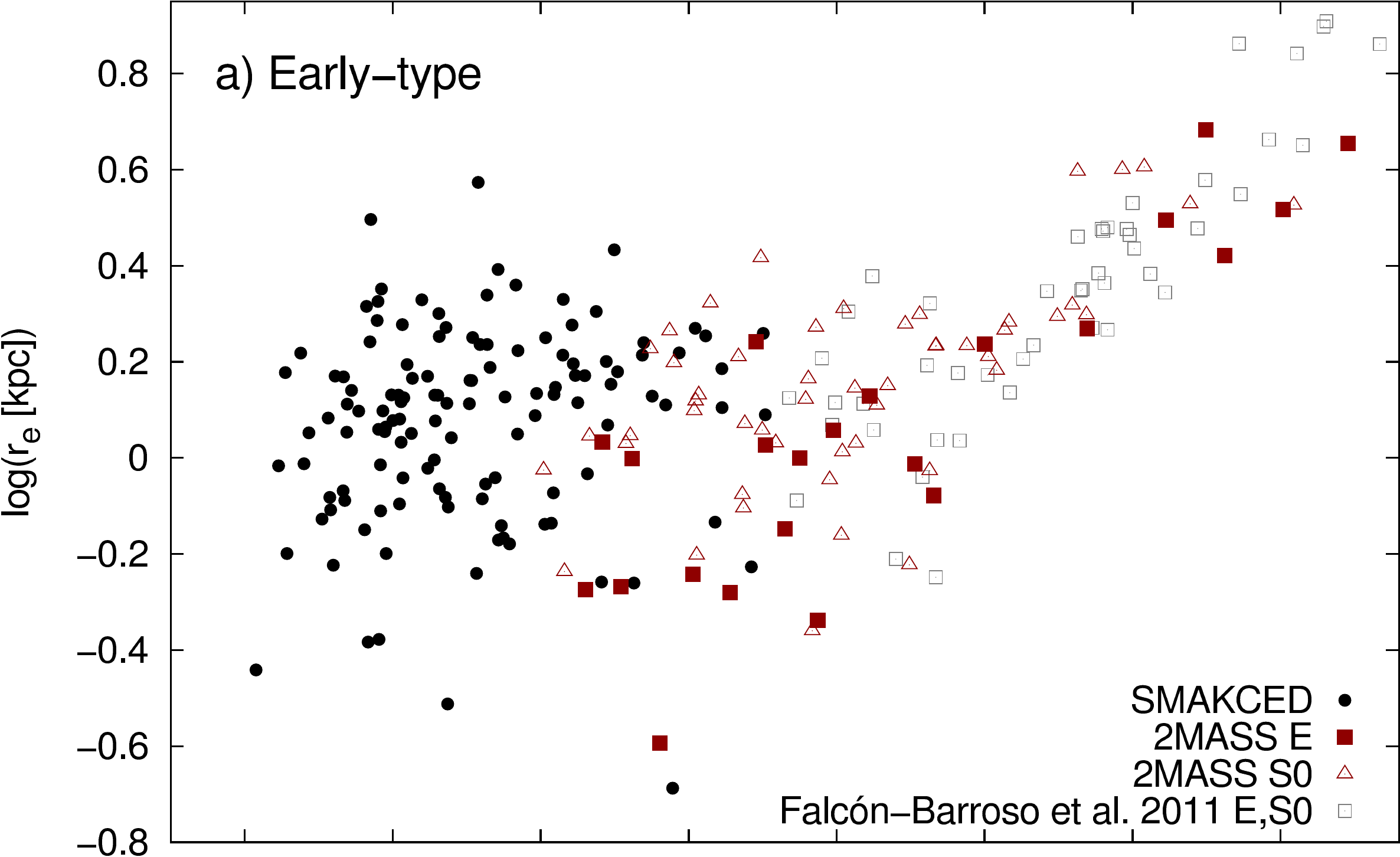}
\includegraphics[height=12cm, angle=-90]{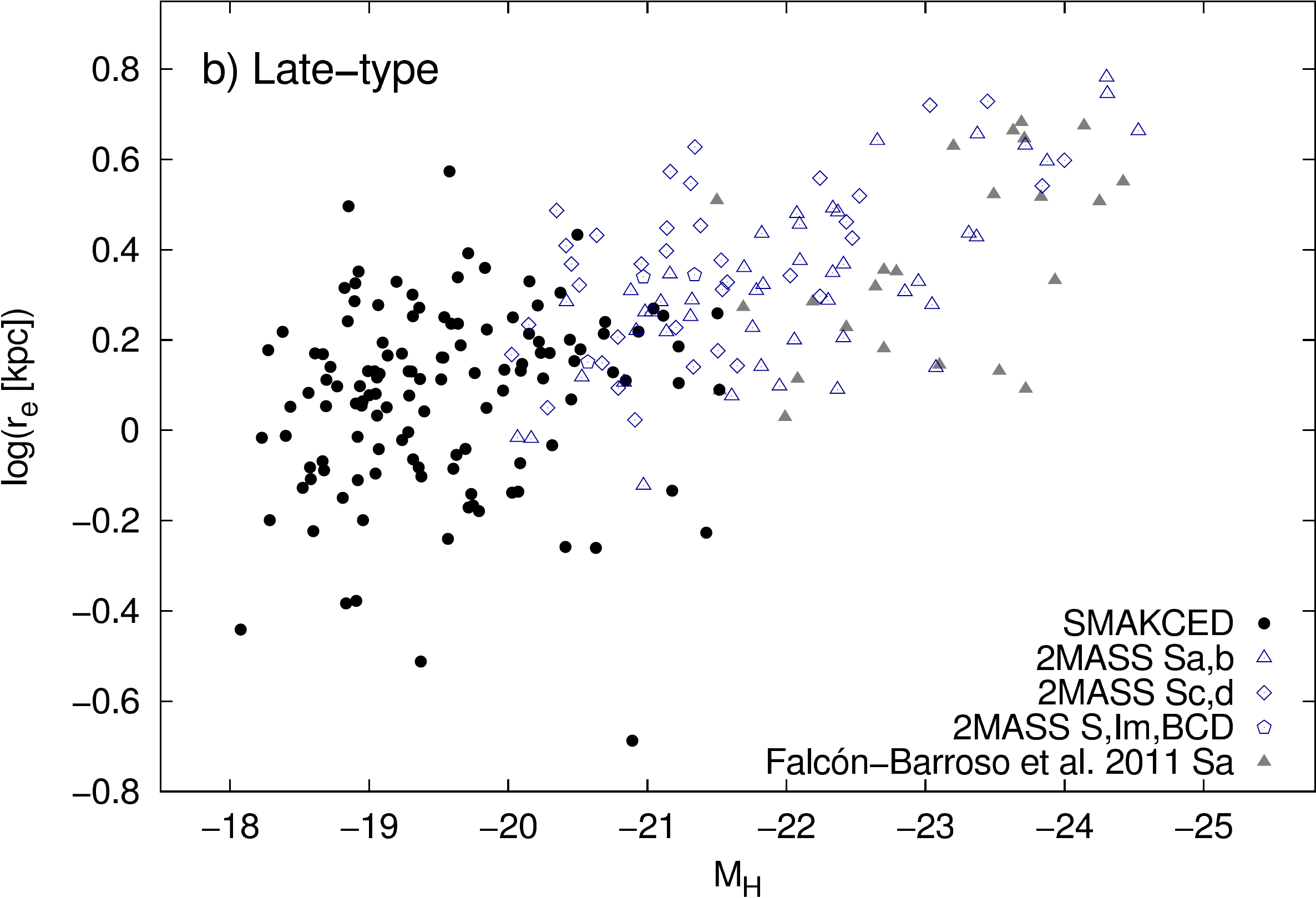}\\
\includegraphics[width=6.cm, angle=-90]{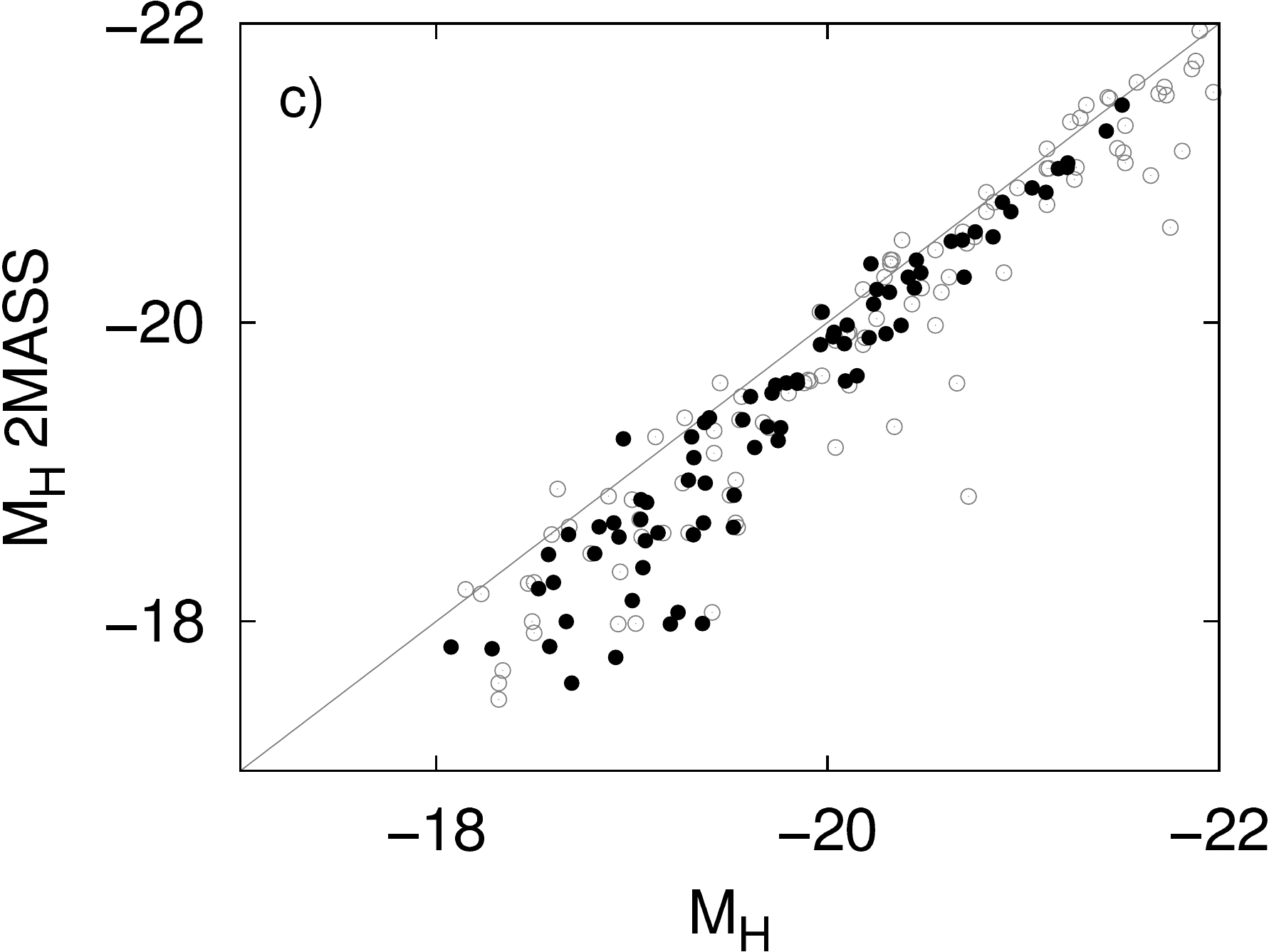}
\includegraphics[width=6.cm, angle=-90]{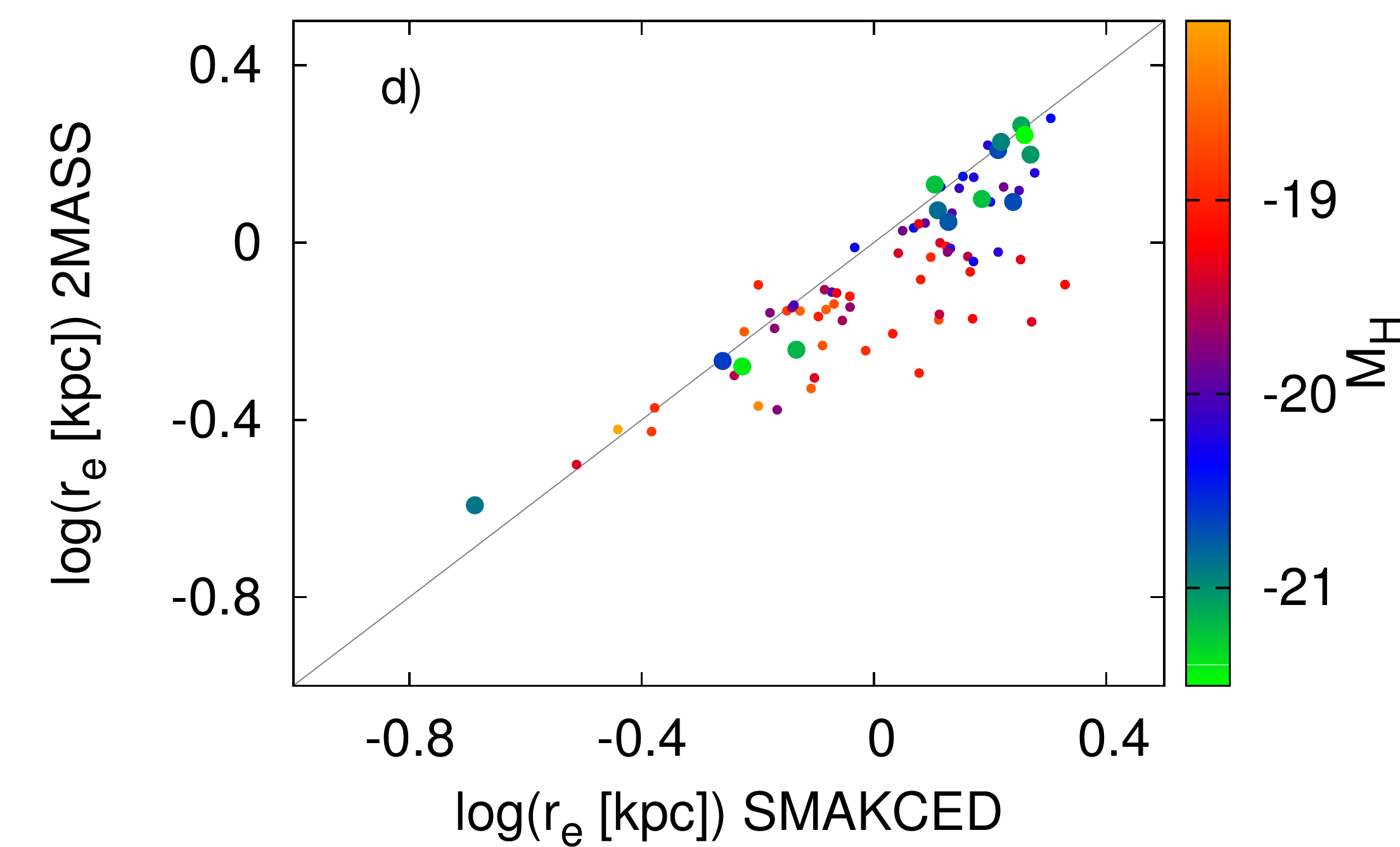}
\caption{Effective radius $r_e$ of the galaxy versus total galaxy brightness $M_H$. The SMAKCED sample galaxies are  compared to bright ($M_H<-20$ mag) early-type galaxies \textit{(a)} and bright late-type galaxies \textit{(b)} in the Virgo cluster. The data for the bright galaxies are taken from the 2MASS extended source catalog. { Additionally, values for bright galaxies from $3.6\mu$ Spitzer data \citep{2011MNRAS.417.1787F} are displayed with gray symbols (with $K-[3.6]=0.1$ mag, \citeauthor{2011MNRAS.417.1787F}, and $H-K= 0.21$ mag \citealt{1999MNRAS.310..703P}). }
\textit{Lower panels:} comparison of photometry with 2MASS for our sample galaxies that are in the 2MASS extended source catalog. Note the good agreement for galaxies  with $M_H\lesssim-20$ mag. In the panel {\it c} the  brightnesses are compared ({\it black} points), and for reference the same comparison is shown also for the \citet{Mcdonald:2011p4445} measurements ({\it gray open circles}). Panel  {\it d} compares the size measurements of SMACKED and 2MASS (for a comparison to \citeauthor{Mcdonald:2011p4445} see Appendix \ref{section:mcd}). In this panel the galaxy brightnesses are colorcoded according to the colorbar, and galaxies brighter than $M_H<-20.5$ mag are displayed with a larger symbol. \label{fig:2MASS}}
\end{center}
\end{figure*}

For the early-type galaxies the Kormendy relation \citep{kormendy_relation}
and  its relatives, with interchanging either the mean effective surface brightness or the  effective radius with brightness, are commonly studied.
In Fig.\ \ref{fig:2MASS} the obtained non-parametric photometric quantities are used to construct the brightness versus effective radius relation
for the galaxies in our sample, whereas the Kormendy relation is shown in Appendix \ref{sec:additional}. 

For comparison we add Virgo cluster galaxies of different morphological types from the 2MASS extended source catalog, using the extrapolated total magnitudes.
2MASS is not very deep,\footnote{Note that 2MASS is 
much shallower than UKIDSS, to which a comparison of our data for one galaxy was shown in \citet{janz+2012}, their Fig.~1.}
 therefore the photometry  for faint galaxies  should not be trusted:  it misses a substantial
fraction of the light so that  both the radius and brightness are underestimated. Furthermore, the extended source catalog does not include all VCC galaxies, though the
coverage of the Virgo cluster is complete. A comparison of the photometry in this study and the 2MASS measurements for  
galaxies common to both samples shows that the 2MASS photometry 
is consistent for the galaxies brighter than $M_H<-20$ mag (lower panels in Fig.\ \ref{fig:2MASS}),
and therefore considered robust for bright galaxies.

 Furthermore, the additional comparison of the magnitudes measured by \citet{Mcdonald:2011p4445}
shows consistently that 2MASS misses some of the flux   for the fainter galaxies. The faintest  galaxies
are missing from the 2MASS extended source catalog. For a further comparison to the study of \citeauthor{Mcdonald:2011p4445}
see Appendix \ref{section:mcd}. When using 2MASS $10\arcsec$  aperture colors to transform the $K$-band magnitudes of \citet[][MAGPOP survey]{2012A&A...548A..78T} into $H$-band, the median difference for the 17 galaxies in common is only 0.03 mag, with a semi-interquartile range of 0.12 mag.\footnote{
Note that we used the correct photometric values of \citet{2012A&A...548A..78T} according to the erratum Toloba et al.~(2013, in press).}

 Fig.\ \ref{fig:2MASS} shows that there is some overlap of the early-type dwarfs with the spiral galaxies (panel {\it b}),  and that they apparently form a more simple continuous relation than with  the bright elliptical galaxies that seem to fall on a different relation (panel {\it a}; see also \citealt{Boselli,2012ApJS..198....2K,2012A&A...548A..78T}).

\subsection{Flattening distributions}
The flattening distributions of the components are of interest, since
they are expected to be different for different distributions of intrinsic shapes
(see e.g.\ \citealt{Lisker:2007p373}).

\begin{figure}
\begin{center}
\includegraphics[height=\linewidth, angle=-90]{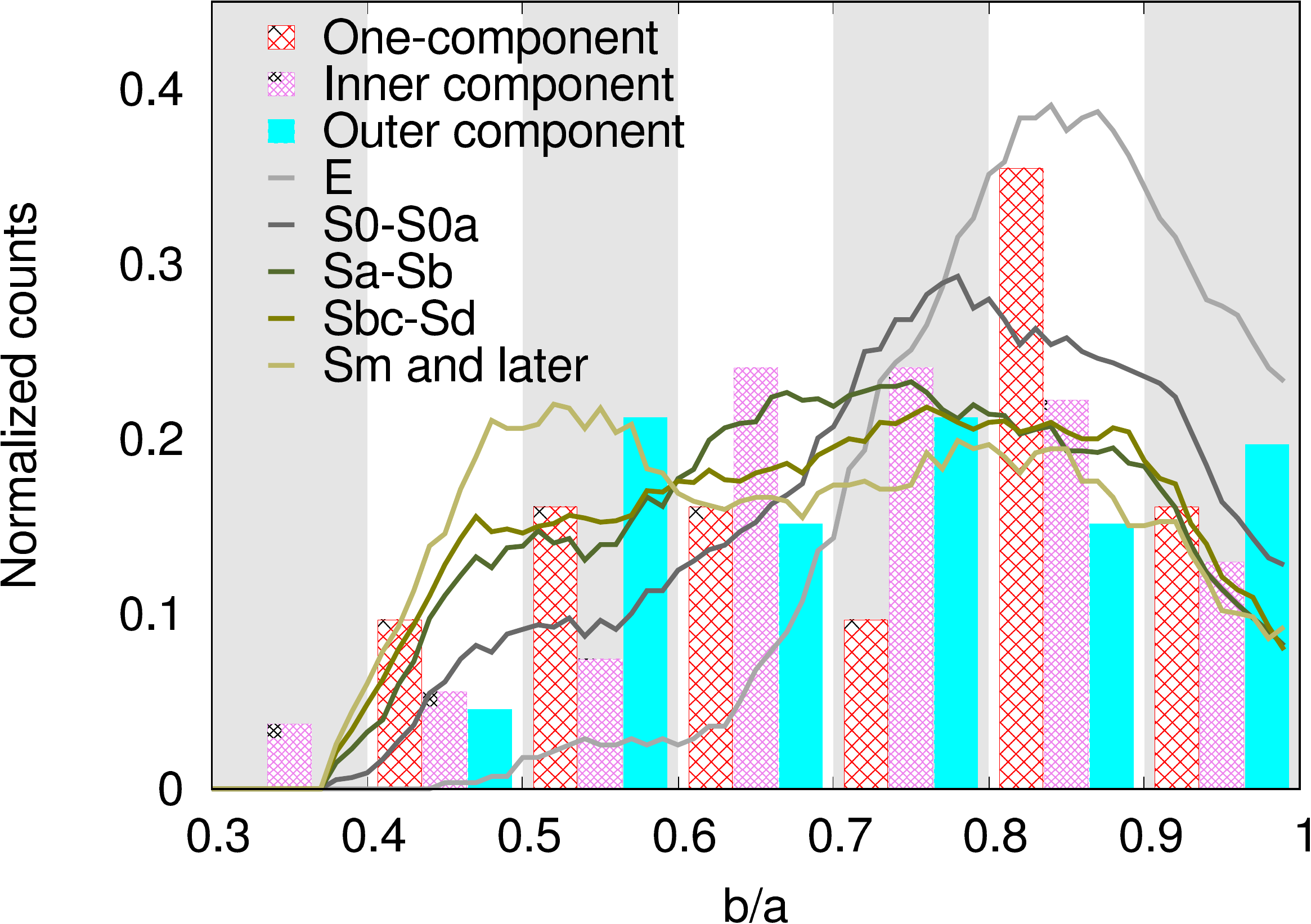}
\caption{Distribution of projected axis ratios of the one-component galaxies and the inner and outer components in the more component galaxies. Our inclination limit falls into the second bin. For comparison the distributions for other galaxy types as found in the SDSS (see text) are shown with running bins of the same bin width of 0.1 and to the same limit of $b/a=0.42$. The histograms are normalized to the number of galaxies in each group. \label{fig:axr}}
\end{center}
\end{figure}

Fig.\ \ref{fig:axr} shows the distributions of the axis ratios of the one-component galaxies, 
and inner 
and outer components of complex galaxies. 
Our inclination limit falls into the second bin ($b/a=0.42$). 
The flattening distributions for the inner and outer components are
very similar, so that they should have similar intrinsic shapes.
The one-component galaxies may have a double peaked distribution, with some contribution of galaxies 
that may be as round as the bright early types.
However, the small numbers make it impossible to firmly state that they have a flattening distribution
different from the other early-type dwarfs. A KS-test cannot reject the null-hypothesis that both the one-component and complex galaxies are drawn from the same   distribution.

For comparison we compiled axis ratios of other galaxy types from the SDSS (DR8, \citealt{sdss}).
We use the galaxies in RC3 \citep{RC3}  with measured redshifts (in SDSS) below $z<0.1$. The axis ratio is calculated from
the adaptive second moments $E1$ and $E2$, measured by the SDSS pipeline in the $r$-band,  with 
\begin{equation}
b/a=\left({{1-e}\over{1+e}}\right)^{1/2}\, ,
\end{equation}
where $e=\sqrt{E1^2+E2^2}$ \citep{hirata}. We do not correct for seeing effects, since the galaxies in RC3 are large compared to the SDSS seeing.
The distributions of axis ratios of the early-type dwarfs and their inner and outer components is closer to that of spirals than to that of early-type galaxies, consistent with the findings of \citet{1995A&A...298...63B}.
\citet{fisdro} showed that in bright galaxies classical bulges are round, while pseudo-bulges have flattenings similar to the outer disk.\\

\subsection{Boxiness}
\label{sec:resultboxiness}
\begin{figure}
\begin{center}
\includegraphics[width=\linewidth, angle=0]{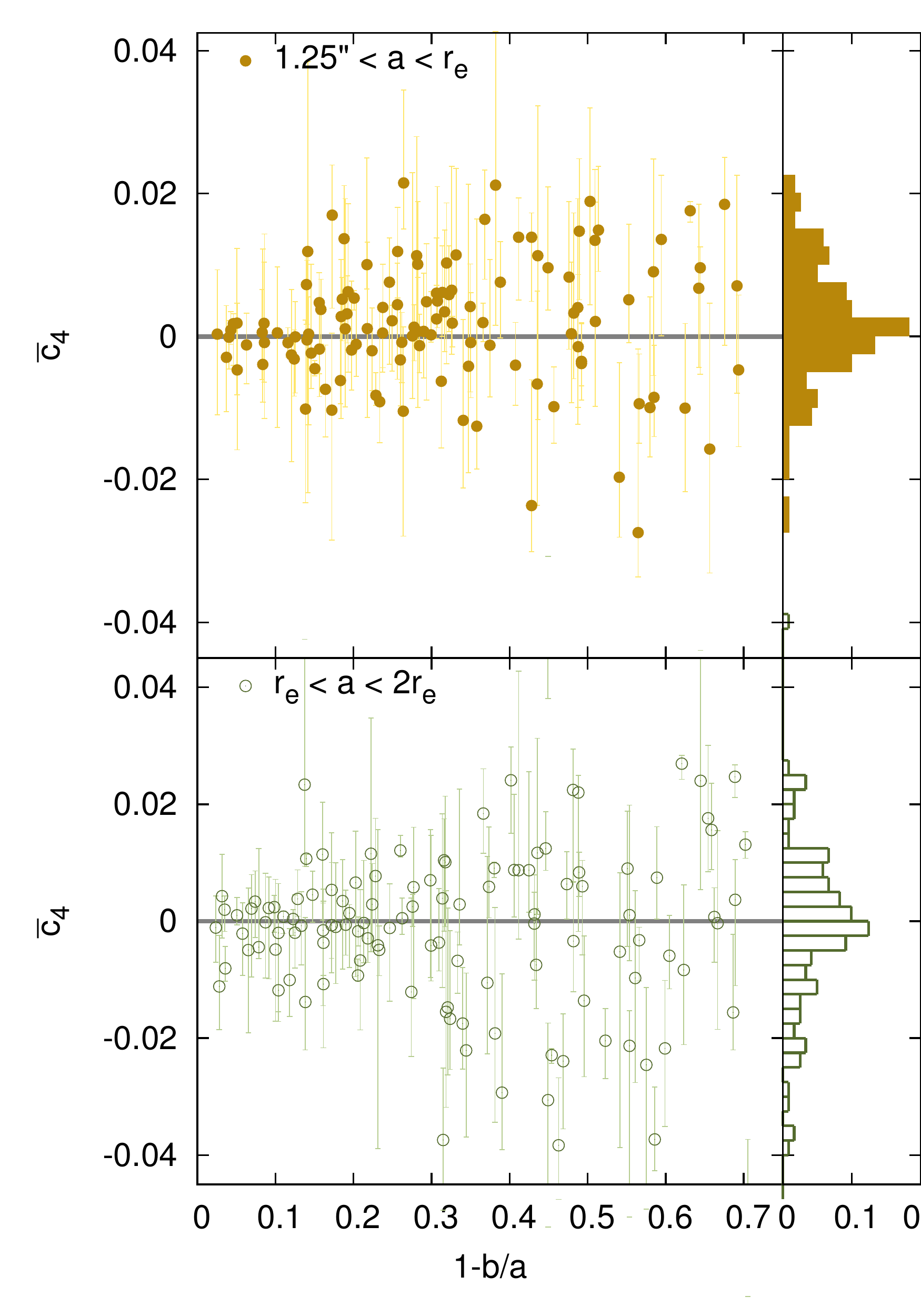}
\caption{The median of the boxiness parameter inside the half-light SMA $r_e$ and within $r_e$ and $2\, r_e$ versus the median ellipticity of the galaxy within the same region. The bars indicate the semi-inter-quartile ranges of the values. Values larger than 0 indicate disky, while those smaller than 0 mean boxy isophotes. \label{fig:boxiness} }
\end{center}
\end{figure}

In Fig.\ \ref{fig:boxiness} we show the median values of the boxiness parameter, $\bar{c}_4$, at radii smaller than the effective radius, $1\farcs25 <a<r_e$, and between  one and two effective radii, $r_e<a<2r_e$. Both  distributions
have a larger scatter for galaxies with higher ellipticities (see also \citealt{Bender:1989wm}). However, some more large boxy values occur  in the outer parts of those galaxies, which have  high ellipticity and are presumably more inclined (lower panel of Fig.\ \ref{fig:boxiness}).

\begin{figure}
\begin{center}
\includegraphics[height=\linewidth, angle=-90]{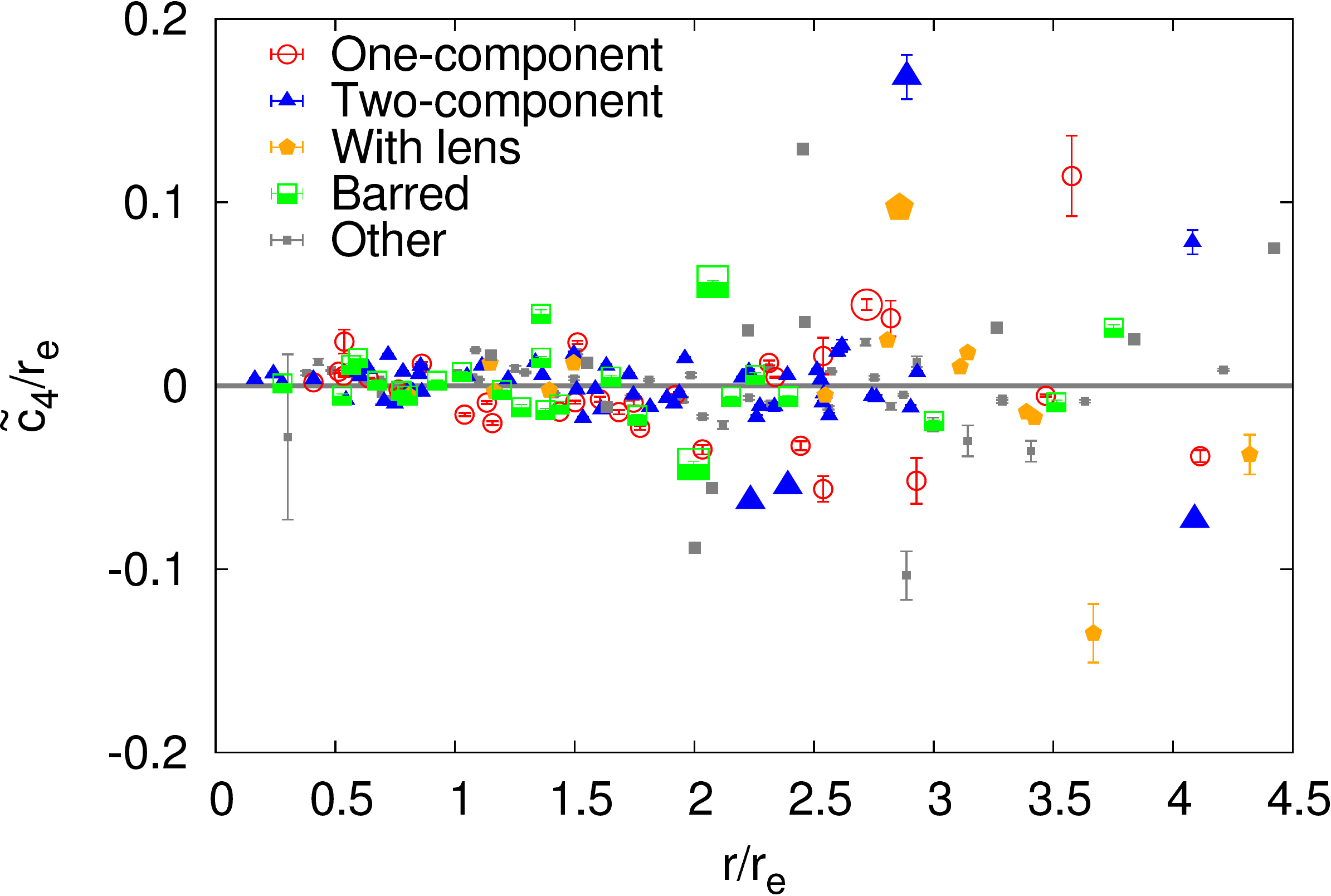}
\caption{Integrated boxiness parameter divided by the effective radius $\tilde{c_4}/r_e$ versus its position in the radial profile of the galaxy. The error bars indicate the formal uncertainties. Larger symbols are used for features that extend more than the effective radius  $r_e$ of that galaxy. Plotted are 198 features in 86 galaxies. Values larger than 0 indicate disky, while those smaller than 0 mean boxy isophotes.
\label{fig:C4}}
\end{center}
\end{figure}

As noted in Section \S\ref{section:boxiredux} the median value of the boxiness parameter might not reveal all boxy or disky sections in the radial profiles. Therefore,  Fig.\ \ref{fig:C4} shows the  $\tilde{c_4}/r_e$  values for all sections in the profiles that deviate more than $\sim$2$\sigma$ from an elliptical shape. The measurements are shown with a symbol according to the structural type of the galaxy.  

Our algorithm detected 198 such features in 86 of 121 galaxies.
Apparently, the
distributions of features are slightly skewed to disky shapes in the inner part within $r_e$, and  to boxy shapes in the outer part beyond $r_e$ until it becomes too noisy.
Interestingly,  also in barred galaxies both disky and boxy features appear. This can be understood as a superposition of two components with a large difference in axis ratios, or alternatively due to the intrinsic bar shape.

\subsection{Positions and velocities within the cluster}
\begin{figure}
\begin{center}
\includegraphics[height=8.5cm, angle=-90]{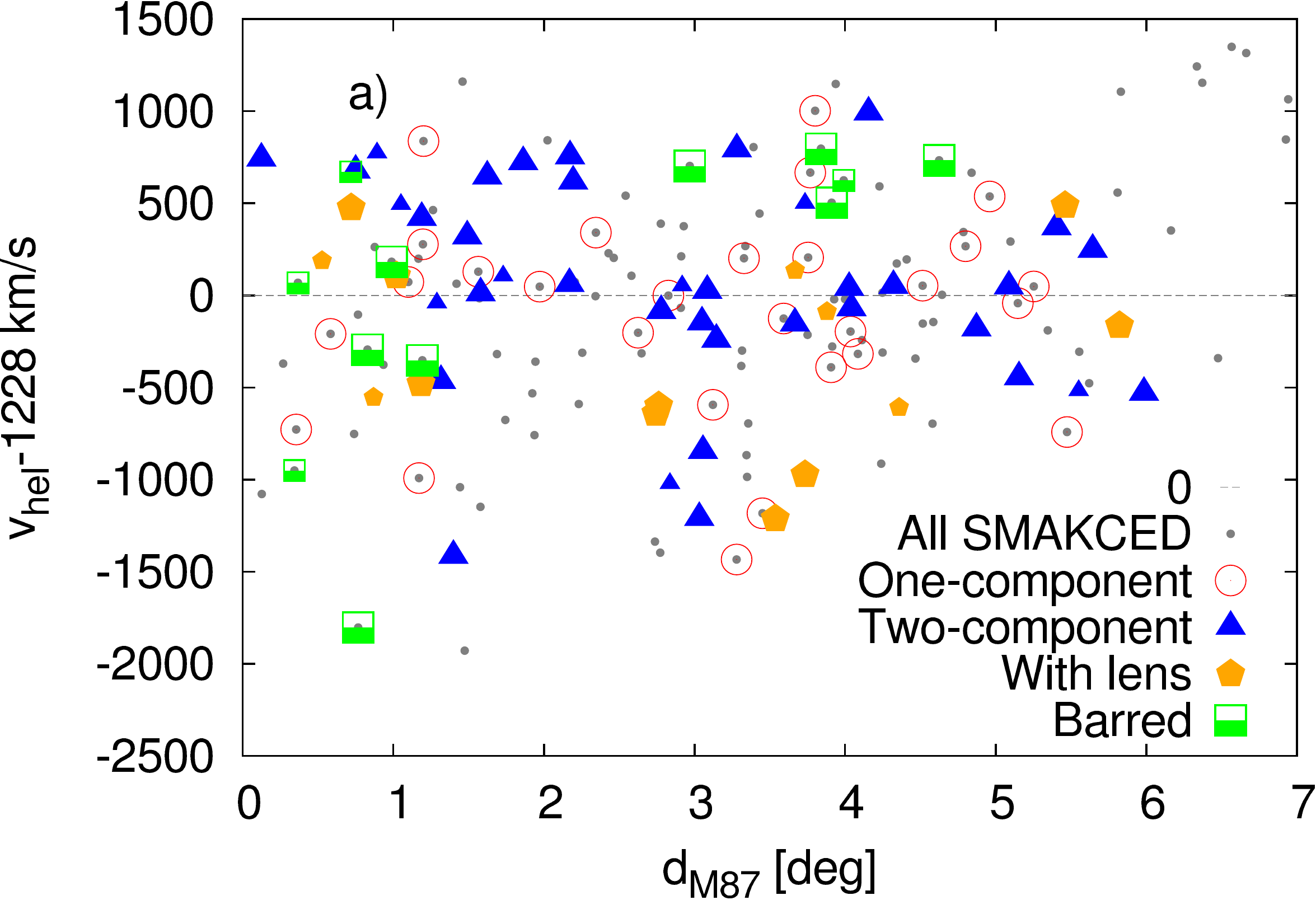}\\
\includegraphics[height=8.5cm, angle=-90]{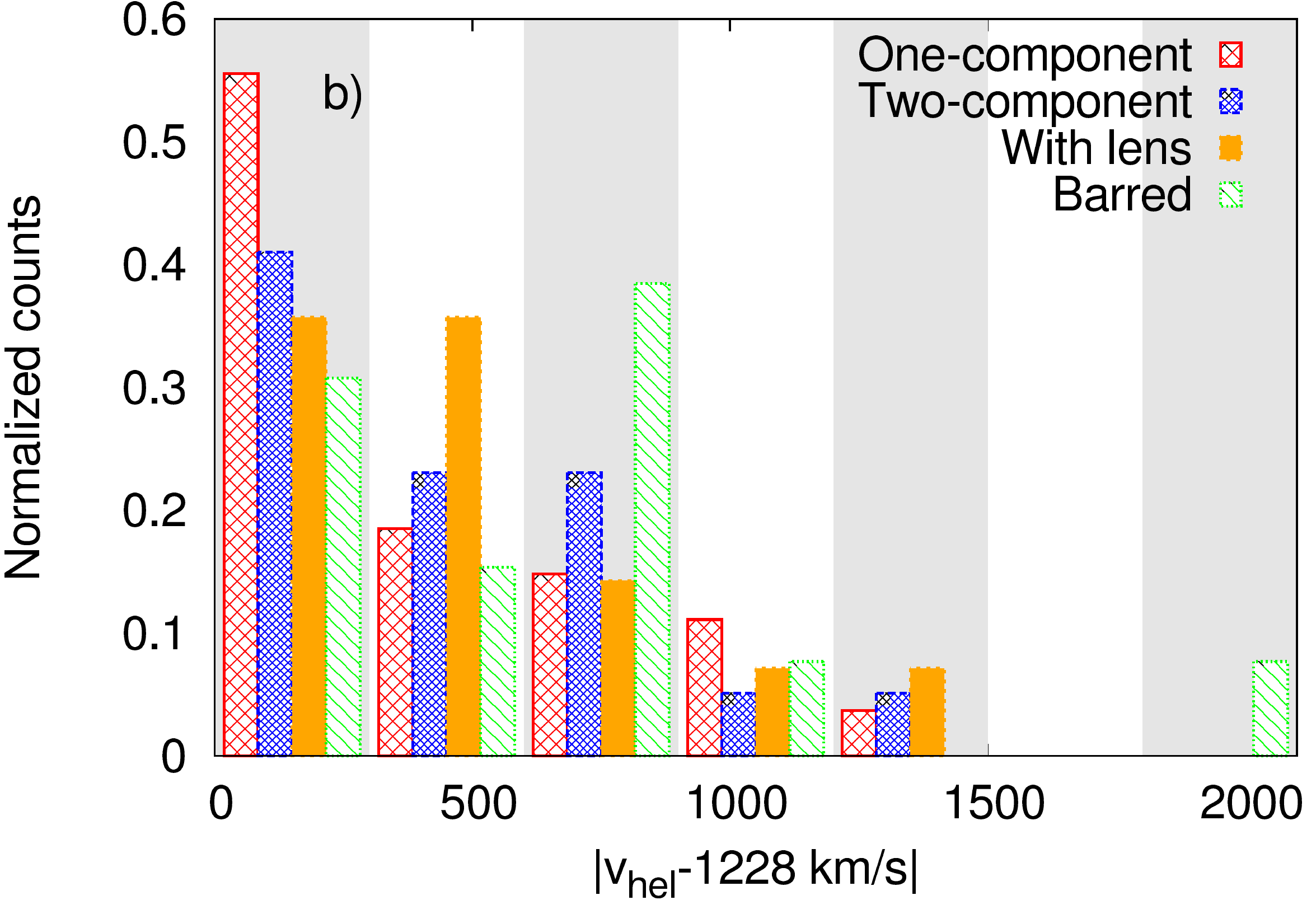}\\
\includegraphics[height=8.5cm, angle=-90]{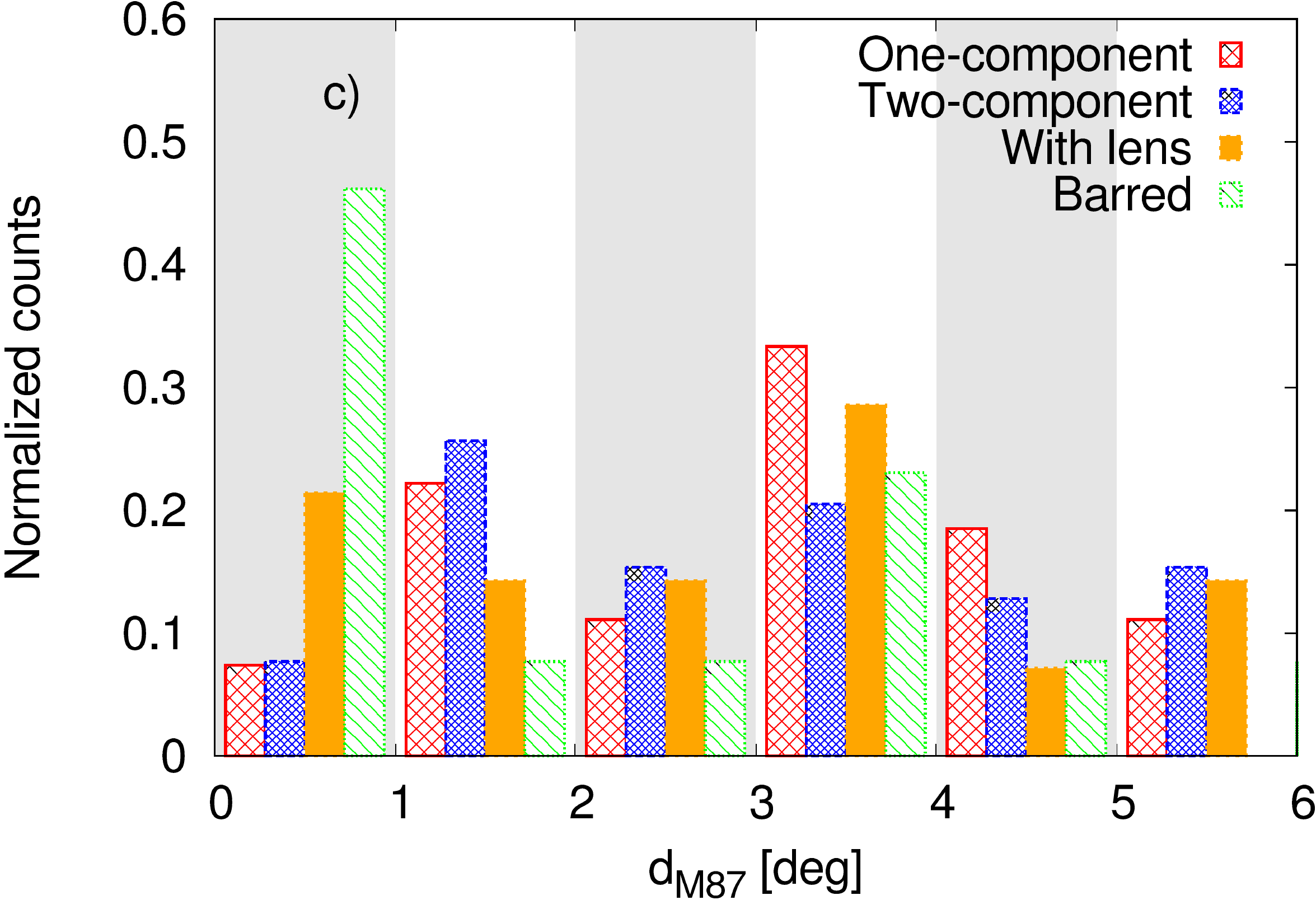}
\caption{\textit{Upper panel:} Projected clustercentric distance versus relative line-of-sight velocity ($1228$ km/s is the median velocity of all dE(N), \citealt{Lisker:2009p3975}). The smaller colored symbols indicate the galaxies with less certain decompositions. \textit{Middle panel:} Normalized histogram of the relative line-of-sight velocities.  \textit{Lower panel:} Normalized histogram of the projected distances to M87. The histograms are normalized to the number of galaxies in each group.} 
\label{fig:helio}
\end{center}
\end{figure}

Fig.\ \ref{fig:helio} shows the projected distance to M87 and the line-of-sight velocities of the galaxies relative to M87 in the cluster center (in Appendix \ref{sec:additional} the diagram is shown with the local projected galaxy density as an alternative parameter for the environment).
The different groups do not clearly separate. 
The biggest difference is found between the one-component and barred galaxies.
In the innermost bin of clustercentric distance ($d_{\textrm{M87}}<1^\circ$ or 0.284 Mpc), these groups strongly differ in their abundance. Furthermore, the velocities of one-component galaxies are peaked towards the cluster velocity, while the barred 
galaxies have a wider distribution.
The two-component galaxies and galaxies with a lens fall somewhat in-between.
A KS-test tells   that the difference between the distance distributions
of one-component and barred galaxies is not significant ($p\sim$10\%).
However, if the   one-  and two-component galaxies on  one hand, and galaxies with a lens or a bar
on the other hand, are combined to two groups the test results in $3\%$, thus showing a significant difference of the two (combined) distributions.

\section{Comparison with more massive galaxies}
In order to judge whether the found components follow any of the scaling relations seen in bright galaxies,
 we use NIRS0S  (Near-InfraRed S0 galaxy Survey, \citealt{Laurikainen:2011ho}) and  OSUBSGS (Ohio State University Bright Spiral Galaxy Survey, \citealt{002ApJS..143...73E}, decompositions in \citealt{Laurikainen:2004kt}) as comparison samples.
 NIRS0S is a magnitude limited ($m_B\leq12.5$ mag) survey of all galaxies ($\sim$200) in RC3 with a Hubble type $-3\leq T \leq 1$ and an inclination less then $65^{\circ}$.
 OSUBSGS is an optical and near-infrared imaging survey of 205 spiral galaxies, also selected  from RC3. The sample was limited to   $m_B \leq 12$ mag, to Hubble types $0 \leq T \leq 9$, and inclinations less than $65^\circ$. 
  The galaxies from these samples are located in many different environments. These studies have the  advantage that their decompositions were done in a similar manner as in this study. However, the decompositions were carried out using a different  code (\textsc{BDBar}).
 In order to ensure the consistency in our comparison, we reran the decompositions of NIRS0S with \textsc{galfit} (Appendix \ref{section:bdbar}).
Some of the galaxies were observed in the $K$-band, for which we use the 2MASS $H$-$K$ colors 
to convert the brightnesses to the $H$-band. 
\begin{figure*}
\begin{center}
\includegraphics[height=\textwidth, angle=-90]{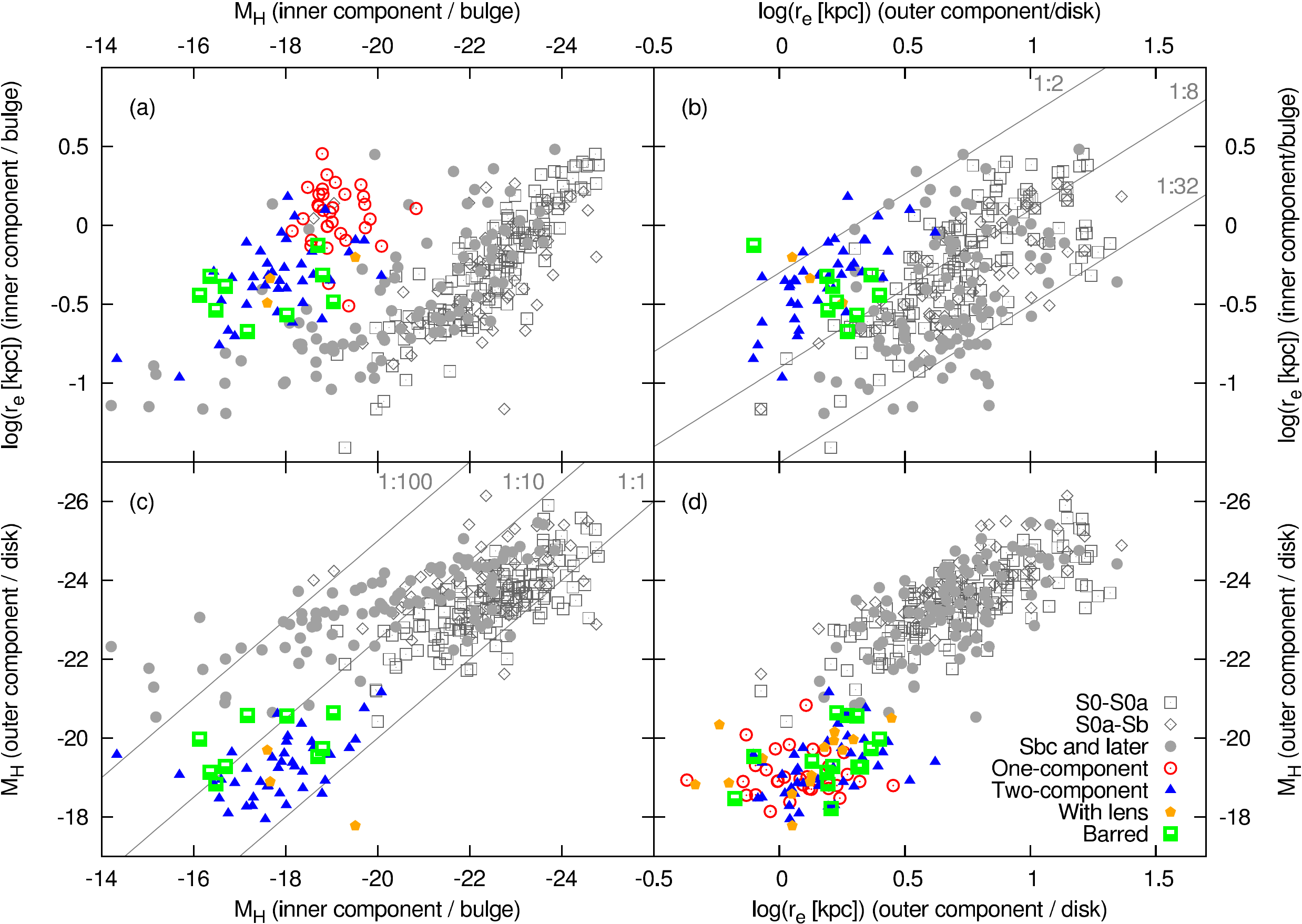}
\caption{The inner and outer components of the galaxies in our sample are shown in the scaling relations of the bugle and disk components of the bright galaxies. The earliest bin of Hubble types is taken from NIRS0S and the two other from OSUBSGS (see text). Panels \textit{(a)} and \textit{(d)} display the relations of brightness and radius of the inner/bulge or outer/disk components, respectively. Panels \textit{(b)} and \textit{(d)} show the scalings for radius and brightness, respectively, versus the corresponding parameter of the other component. In panel \textit{(b}) the gray lines show a ratio of the effective radii of bulge and disk (or of the inner and outer component) of 1:2, 1:8 and 1:32, while in panel \textit{(c)} they show a bulge to disk (or inner to outer component) ratio of 1:1, 1:10 and 1:100. The one-component galaxies of our sample are plotted in both the bulge and disk panels, { while the
for galaxies with a lens in addition to only one `global' component, we add that global component to the disk panel.} Galaxies with lens or bar are plotted in the respective panels, if there was an inner S\'ersic or outer component fitted in the decomposition.\label{fig:nirs0s}}
\end{center}
\end{figure*}

In Fig.\ \ref{fig:nirs0s} we plot the inner and outer components of the SMAKCED early-type dwarf galaxies { (see also Table~5)},
and compare them with the properties of bulges\footnote{``Bulge'' refers here to the
bulge component in the photometric decomposition and does not imply any  physical meaning.  }  and disks 
 in the bright galaxies.
Again, the photometric scaling relations are shown for the effective radii and galaxy brightnesses, and the corresponding figure
with mean effective surface brightness instead of brightness can be found in Appendix \ref{sec:additional}.

The bulges of bright S0$-$Sb  galaxies form a rather steep relation in effective radius\footnote{Note that, the effective radii (that is, half-light aperture SMAs) of components are also referred to by $r_e$, although they are obtained by fitting profiles to components and are not non-parametric quantities (\S\ref{section:fitting}).} versus
brightness (Fig.\ \ref{fig:nirs0s}a).
The later types show more scatter, a few even fall into the region occupied by the inner components of the dwarfs.
The latter follow a  relation with a similar slope like the bulges in bright galaxies, but  offset towards larger radii.
However, this relation might partly be  
 due to the limited range of total brightness:
for galaxies with similar total brightness, the inner components of the complex galaxies have to 
be smaller and fainter than the one-component galaxies, i.e.\  whole galaxies. 
Interestingly, the one-component galaxies just seem to scatter, while for the
two-component galaxies the size and brightness of the inner component are correlated (Spearman's $r$ and $p$: $r=0.07$ and $p=70\%$,  $r=-0.51$ and $p=0.7\%$, respectively).

For the bright galaxies also the disk size scales with its brightness (Fig.\ \ref{fig:nirs0s}d), although the relation is broader than for the bulges. \
If the relation of the bright galaxies is extrapolated the dwarfs do not fall on it. Their outer components
are larger. 
However, the about 10  faintest non-dwarf galaxies 
might hint at some turn of the relation towards larger radii.
Due to the sample selection
faint late types do not form part of OSUBSGS. A different comparison sample would be needed to confidently judge  whether or not the relation turns to the outer components of the dwarf galaxies.  

Fig.\ \ref{fig:nirs0s}b illustrates that  the sizes of  the bulges and disks  of bright galaxies are correlated as well.
The inner and outer components of the dwarfs show a similar relation but are offset towards a ratio of the sizes of the inner and outer component  closer to unity. (The average value for the late-type spirals is consistent with the findings of \citealt{1996ApJ...457L..73C}).
This could be a hint that the components are not of the same nature as the bulges and disks in bright
galaxies.

In Fig.\ \ref{fig:nirs0s}c  the brightnesses are plotted, thus effectively illustrating the 
bulge to disk ratio, and the 
ratio of light in the inner and outer components.
For the bright galaxies the Hubble sequence becomes visible: the later types
have less prominent bulges. The clearest relation exists for the bulges of the S0 galaxies.
The later types are offset to smaller bulge-to-disk ratios.
The ratio of inner-to-outer component in the dwarf galaxies fall about on the extension of the relation for early-type disk galaxies.
Also here a comparison sample for the latest type  galaxies is missing. 
{ We conclude that the common bulge-plus-disk picture of bright galaxies is not applicable to the combination of inner and outer components of early-type dwarfs.} 

~\\

\section{Discussion}
\subsection{Multicomponent structures}
{ In this study nearly two thirds of the early-type dwarfs are found to have complex structures.}
The fraction of one-component galaxies is found to be highest in those early-type dwarf galaxies, where no disk features or central star formation was detected  \citep{Lisker:2006p385,Lisker:2006p392}. 
This might not be so surprising, because of the way disk features were detected by \citeauthor{Lisker:2006p385}. Their unsharp masking  reveals spiral arms, bars, and possibly edge-on disk components. Some of the spiral arms that we  see in our images are found in one-component galaxies. All the other features, however, are likely to be related to multi-component structures.
The strongest correlation with the decomposition characterization is found with galaxy brightness: at the bright end of the sample the fraction of one-component galaxies is about 0\% and it increases steadily towards fainter galaxies. The trend seems plausible, since the low mass galaxies are expected to be more easily affected by dynamical heating that can wash out structures or prohibit their formation. 

Already \citet{binggeli_cameron} defined 5 different surface brightness profile types (their Fig.\ 6).
Their group I comprises profiles that deviate globally from an
exponential shape. They did not consider the S\'ersic function for the profiles at that
time, and many of these may in fact be fitted with a single S\'ersic function.
Their groups IVa and IVb are profiles following an exponential function, with and without a
nucleus, respectively. Therefore, the profiles of groups I, IVa, and
IVb correspond to our models using a single S\'ersic function or a S\'ersic
function plus nucleus.
The profiles in their groups II, IIIa, and IIIb show a break:
the outer part can be described with an exponential function. However, the inner
part does not follow this outer exponential.
It either follows an exponential function with a shorter  scale
length (IIIa and IIIb with and without an additional nucleus) or
is steeper than an exponential (which we would describe with a S\'ersic 
function with $n>1$). Therefore, these three groups correspond to
our two component galaxies.
Additionally \citeauthor{binggeli_cameron}  have a type V, which has a deficiency of flux in the center compared to an exponential profile
(also see \citealt{allen}).
These models might be to some degree represented in our models with an outer component with $n<1$.
Thus  \citeauthor{binggeli_cameron}'s results did already indicate part of the structural
complexity of the early-type dwarf galaxies found in this study, but with the S\'ersic profile becoming popular,
their profile classifications were largely forgotten.

\citet{gavazzi+profiles} fitted the near-infrared light profiles of over 1000 galaxies of all types, including galaxies in the Virgo cluster, with a de Vaucouleurs $R^{1/4}$ law, an exponential function, or a bulge+disk combination. A handful of objects were early-type dwarf galaxies, and the authors  found that not all of them follow a simple exponential profile. 
\citet{2003A&A...407..121B} studied Virgo cluster early-type dwarf galaxies in the optical ($B$- and $R$-band). They
state in the abstract that `the observed profiles of the brightest cluster dwarfs [in the Virgo cluster] show significant deviations from a simple S\'ersic model, indicating that there is more inner structure than just a nucleus.' They stress that radial profiles  of parameters like the ellipticity and position angle support this finding. The features  in the profiles are not restricted to objects like IC3328, where unsharp masking or model subtraction can reveal spiral arms, but occur also in `many apparently normal dEs'.
Based on HST data from the ACS VCS, \citet{Ferrarese:2006p586} reported the detection of stellar disks on nuclear and kiloparsec scales, as well as bars, in more than half of the Virgo early-type galaxies with `intermediate luminosity'.
More recently, \citet{Mcdonald:2011p4445} did profile decompositions using  near-infrared images for  286 Virgo cluster galaxies, of which about 65 are early-type dwarfs.
They find a similar fraction of galaxies that are fitted by simple exponential profiles as opposed to more complex galaxies with two-component profiles (see also Appendix \ref{section:mcd}).  
For individual galaxies a different decomposition might be reported in the mentioned studies, since they use one-dimensional,  sometimes automated profile fitting, different model assumptions such as the possibility of separately fitting nuclei or restricting the single component models to exponental profiles (as it is the case in   \citealt{Mcdonald:2011p4445}). Also, the decompositions presented in this study rely on a higher image quality.
However, qualitatively  our results are consistent with theirs.

In the following we discuss studies in which the profiles and morphologies of early-type dwarfs in the Coma cluster are studied, mostly using HST data  (for Fornax see e.g.\ \citealt{SmithFornax} and \citealt{2012ApJS..203....5T}).
The Coma cluster is denser and richer than the Virgo cluster.  \citet{graham_guzman} fitted light profiles and  in obvious cases included an outer exponential component in addition to a S\'ersic component (and a nucleus when present).  
They found 3 galaxies out of 18 to be complex.
\citet{2005AJ....130..475A} fitted about 100 light profiles of early-type galaxies within the magnitude range  $-18\leq M_B\leq -16$~mag. Using the largest deviation of the fit from the observed profile as a criterion, they found 34\% of the galaxies, with reliable photometry, to be complex.
\citet{MA} focused on detecting bars in early-type dwarf galaxies. Their bar fraction and its decrease towards lower galaxy brightnesses is consistent with our findings.
\citet{2012ApJ...746..136M} used unsharp masking to identify disk signatures like spiral arms and bars for a larger sample of 333 galaxies ($M_V>-18$ mag). Only 13 galaxies host such features. However, they conclude that this is in good agreement with \citet{MA} and the decline of the bar fraction towards fainter galaxies. Interestingly, for bright S0 galaxies they find a similar bar fraction as in Virgo, despite the higher galaxy density. They suspect an interplay of two effects of the higher galaxy density to cause the similar bar fractions in the different clusters, similar to \citet{janz+2012}'s suggestion for the 
 early-type dwarfs in the Virgo and the Coma cluster: encounters are more frequent, but the galaxies are more stable against bar instabilities. 
 \citeauthor{2012ApJ...746..136M} argue that the higher stability of galaxies in the dense environment is due to a more effective dynamical heating of the galaxies, a smaller gas fraction, and  encounters being less efficient due to higher relative velocities.

\subsection{Isophotal shapes and kinematics}

{  The isophotes of elliptical galaxies show small deviations from ellipses: 
galaxies with disky and/or boxy isophotes are common. \citet{1988A&A...193L...7B}
 showed that disky galaxies ($c_4 > 0$) are always consistent 
with being rotationally flattened, while boxy galaxies frequently are 
supported by anisotropic velocity distributions and rotating slowly. 
A larger dataset, based on SDSS data, later published by  \citet{Pasquali}, 
shows the same effect, and that the fraction of rotating objects
increases going to lower magnitudes.
We find that this trend breaks down for
$M_r > -19$ mag, i.e.~in the dwarf regime. Measuring the boxiness
as indicated in Section  \S\ref{section:boxiredux} we find a considerable scatter in the boxiness
at every magnitude, both inside $r_e$ and between $r_e$ and $2r_e$ (Fig.~\ref{fig:boxiness_all}). 

\begin{figure*}
\begin{center}
\includegraphics[height=0.48\textwidth, angle=-90]{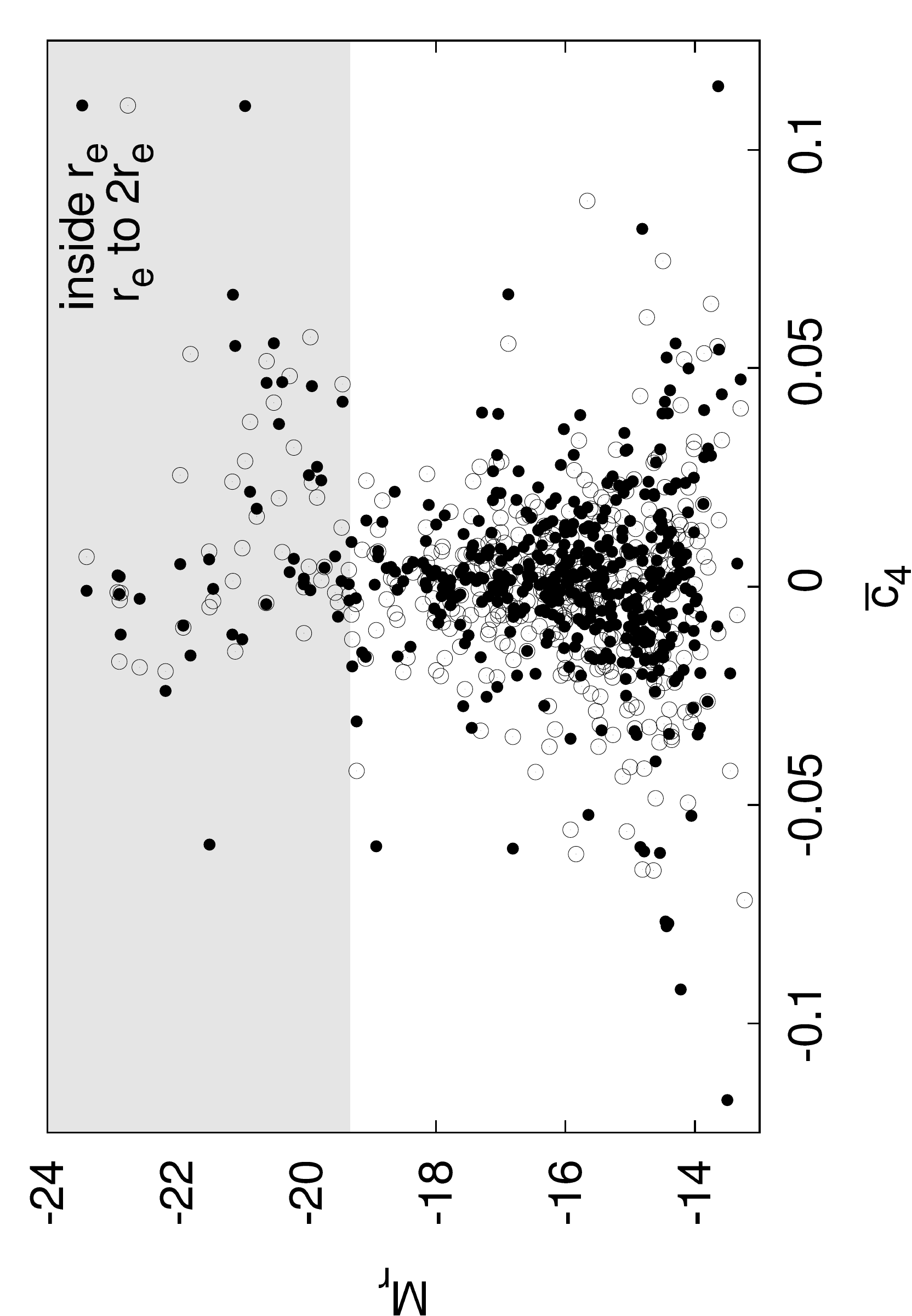}\ \ \ \ \ \ \ \ \ 
\includegraphics[height=0.48\textwidth, angle=-90]{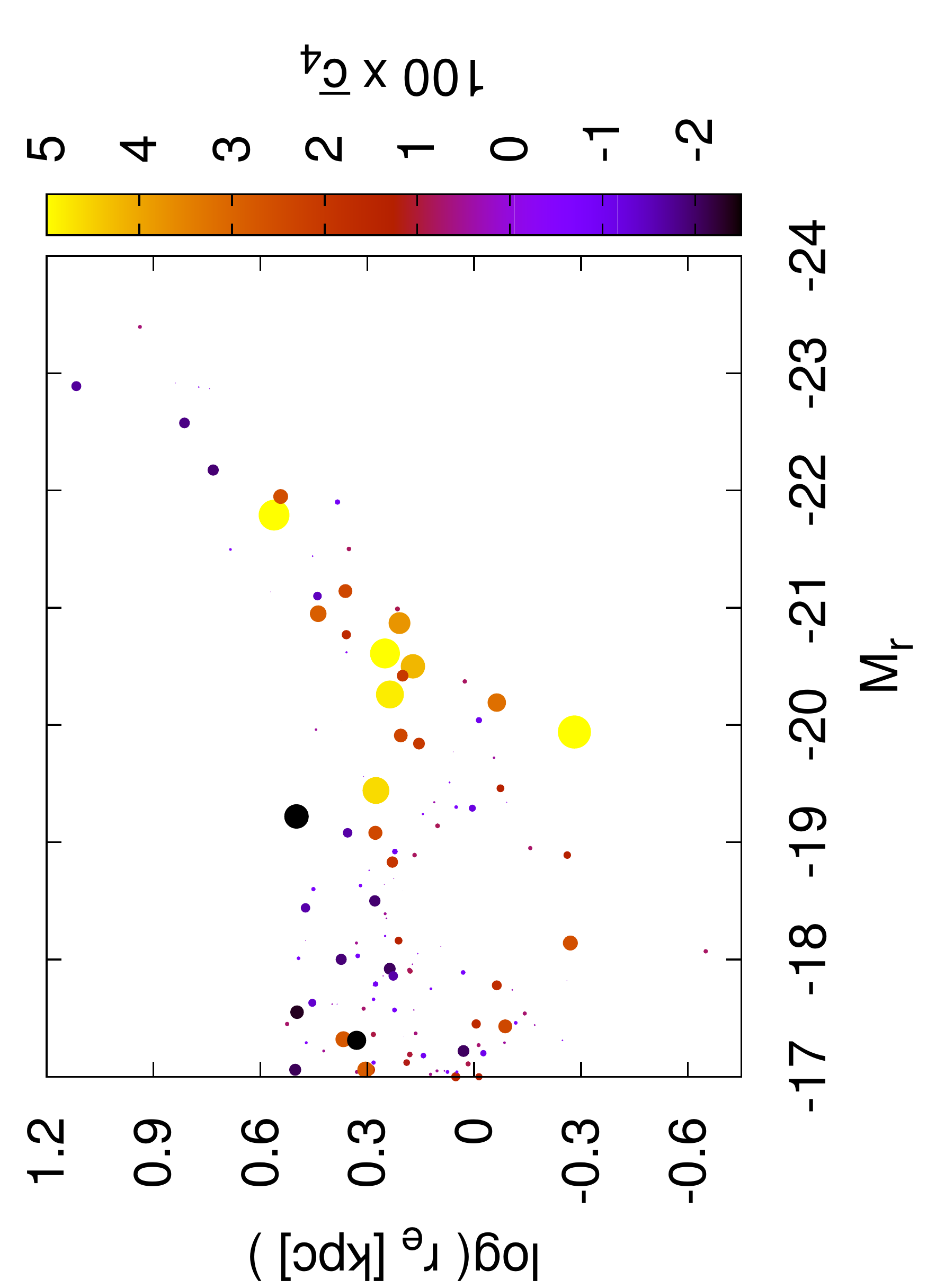}
\caption{\textit{Left panel:} boxiness versus galaxy brightness for all early types in Virgo, using $r$-band images from SDSS; data from profiles fits in \citet{Janz:2008p151}. The gray area indicates the bright galaxies, for which there is the trend of  galaxies being more disky towards fainter galaxy brightness. Filled and open symbols mark the boxiness inside $r_e$ and between $r_e<r<2r_e$, respectively. Values larger than 0 indicate disky, while those smaller than 0 mean boxy isophotes.
\textit{Right panel:} the value of the boxiness is shown in the 
optical brightness versus radius relation \citep{Janz:2008p151} for early-type dwarfs and bright early types. The boxiness is color coded and the size of symbols reflects the absolute value so that the galaxies with nearly elliptical isophotes disappear.
\label{fig:boxiness_all}}
\end{center}
\end{figure*}

\citet{Toloba:2009p3937} found that galaxies in the center of the Virgo cluster 
are less supported by rotation that galaxies in the outer regions. At the same 
time they were older and more boxy (see their Fig.~2, and see also 
\citealt{Toloba:2011p3986}). Here we try to reproduce the relation between kinematical
support and boxiness. 

For the kinematics we made a compilation of the available literature data  \citep{002A&A...384..371S,002MNRAS.332L..59P,004AJ....128..121V,Chilingarian,Toloba:2011p3986}. 
For galaxies covered by more than one
published dataset, we gave preference to \citeauthor{Toloba:2011p3986} and
choose the one with the largest radial extent of the rotational curve otherwise. These studies
measure the velocity dispersion $\sigma$ in different ways. One should realize that
our literature compilation is a heterogeneous set of measurements, with
in which $\sigma$ has been determined in different ways, and
which does not allow an in-depth investigation. For that reason the 
largest dataset, the one of \citet{Toloba:2011p3986}, is indicated in larger symbols in Fig.~\ref{fig:vsigma},
 where the ratio of rotational velocity and central velocity dispersion
is shown as a function of ellipticity. Since for most galaxies
the rotation curves are still rising, the $v/\sigma$ values should
be considered as lower limits. Overall, we do not find any differences in the kinematical characteristics of the galaxies from
different structural groups. This statement holds also when
restricting the analysis to the homogeneous data of \citeauthor{Toloba:2011p3986}. Interestingly, also the one-component galaxies
with kinematics show no evidence of being slow rotators. The
numbers of galaxies with kinematics might, however, be still
too small to draw firm conclusions. A homogeneous analysis
of kinematics through SMAKCED spectra for a larger
sample is under way (Toloba et al., in prep.).

\begin{figure}
\begin{center}
\includegraphics[height=\linewidth, angle=-90]{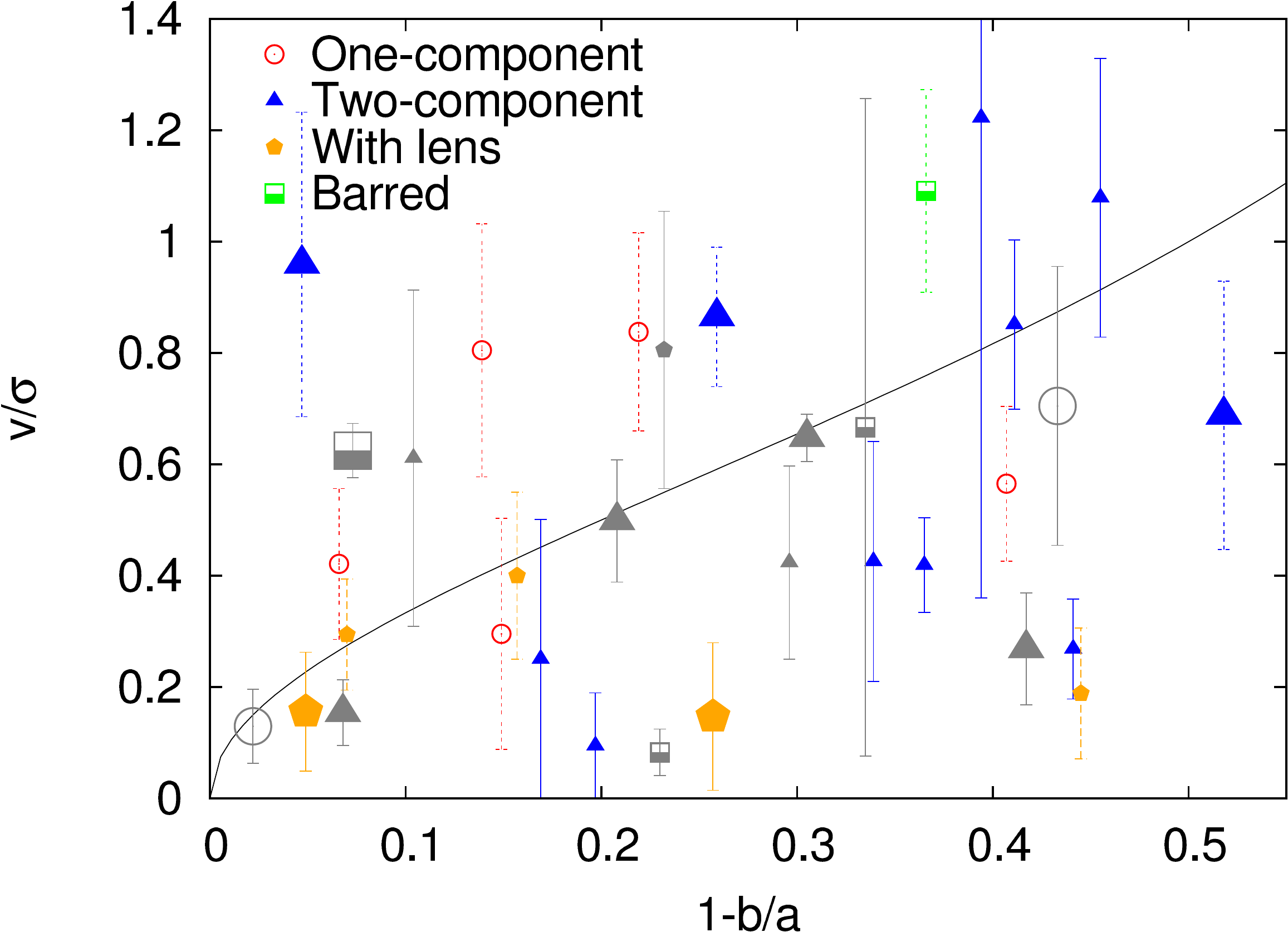}
\caption{$v/\sigma$ diagram. The axis ratio $b/a$ is measured at $2r_e$. Galaxies for which the kinematical data reach less than 2/3 of $r_e$ are shown with gray symbols. The gray solid line indicates an edge-on, oblate, isotropic galaxy flattened by rotation. { Data from \citet{Toloba:2011p3986} are indicated with bigger symbols.} 
\label{fig:vsigma}}
\end{center}
\end{figure}

When looking at the relation between rotational support and boxiness
we use $(v/\sigma)^*$, i.e.~  $(v/\sigma)/\sqrt{\epsilon/(1-\epsilon)}$. Although this parameter does not 
contain much information for galaxies with $\epsilon \sim 0$, it allows
us to investigate in a rough way whether boxiness correlates with rotational
support. Fig.~\ref{fig:boxy_vs_kin} shows both the median boxiness and all the individual
boxiness features identified in the radial profiles (see \S\ref{section:boxiredux}).
For neither panel a correlation between disky isophotes and rotational
support is seen. If any relation at all is present, it goes to the other direction.
When trying to understand this, and why \citet{Toloba:2011p3986} do find a correlation between
disky isophotes and rotational support, one has to realize that we
determine $c_4$ differently from \citeauthor{Toloba:2011p3986}. They assigned galaxies
with clear disky features in the profile the largest peak value for $c_4$, regardless
of the behavior of the rest of the profile and of its median value.
In case no significant disky feature was found, the median value of the profile 
was used even if there were boxy subsections. In our analysis here we treat boxy and disky 
profiles in the same way. 
}

\begin{figure}
\begin{center}
\includegraphics[height=\linewidth, angle=-90]{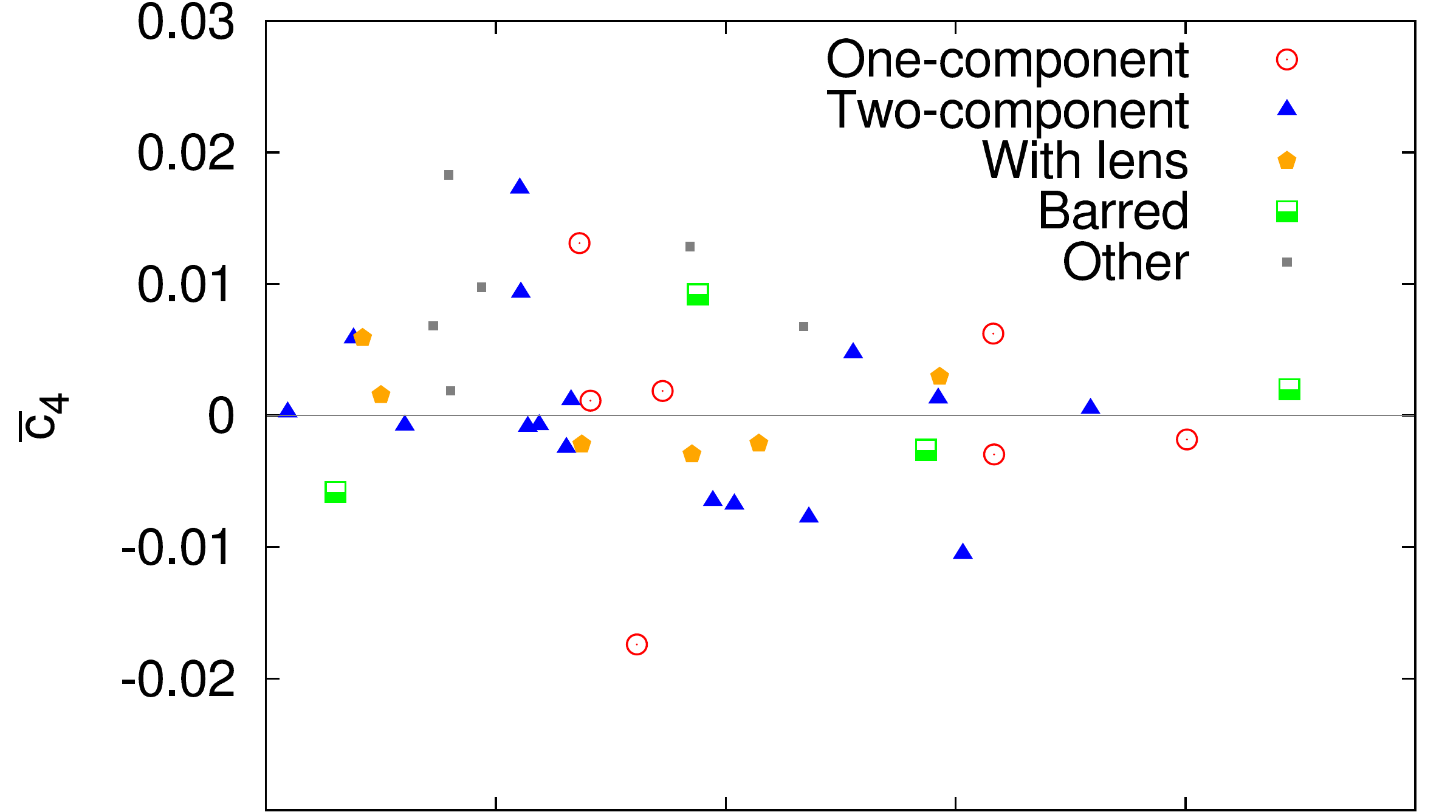}\\
\includegraphics[height=\linewidth, angle=-90]{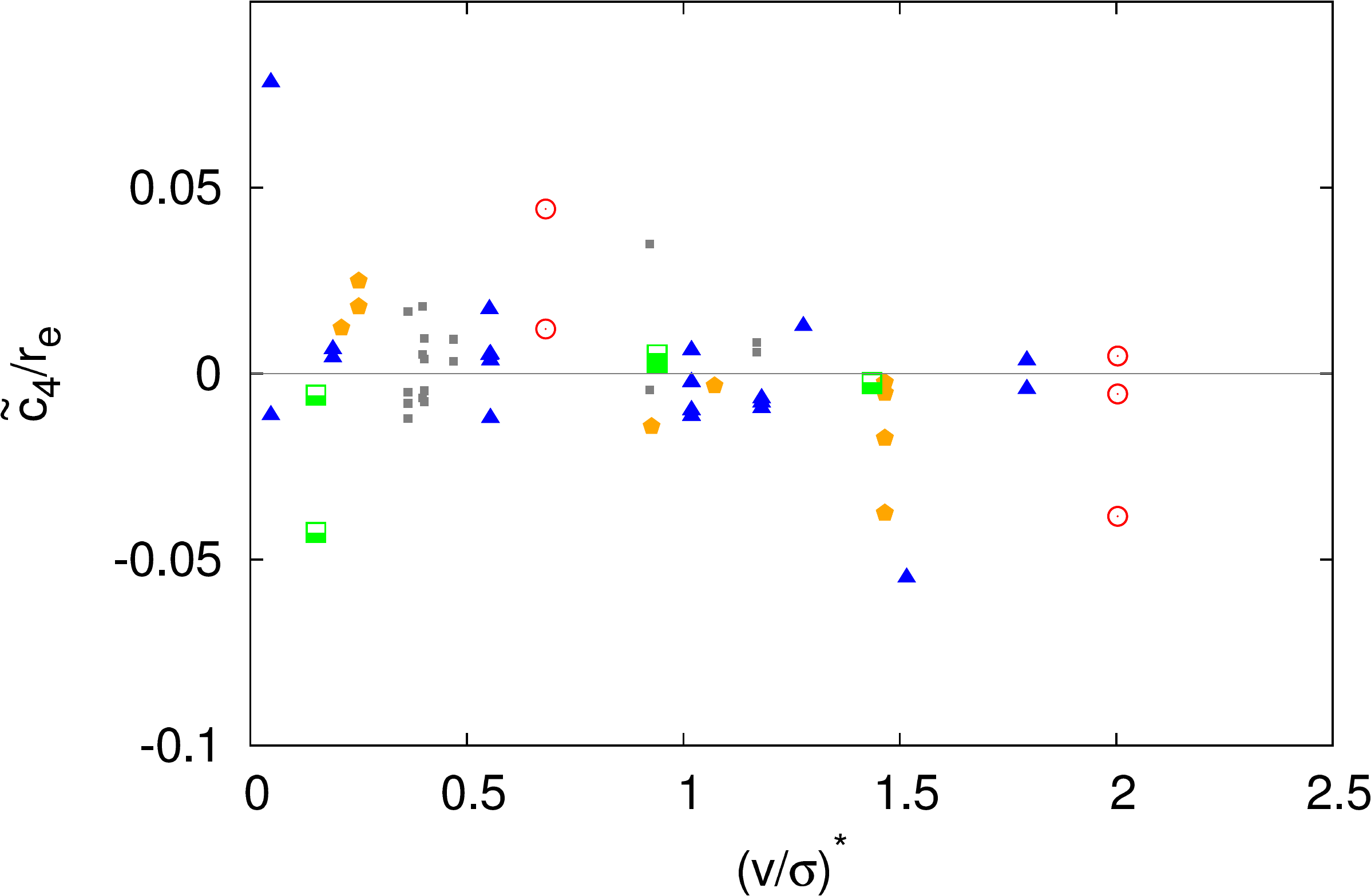}
\caption{Boxiness versus  $(v/\sigma)^*$, with kinematic values from literature. In the \textit{upper} and \textit{lower} panel the boxiness is  quantified by the median boxiness ($\bar{c_4}$) and the integrated values of boxy or disky features  ($\tilde{c_4}/r_e$), respectively. Values larger than 0 indicate disky, while those smaller than 0 mean boxy isophotes. \label{fig:boxy_vs_kin}}
\end{center}
\end{figure}

While the quality of the kinematic measurements (uncertainties
shown in Fig.~\ref{fig:vsigma}) only allow us to look for general
trends, the same trend that is seen in Fig.~\ref{fig:boxy_vs_kin} with $(v/\sigma)^*$
would be seen when plotting instead $v/\sigma$ or only v on the
ordinate, and also when removing the faintest galaxies and
the ones with low radial extent from the sample. At this
point it should be mentioned that \citet{2013MNRAS.428.2980R}, using two-dimensional
spectra for a small sample of Virgo early-type
dwarfs, recently reported that kinematically the dwarfs form
anything but a homogeneous group of objects. { Furthermore,
it can be noted that \citet{Emsellem:2011p4054} did not find the
strong correlation between boxiness and rotational support
of \citet{1988A&A...193L...7B} for bright galaxies, using a sample
spanning a larger range in galaxy brightness than \citeauthor{1988A&A...193L...7B}.}

Generally, disky isophotes are thought to be related to disks. 
For boxy shapes in the dwarf systems one suggestion is that 
they are so-called naked bars \citep{Mastropietro:2005ba}: a 
strong bar is induced in a disk  by an encounter. Subsequent 
encounters harass the galaxy further and remove the outer disk 
{ so that only a bar-like shape remains, which might be boxy. }
In the most extreme case they might even be caused by
mergers of two disk galaxies with appropriate orbital geometry
(see discussion of the boxy shape in  \citealt{leda}).
All  these interpretations attribute the boxy shape to disk phenomena.
As seen before, especially  galaxies with high ellipticity, 
presumably inclined galaxies, have  boxy sections in the 
outer parts of the profiles, which is consistent with the aforementioned 
explanations.

\subsection{Are early-type dwarf galaxies dwarf versions of ellipticals?}
It is still debated, whether it can be concluded from photometric scaling relations  (e.g.\ Fig.\ \ref{fig:2MASS}{\it a}) that 
the early-type dwarfs and bright early-type galaxies
are distinct galaxy types, or that they follow  one curved relation.
 \citet{2009ApJS..182..216K} 
suggest that they are different,
and that compact ellipticals like M32 are the true low brightness ellipticals \citep{1984ApJ...282...85W}.
However, \citet{graham_guzman} show that scaling relations including  the effective radius $r_e$ or the effective surface brightness 
$\left<\mu\right>_e$ have to be curved, if the galaxy profiles follow S\'ersic functions 
and the S\'ersic parameters $\log(n)$ and $\mu_0$ form linear relations with  the galaxy brightness.
\citet{Janz:2008p151} find  deviations from this expected curved behavior in the size brightness diagram (their Fig.\ 2): 
the brightest dwarf galaxies are too large,
while the faintest non-dwarf galaxies at a similar brightness are too small.

This is tentatively confirmed with our NIR data (see Fig.~\ref{fig:2MASS}{\it a}).
Our multi-component analysis gives a possible explanation for why the
galaxies do not follow { exactly} the curved relation, even though \citeauthor{graham_guzman}'s 
argument is  a rigid, mathematical one. Namely, we  find that  the assumption that all the galaxies
can be described with a single S\'ersic profiles is not fulfilled. 
The multicomponent structures and their higher frequency   in brighter galaxies 
cause a relation between the concentration parameter and  the galaxy brightness (see Appendix \ref{sec:additional}).
When fitting with single S\'ersic functions the higher concentration results in a higher S\'ersic $n$ (Fig.\ \ref{fig:profiles}, see also Appendix \ref{sec:sersic}) 
and thus also to a  relation between the effective S\'ersic $n$ and galaxy brightness.
 However, the multi-component nature by itself may not invalidate the existence of a general curved relation of 
 effective radius versus galaxy brightness for the \emph{global} galaxy properties.

{ A similar thing can be seen in Fig.\ \ref{fig:boxiness_all} with the boxiness:
from bright to fainter galaxies the bright early types change from being boxy to becoming more and more disky.
On the other hand, the early-type dwarfs  have both boxy and disky shapes.
We note that at intermediate brightness, where galaxies classified as dwarf and as bright early types overlap, the difference in radius is largest between early-type dwarf and E galaxies. In boxiness, the difference is instead largest between early-type dwarf and S0 galaxies.
Also,  the smallest elliptical galaxies scatter around zero boxiness.}
Thus we refrain  from a statement about the true low-mass ellipticals. 
The compact  ellipticals may be the extension of the elliptical galaxies towards lower mass \citep{2009ApJS..182..216K}. Alternatively, a population of true low mass ellipticals may exist and exhibit similar radii for a given brightness 
as the majority of early-type dwarf galaxies, as predicted if they have S\'ersic profiles and follow the linear relations of the S\'ersic parameters with the profiles \citep{GrahamWorley,2011arXiv1108.0997G}. In this case the compact ellipticals may be tidally stripped galaxies in the vicinity of massive galaxies (e.g.\ \citealt{SmithCastelli:2008io,2009MNRAS.397.1816P,Chilingarian}, the latter two also report multi-component structure for the compact ellipticals, as it was found for M32 by  \citealt{grahamM32}). \citet{2011MNRAS.414.3557H} found evidence of stripping for a number of compact ellipticals. 
Or, as a third possibility, true low-mass ellipticals might be rare objects.

In the context of comparing the early-type dwarfs with bright ellipticals and early types (E\&S0) it should also be noted
that { the existence of structures beyond light distributions following simple S\'ersic models alone does not make the dwarfs
necessarily different from bright ellipticals. Recently,  \citet{huang} used three to four S\'ersic functions to describe the overall
light distributions of bright ellipticals.
Also, it is noteworthy  that the SAURON/ATLAS3D project found  the red  low-mass galaxies,  despite their classification as ellipticals, 
to be relatively fast rotating in general  ($v/\sigma \sim 1$; \citealt{Emsellem:2011p4054}), similar to many early-type dwarfs.
However, among those there are also slowly rotating ones \citep[see e.g.][]{Toloba:2011p3986}.}

\subsection{Nature of inner components and one-component galaxies?}
The inner components that we find in dwarf galaxies do not resemble bulges in bright galaxies.  Their profiles are close to exponentials, i.e.\ the median value of the S\'ersic index is $n=1.15$, 75\% of the inner components have $n<1.6$.
Their flattening distribution is similar to that of the outer components. 
In fact, the flattening distributions of the inner and outer components
are similar to that of spiral galaxies. 
Also, the shapes of bulges and disks  (identified with decompositions)  in low mass late-type galaxies  might be similar (see e.g.\ 
\citealt{2010A&A...521A..71M}). Most of them are thought to arise in the  secular evolution of the galaxy and are then called pseudobulges (\citealt{Kormendy:2004ka}).
However,  the inner  components  in this study are clearly offset from the scaling relations of bulges in bright galaxies.

\begin{figure}
\begin{center}
\includegraphics[width=5.8cm, angle=-90]{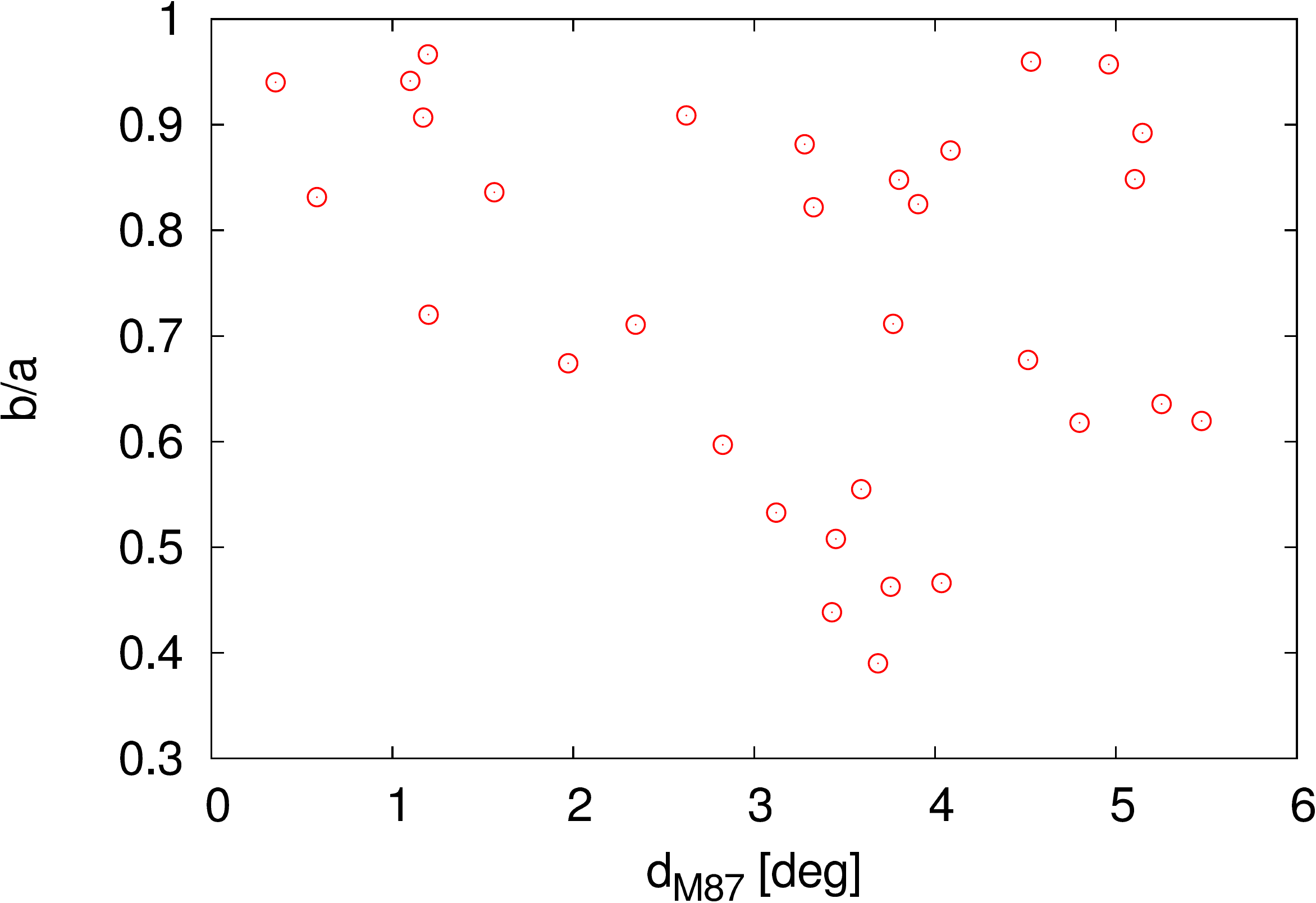}
\caption{Axis ratio of one-component galaxies versus clustercentric distance. 
One-component galaxies with flat shapes do not exist in the cluster center.
\label{fig:onecomp}}
\end{center}
\end{figure}

The photometric parameters of one-component galaxies  are apparently more similar to those of  the outer components of the complex galaxies.
They are also not significantly rounder. The few one-component galaxies with kinematics show significant rotation. The last point may be a selection effect, since round galaxies might not have been the preferred objects in studies of rotation.
If one assumes that some process causes the multi-component structure, then the galaxies with one component could  either be not yet processed, or the multi-component structure could be destroyed. The flattening distribution of one-component galaxies appears to be double-peaked (see Fig.\ \ref{fig:axr}).  
The flatter one-component galaxies  are  found only at larger projected clustercentric distances (Fig.\ \ref{fig:onecomp}; see also \citealt{Lisker:2009p3975})  and tend to have smaller concentration indices than the round ones. 
 When selecting these one-component galaxies with large axis ratios, a trend is seen for them to have higher concentration indices in dense regions (which is not simply caused by a relation between the concentration and the galaxy brightness).
While these statements about the one-component galaxies are clearly tentative, they hint at   the population of one-component galaxies being complex.

 The fact that the fraction of one-component galaxies strongly increases towards
objects of fainter magnitude, i.e.\ lower stellar mass, may be
interpreted in two ways. Either the continuous tidal influence has been so strong for lower-mass galaxies that most structures that had formed have already been smoothed out again, or they did not form in the first place.  The latter could be due to a possibly thicker shape of the progenitor galaxy at lower masses  \citep{2007MNRAS.382.1187K,2010MNRAS.406L..65S}, thus being dynamically hotter, or due to the lower binding energy that makes it less likely to develop internal structures. These scenarios could be tested with dedicated simulations, and
would represent a striking example for the ``fossil record'' of galaxy
evolution mechanisms that is imprinted in early-type dwarfs.

\subsection{Late-type galaxies as possible progenitors}
While a few  of the latest types in the OSUBSGS comparison sample have bulges similar to the inner components in the dwarfs,
the inner components are clearly offset from an extension of  the bulge relation in bright galaxies (see Fig.\ \ref{fig:nirs0s}).
Given that the inner structures are  so different, can the early-type dwarf galaxies be transformed late-type galaxies?

It is worth mentioning that the increasing number of multi-component galaxies towards higher brightness resembles the increasing complexity in the late-type galaxy mixture. 
Using profile decompositions, \citet{gavazzi+profiles} found  a similar trend from  simple exponential galaxies  to two-component late types towards brighter galaxies.  The fraction of simple galaxies in their study is similar to that in this study: among Sc galaxies it is slightly larger, and  smaller among Sd/Sm galaxies. 
This calls for a comparable multi-component study of low mass late-type galaxies, in order to see whether our results can be reproduced by pre-existing multicomponent structures (see also \citealt{GrahamWorley}) that survive the disk thickening during the transformation process (\citealt{Gnedin,Smith10} -- tidal heating, \citealt{Smith11} -- gas depletion from ram pressure stripping).
In this context it is worth noting that the old stellar population of the blue compact dwarf galaxies is morphologically similar to the more compact $\sim$half of the early-type dwarf galaxies \citep{meyer}.

Another option is that the structures are created during the transformation process (for a review of various transformation processes, see e.g.\ \citealt{bosgav06}).
The complete transformation includes removal of the gas and  dynamical heating of the galaxy.
In a massive cluster the hot gas is assumed to effectively strip the internal gas of a galaxy via ram pressure.
With the Gunn \& Gott criterion  \citep{1972ApJ...176....1G} it can be 
shown that the more massive early-type dwarf galaxies should be able to retain some gas in the galaxy center, 
with typical numbers for a passage through the cluster avoiding the very center.
The central star formation can continue for a longer time or might even be enhanced. \citet{Roediger:2009jf} shows that the Gunn \& Gott criterion works reasonably well for most geometries. She also discusses that the ram pressure may or may not enhance the star formation in those regions of the galaxy, where the gas is not yet stripped (see also \citealt{vollmer,bekki}).
A few early-type dwarf galaxies in the Virgo cluster  are detected in \textsc{h i} \citep{Hallenbeck:2012gd}.

In case harassment is to play a role in the dynamical heating of the galaxy, the high speed encounters can induce bar formation in the perturbed galaxy \citep{Mastropietro:2005ba}. If there is still gas in the galaxy, the bar can effectively transfer it to the central regions of the galaxy \citep{shlosman1,shlosman2}. Even without any gas the bar can redistribute the stellar mass and lead to a profile steepening,  as illustrated by the simulations of \citet{Mastropietro:2005ba}. This also demonstrates that the different components that we found can simply be different parts of the disk.

In case the increase of concentration and changes to the light profile are  inherent to the disk,  no significant differences in the shapes and other parameters  
of the inner and outer components are expected. 
Also the similar flattening distributions of the early-type dwarf galaxies and their possible progenitors (see also \citealt{ferguson_sandage,1995A&A...298...63B,sung}) are explained in such scenarios.

\begin{figure}
\begin{center}
\includegraphics[width=6.1cm, angle=-90]{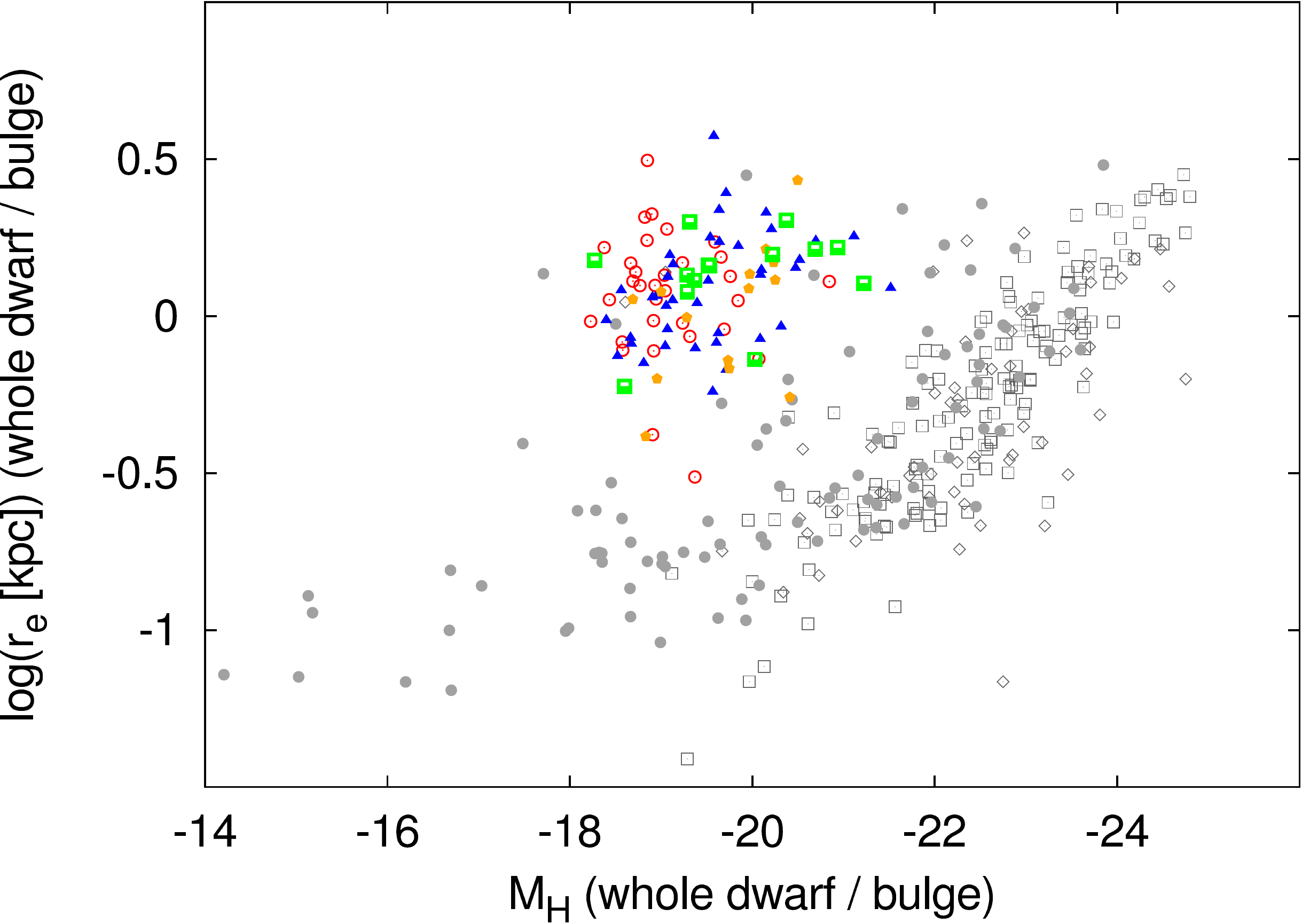}\\
~\\
\includegraphics[width=6.1cm, angle=-90]{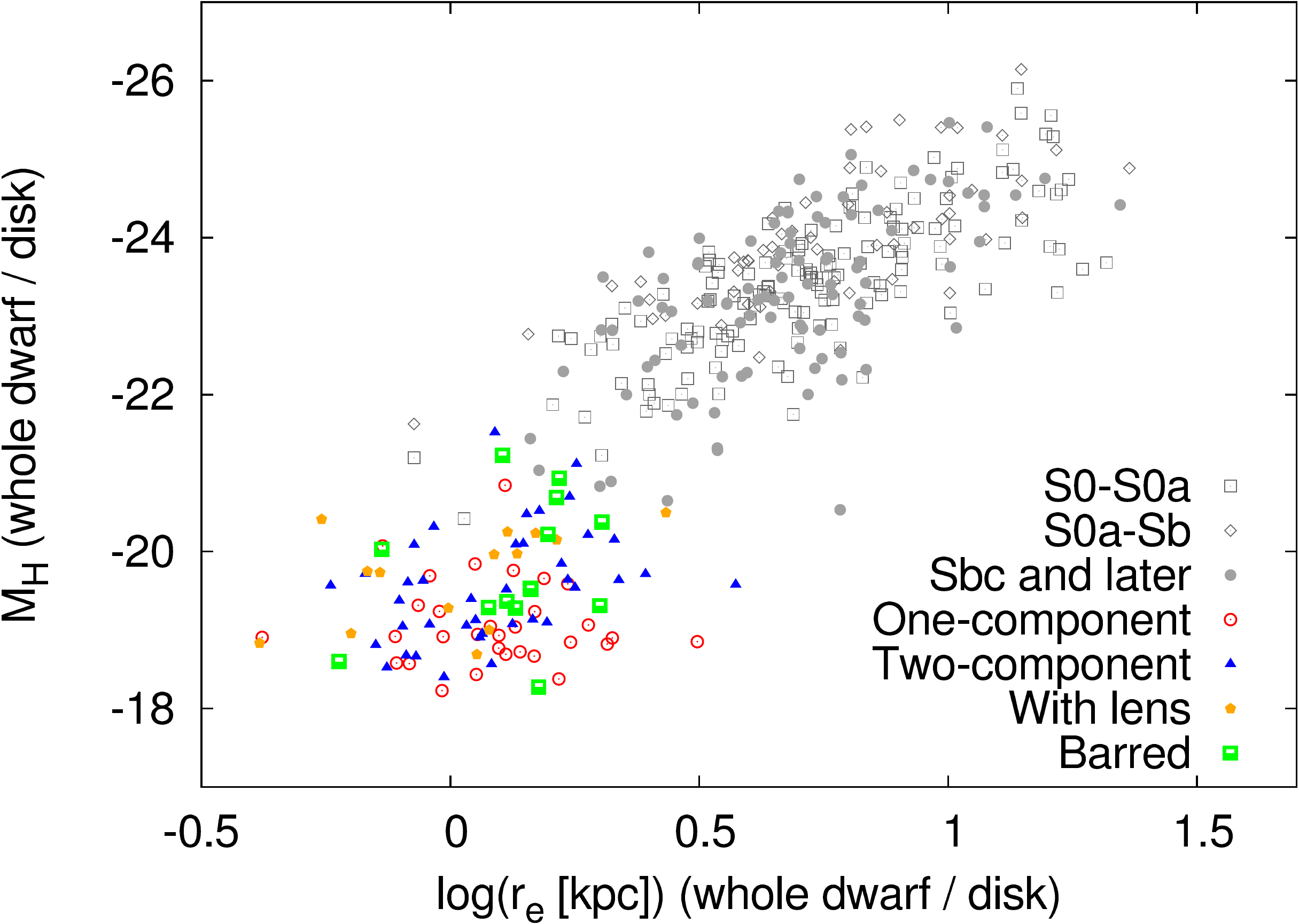}
\caption{Same as Fig.\ \ref{fig:nirs0s} panels \textit{a,d}, but plotting for the dwarf galaxies instead of the inner and outer components the photometry of the whole dwarf galaxy.\label{fig:nirs0s2}}
\end{center}
\end{figure}

In this context the spiral structures and disk signatures detected with image enhancements in some of the early-type dwarf galaxies fit into the above picture
(\citealt{Jerjen:2000p1547,Lisker:2006p392,2009A&A...501..429L} for  the Virgo, \citealt{graham_guzman} for the Coma cluster, \citealt{derijckedisks} for the Fornax cluster).
Recently, 
\citet{vdB}'s idea of parallel sequence classification was substantiated observationally  \citep{Laurikainen:2010p4602,2011MNRAS.416.1680C,2012ApJS..198....2K}. In particular, \citeauthor{2012ApJS..198....2K} place the early-type dwarf galaxies at the end of the S0 sequence on the side opposite to the bright ellipticals, with late Sc, Sd, and Im galaxies being their late-type relatives. This is in contrast to  \citet{Sandage:1984hg}, who categorized the early-type dwarf galaxies as low mass ellitpticals (although they anticipated a possible evolutionary connection to late-type galaxies).
Detailed models and observations still need to show  that  the multi-component structures in low mass late-type galaxies and early-type dwarfs are compatible or, alternatively, that the structures are created in the transformation process. Additionally, it has to be shown that the abundance of nuclei  (see e.g.\ \citealt{cote2}) and the larger number of  globular clusters in the early-type dwarf galaxies are not a problem in the  scenarios, in which  late-type galaxies are transformed to early-type dwarfs. \citet{Boselli} suggest that these two issues can be explained by the fading of the transformed galaxy relative to the unperturbed one (see however \citealt{sja}
 and \citealt{2013MNRAS.429.1066S}). Moreover, \citet{lisker_sam} point out that the progenitor galaxies, which were possibly transformed into the early-type dwarfs we observe today, are not necessarily similar to present-day late types. 

{
If the whole galaxy photometry of the early-type dwarfs is compared to the bulge and disk scaling relations of the bright galaxies, they  behave more like the disks than the bulges (Fig.\ \ref{fig:nirs0s2}), as also pointed out by \citet{2012ApJS..198....2K}.  We conclude  that the early-type dwarfs, based on their structural appearance, may be descendants of bulgeless starforming disk galaxies from earlier epochs. }

\section{Summary}
We presented the infrared imaging part of the SMAKCED survey of 121 early-type dwarf galaxies in the Virgo cluster.
The images are deep, some are probably close to what can be achieved for a large number of galaxies from the ground with today's instrumentation. The sample is a nearly complete sample of all early-type galaxies in the Virgo cluster  in a brightness range of  $-19 < M_r < -16$ mag.

For the first time two-dimensional multi-component decompositions for the early-type dwarf galaxies in the Virgo cluster have been carried out. 
 While the global profiles of most galaxies could be approximated with a single S\'ersic functions, more detailed inspections revealed that most of them also show  deviations from that,
which potentially  offers valuable insight in the galaxies' past.
The fraction of galaxies well described  by a single S\'ersic profile is only about one third. Furthermore, we find a fraction of galaxies with lenses of 14\%, and a fraction of barred galaxies of~16\%. This  is qualitatively consistent with earlier profile decomposition studies, but more robust, since it is based on higher quality  near-infrared images. Furthermore, the two-dimensional approach allowed for quantitative comparisons between the different structural components, and with the bulges and disks of bright galaxies.

The overall photometry places the galaxies in scaling relations as predicted by \citet{graham_guzman}, but with the deviations described in \citet{Janz:2008p151}. { If the dwarfs nonetheless relate to giants, they do so in a non-homologous fashion.
However, dwarfs and spirals do show no sign of a discontinuous relation in the same diagram.} Moreover, the frequent multi-component structures
argue  in favor of a disk-like nature of the early-type dwarf galaxies.
Interestingly, we find that, while the brightest early types have on average boxy isophotes, contrary to the disky early types at intermediate brightness, the dwarfs show as often boxy as disky shapes, but have on average elliptical isophotes.
We speculate that for the dwarf systems both disky and boxy shapes  are probably related to disks.

In addition to the overlap with late-type galaxies in  the photometric parameters of the whole galaxies, the outer components may fall on the extension of the late-type (Sbc and later) spiral disks.
However, the inner components are offset from the relations of bulges in the bright galaxies. 
We argue that they are not likely to be bulges, but 
{ potentially from parts of initially exponential disks, which would have  gained a higher concentration during the transformation process.}

\appendix{
\section{Flux Calibration \& Illumination correction}
\label{sec:zp}
The flux calibrations were determined by comparing sources in the field-of-view to the entries in the 2MASS and UKIDSS catalogs (which did not show systematic offsets when transformed to the same photometric system).
Additionally, we compared the galaxy photometry in circular apertures of 5, 7, 10, 15, and 20 arcsecs.
The 2MASS  values were taken directly from the extended source catalog, and the zeropoints obtained with the galaxy photometry are consistent 
with the stellar comparisons for the brightest galaxies. For fainter galaxies, both the scatter increases, and an increasing systematic offset due to the 
shallow image depth of 2MASS is found.
The UKIDSS pipeline reduction  has apparently a problem
with the sky subtraction, since the galaxy photometry is clearly offset from the stellar photometry, and this offset increases systematically with larger galaxy apertures.
The situation has slightly improved, when determining the sky as a median of the pixel values after masking sources, instead of using the pipeline-provided value.
However, a significant systematic offsets of the zeropoints in comparison to those obtained with the stars or 2MASS galaxy aperture values remains, 
including  the systematic trend with the apertures size. 

Therefore, we decided to use the stellar photometry for the calibration using both 2MASS and UKIDSS stars. UKIDSS is a bit deeper and the photometry of fainter stars
is still reliable. However, the UKIDSS  Virgo coverage is incomplete (and for some stars the combination offers multiple measurements, which decreases statistical uncertainties and makes the comparison less sensitive to systematic brightness variations). 
Each object was measured within an aperture appropriate to the FWHM of the PSF, and several apertures  of  sizes increased or decreased in steps of one pixel were used to assess the stability of the measurement against variations in the aperture size.

For the zeropoint determination the objects are required \textit{(i)}  to have a brightness fainter than $m>11$ mag in order to stay in the linear regime of the detectors. \textit{(ii)} The uncertainty of the comparison needs to be small, which effectively introduces a lower brightness limit for the comparison objects. 
Furthermore, we exclude  objects, \textit{(iii)} for which the different aperture values are not consistent with each other, \textit{(iv)} for which the star probability as given by \textsc{Sextractor} is smaller than 85\%, \textit{(v)} which are clipped away by a lower clipping 0.2 mag below the score at 75\%, $\textrm{ZP}_{75}-0.2$ mag, or \textit{(vi)} at a subsequent 2.33 $\sigma$-clipping.
The final zeropoint is calculated as the weighted average of the remaining comparison objects. The median value of the   uncertainty is of the order of 2\%, to first order independent of the airmass and the conditions of the observation.
The errors are comparable to those quoted for the traditional method of observing standard stars \citep{dJ96,Gavazzi+1996,MacArthur+2003},
and similar to the uncertainties determined by
\citet{Mcdonald:2011p4445}  via Monte-Carlo simulations for their similar calibration method for a typical number of more than 10 comparison objects.
We saw that this calibration accuracy is achieved only if the illumination corrections for two of the three detectors are taken into account, the coaddition process does not involve any $\sigma$ clipping (see \S\ref{sec:distortion}), and  the 
photometry of the comparison objects relies on our sky  subtraction instead of a local determination. 

The illumination correction is illustrated in Fig.\ \ref{fig:illumination}.
It is derived from  the brightness variation of a standard star across the field of view \textit{(upper panels)} using aperture photometry.
The standard star observations were done for TNG and NOT in 2012. For NTT we did not observe standard stars for that purpose.
For the 13 galaxies observed with NTT (and for the archival data from NTT) no illumination correction was applied.
However, archival illumination correction frames from ESO show that the illumination corrections for NTT are significantly smaller,
only a few percent for the extremes.
The illumination corrections might show long term evolutions. Therefore, we compare surfaces fitted to the
variations to those determined on stars in the field of view in the science frames of two photometric nights two years earlier 
 \textit{lower panels}. These measurements are a lot noisier, but qualitatively follow the same patterns.
 
\begin{figure}[h]
\begin{center}
\includegraphics[height=8cm, angle=-90]{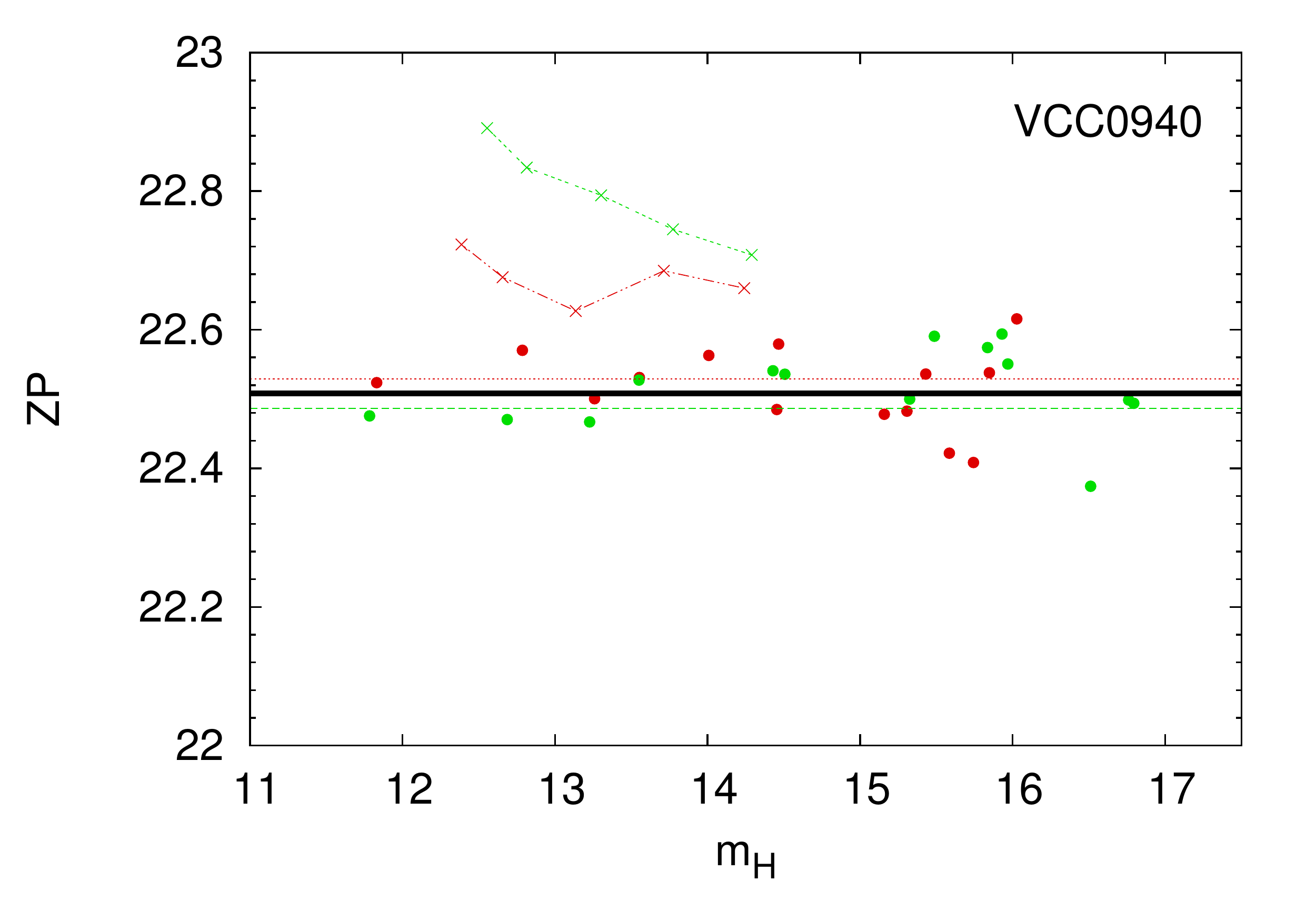}\ \ \ \ \ \ \ \ 
\includegraphics[height=8cm, angle=-90]{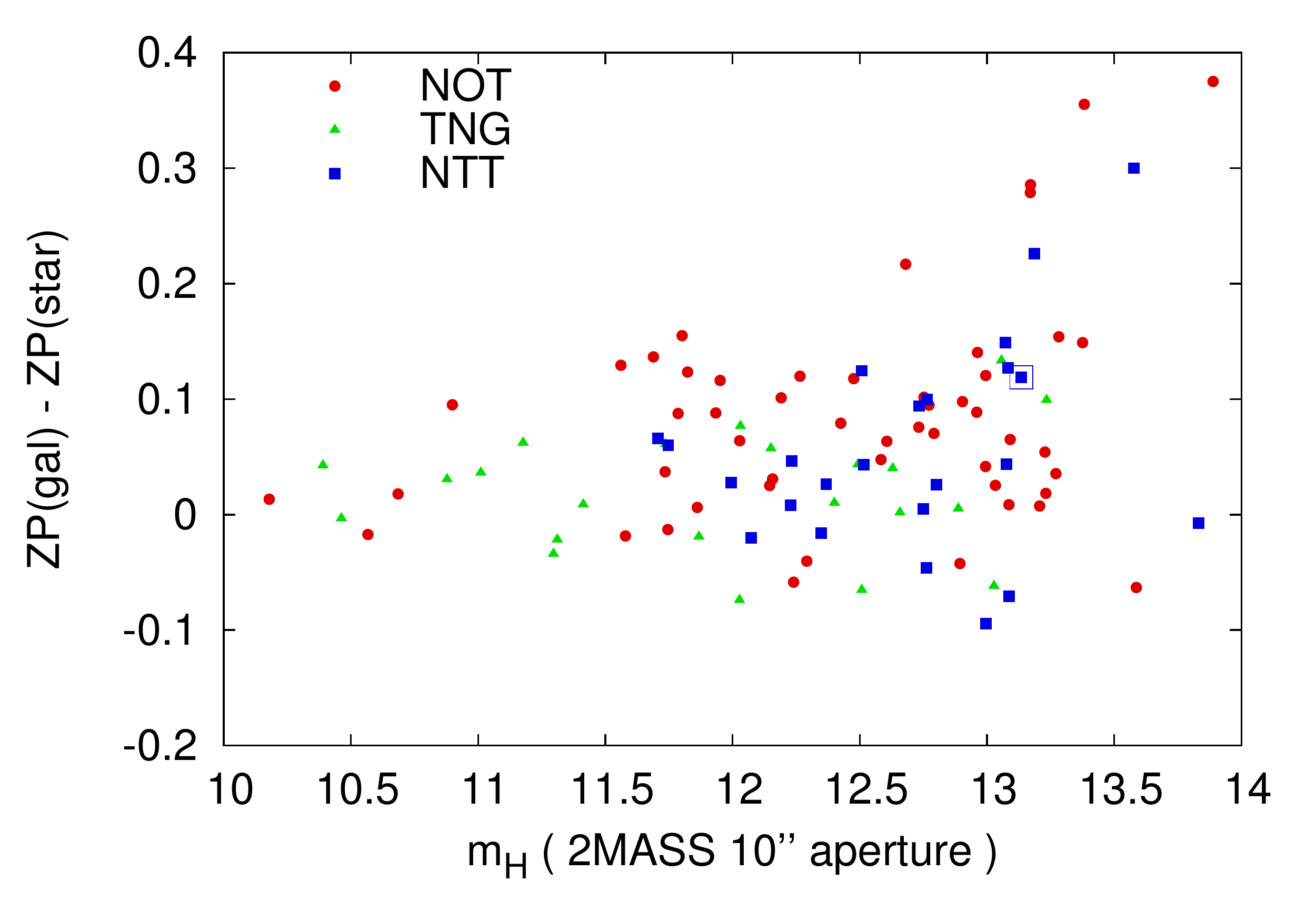}
\caption{\textit{Left panel:} Example for the zeropoint determination by comparison of point sources to 2MASS \textit{(red)} and UKIDSS \textit{(green)}  brightnesses, shown for VCC0940. The points are the individual stars, while the crosses mark the zeropoints obtained from  galaxy aperture photometry for comparison. The 2MASS galaxy measurements show a rather constant systematic offset for all apertures, while the UKIDSS values show an offset that  systematically depends on  the aperture size.
The scatter of the zeropoints determined with the stars is 0.064 mag and the resulting formal error 0.012 mag. \textit{Right panel:} Comparison between the star and galaxy aperture values for 2MASS, showing a systematic trend of larger offsets for fainter galaxies and the increasing galaxy aperture uncertainty. The highlighted point shows the field of VCC0940, which is shown in the left panel. \label{fig:zp}}
\end{center}
\end{figure}

\begin{figure}[h]
\centering
\includegraphics[height=0.45\textwidth,angle=-90]{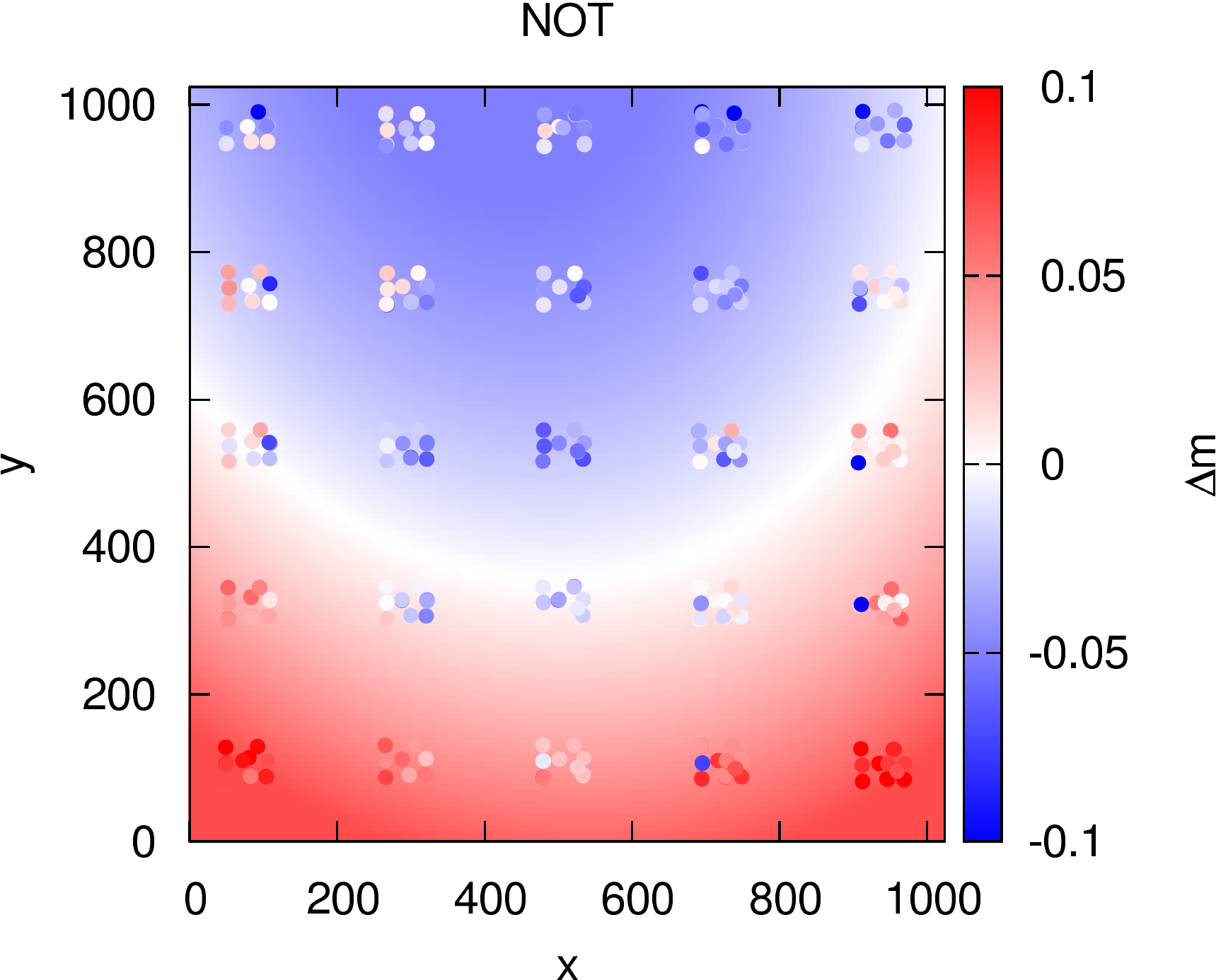}\ \ \ \ \  \includegraphics[height=0.45\textwidth,angle=-90]{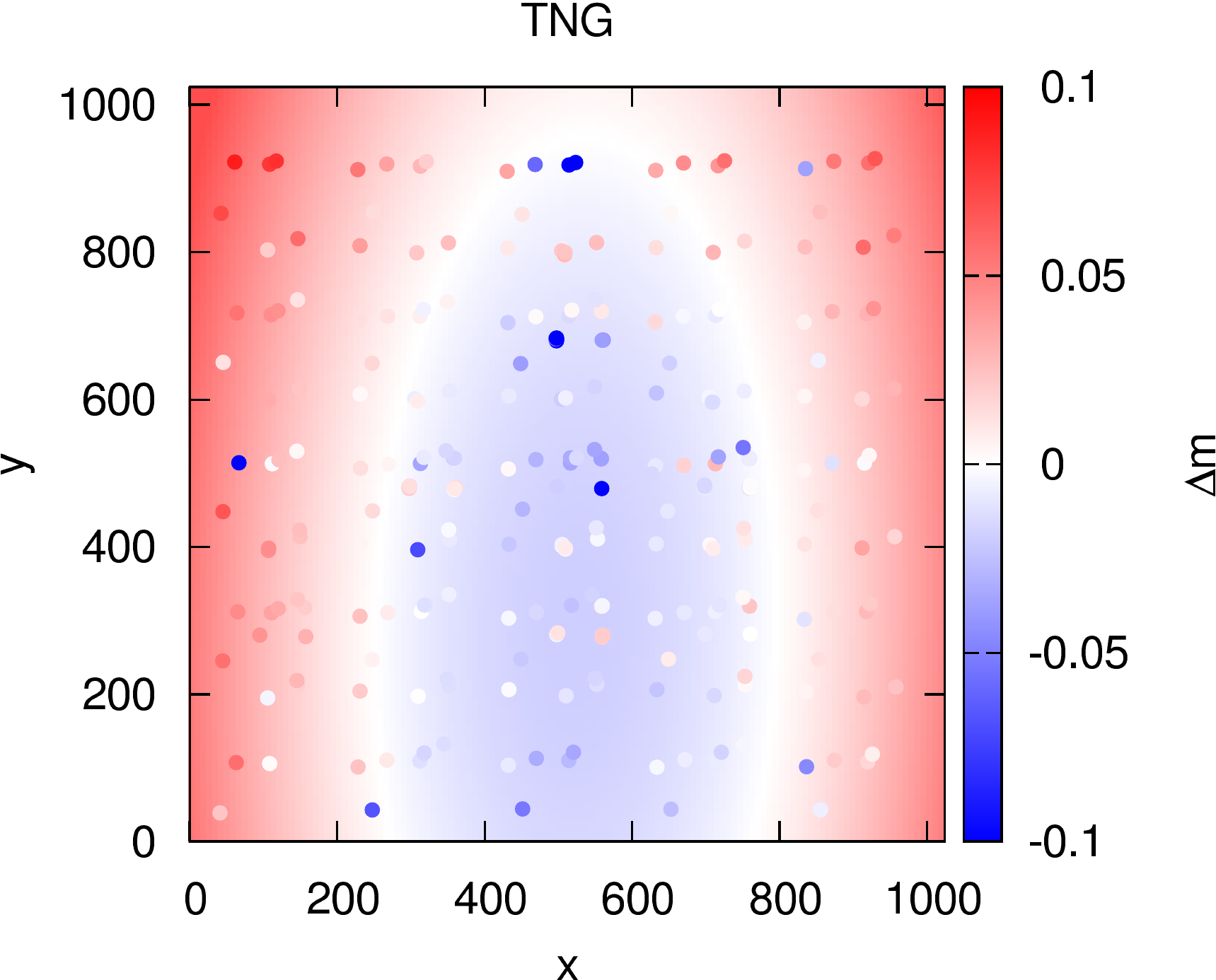} \\
~\\
\includegraphics[height=0.45\textwidth,angle=-90]{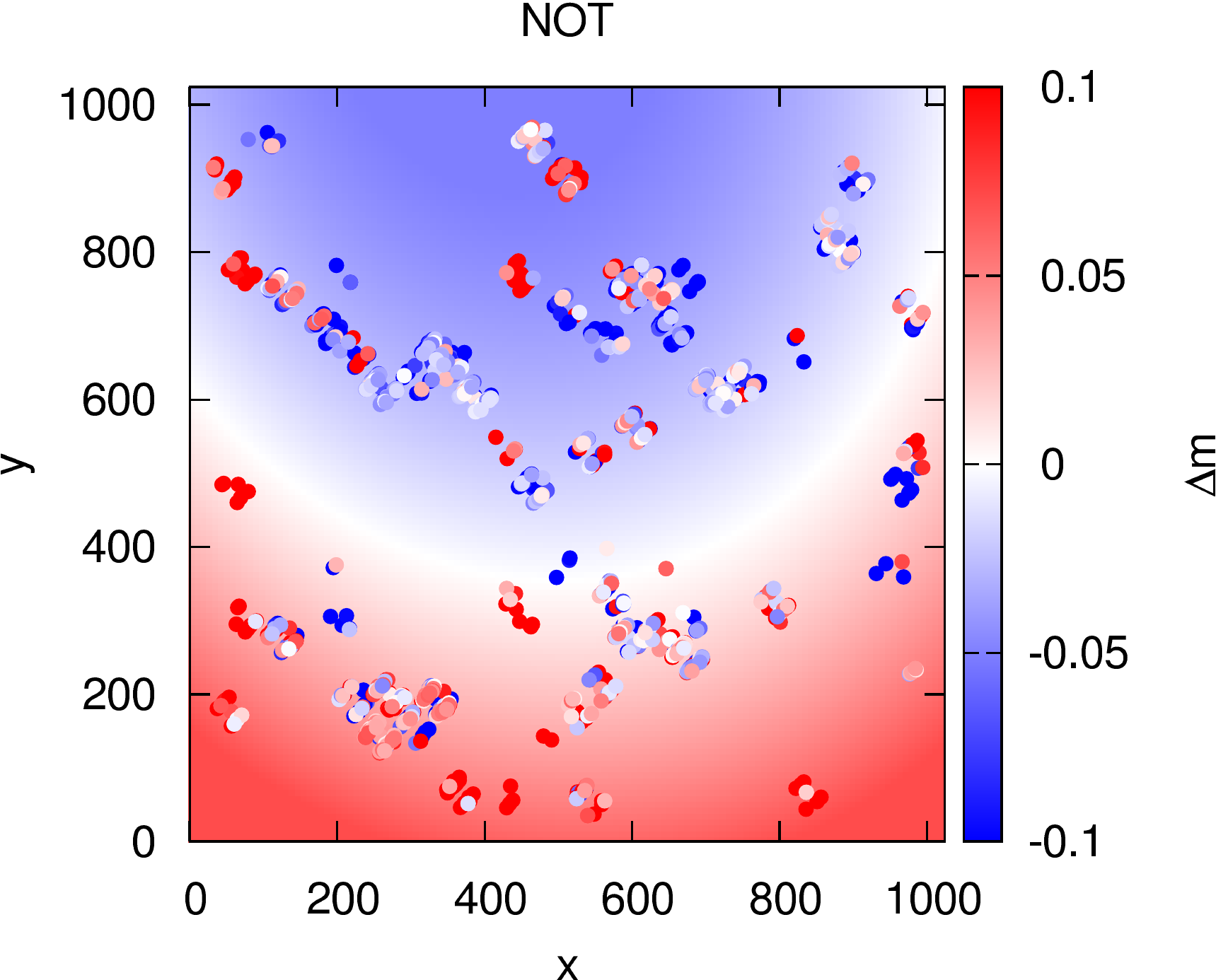}\ \ \ \ \  \includegraphics[height=0.45\textwidth,angle=-90]{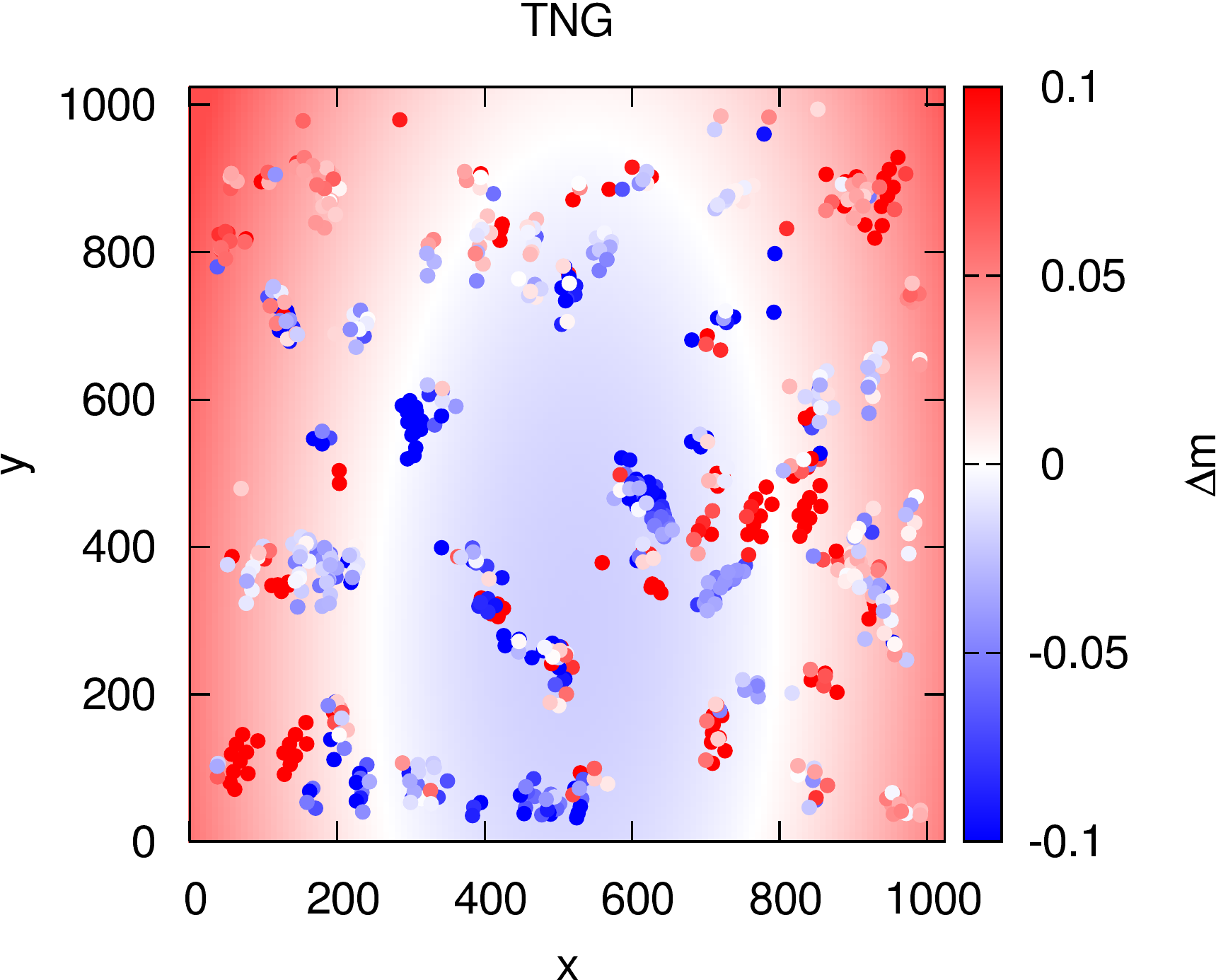} 
\caption{Variations of the brightness of stars measured across the field of view. The \textit{upper panels} display the  dedicated standard star measurements in 2012. The background color shows the fitted surface. The \textit{lower panels} show the same surface and on top the measurements of stars in the science frames of two photometric nights in 2010. The \textit{left} and \textit{right panels} show the data for NOTCam and NICS, respectively.\label{fig:illumination}}
\end{figure}

\section{Further quantification of fit quality}
\label{section:areas}
\textsc{GALFIT} uses the $\chi^2$ in the fitting process.
Therefore, we compare the $\chi^2$ after fitting a simple model (single S\'ersic + possible nucleus) to one-component and complex galaxies. 
We restrict the calculation of the  $\chi^2$ to an
elliptical aperture with a SMA of $1.5\, r_e$. At large radii both the model and the galaxy approach flux values of 0 and the signature of enhanced $\chi^2$ can be diluted.
The distributions  of  $\chi^2$ for one-component and complex galaxies are different:
 75\% of the one-component galaxies have reduced $\chi^2<1.1$, while 65\% of the complex galaxies have reduced $\chi^2>1.1$. 
 Note that, a galaxy with spiral arms 
has an enhanced $\chi^2$, 
but  is not necessarily significantly better fitted with a more complex model. If the improvement is marginal, we  were satisfied with the simple model. However, the $\chi^2$ is not an optimal  measure of the quality of two-dimensional fits (e.g. \citealt{chi}).

Therefore, we use two other quantifications. The result of using the RFF and EVI indices were described in Section \S\ref{EVIRFF}.
Furthermore, we also calculate  the tessellated excess residuals:
the galaxy image is divided from a SMA of 10 pixel to the outermost isophote into 9 elliptical rings. Each of these rings we 
subdivide into 32 sectors. In these patches the mean residual in units of the noise is  calculated:
\begin{equation}
R = \sum_i^{N_{pix}} {{(O_i-M_i)} \over{\sigma_i}}\,,
\end{equation}
with the observation $O$, the model $M$, the $\sigma$-image $\sigma$ and summing over all unmasked pixels.
 If $\left|R\right|>1/\sqrt{N_{pix}}$ we consider this patch to be not well modeled, since on average the pixels deviate more than expected. Subsequently, we calculate the area
 of the galaxy for which this criterion is met: 
\begin{equation}
A = \left(\sum_{\textrm{\footnotesize failed patches}} N_\textrm{pix}\middle) \middle/  \middle(\sum_{\textrm{\footnotesize all patches}}N_\textrm{pix}\right)\,.
\end{equation}
The obtained quantity  measures how much of the galaxy a model can reproduce in terms of the solid angle.
In the left panel of Fig.\ \ref{fig:areas} we show a histogram of the fraction of a galaxy that is not well modeled by a simple model for simple and complex galaxies. The right panel indicates the improvement for the complex galaxies in the final model.

\begin{figure}[h]
\begin{center}
\includegraphics[height=8cm, angle=-90]{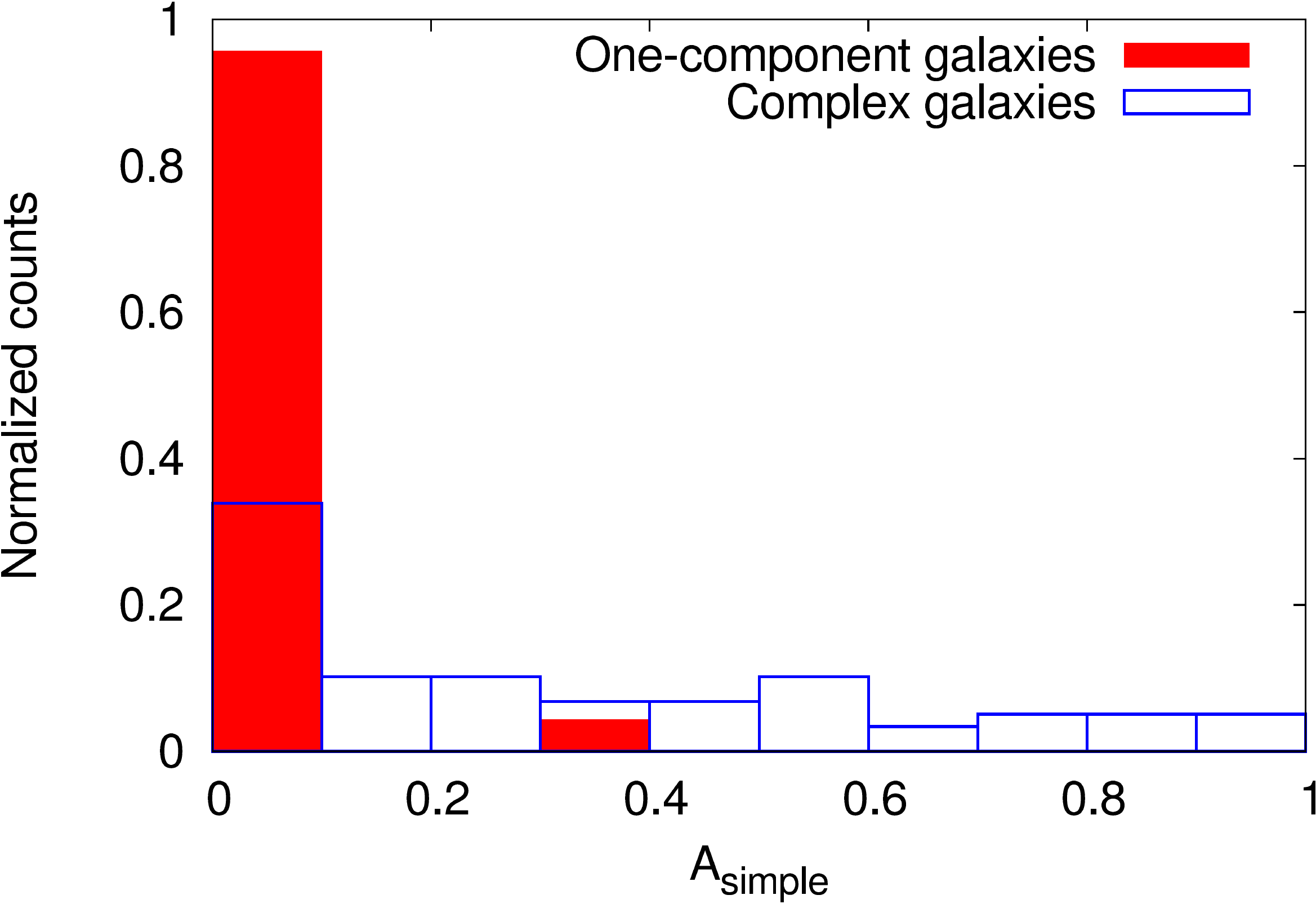} \ \ \ \ \ \ \ \ \ \ \ \ \ 
\includegraphics[height=8cm, angle=-90]{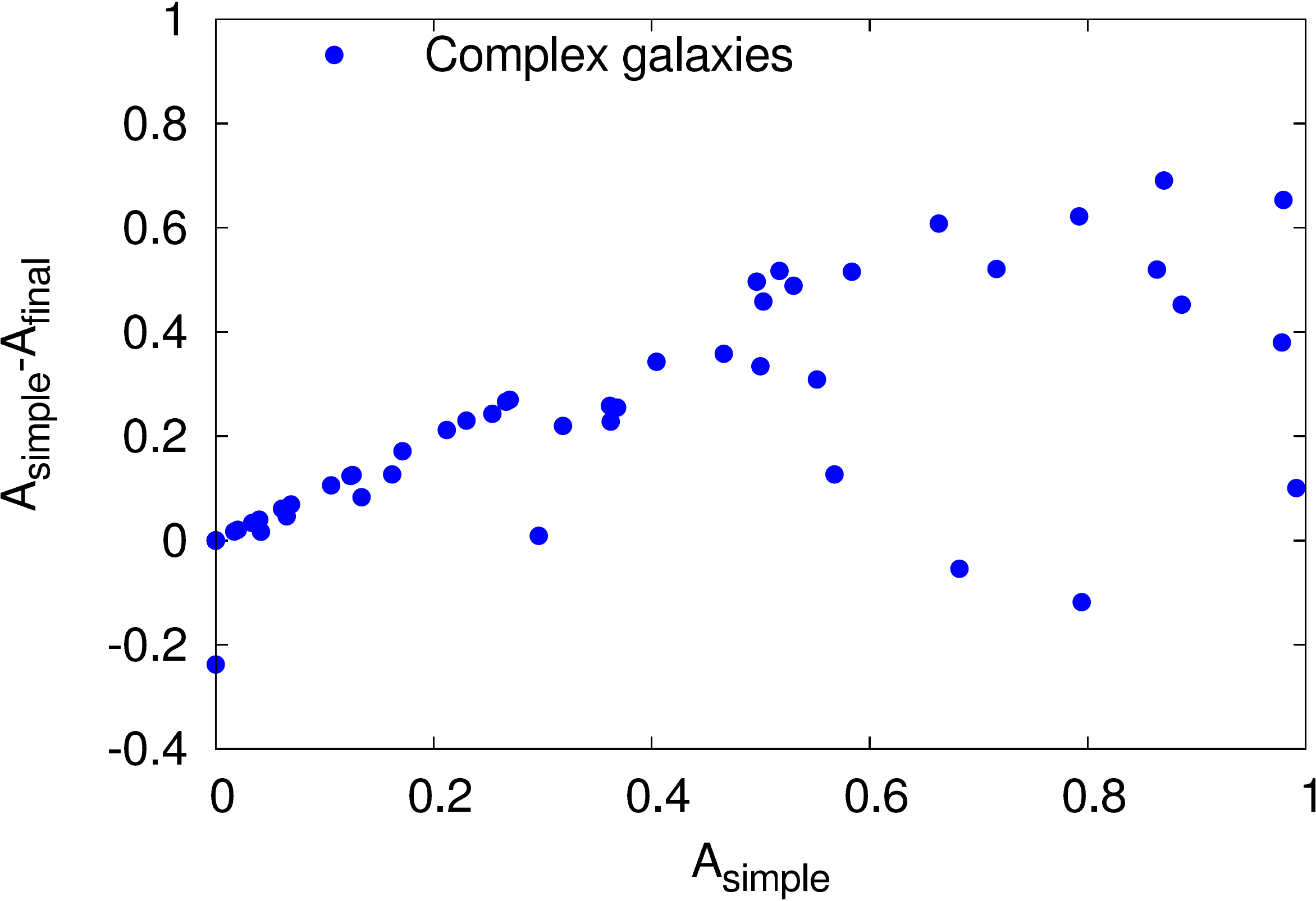}
\caption{\textit{Left panel:} Normalized histogram for the fractions of area that is not well modeled by a simple model for one-component and complex galaxies. \textit{Right panel:} Improvement of the final model compared to a simple model versus the fraction of the galaxy that is not well reproduced in the simple model. \label{fig:areas}}
\end{center}
\end{figure}

Both quantifications  agree with our visual choices reasonably well. However, for a few individual galaxies,  the parameters might argue in favor of a different model. 

\section{Notes on S\'ersic profiles \& Concentrations}
\label{sec:sersic}
The  non-parametric photometry uses  concentration measurement to calculate the correction for missed flux outside the apertures of two Petrosian radii.
This implicitly assumes that the profile follows a S\'ersic function. If the profile is instead a superposition of two functions, there will be a small error introduced. In Fig.\ \ref{fig:error}  this error for a superposition of two exponential functions is shown as a function of the ratio of their effective radii and the difference of their surface brightnesses at $r_e$.
For most of the galaxies the introduced error is below 0.05 mag. 
If there is a bright inner component with a smaller size, the concentration increases (Fig.\ \ref{fig:sersicprofiles}, left panel). The high concentration corresponds to a large S\'ersic index $n$ (Fig.\ \ref{fig:sersicprofiles}, right panel, see also \citealt{2005PASA...22..118G}), which makes the assumed correction large, even though it should be small for the superposition of two exponentials. That can be seen as the feature in the lower left corner in Fig.\ \ref{fig:error}.

Fig.\ \ref{fig:conc} shows that the complex galaxies have higher concentrations. The fraction of one-component galaxies decreases and there is a trend of increasing concentration  towards higher brightness 

\begin{figure}
\begin{center}
\centering
\includegraphics[height=7.5cm, angle=0]{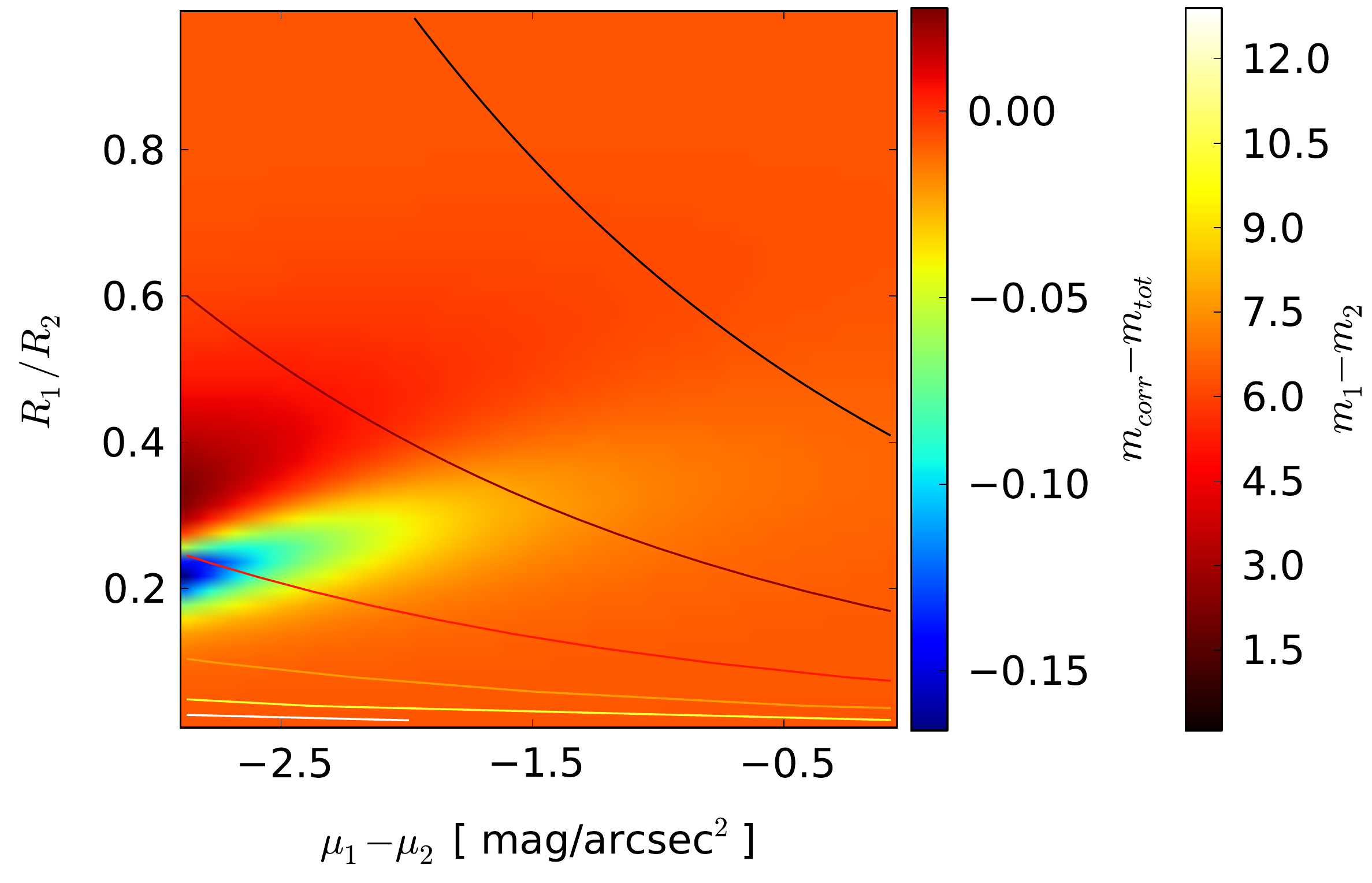}
\caption{The error of the total brightnesses (colors, color bar in the middle), if the profiles do not follow S\'ersic functions, but two exponentials, is shown as a function of the ratio of scale lengths of the two components and their difference of surface brightness at $r_e$. The lines indicate the difference in brightness of the two components (right color bar).\label{fig:error}}
\end{center}
\end{figure}

\begin{figure}
\begin{center}
\centering
\includegraphics[height=6.5cm, angle=0]{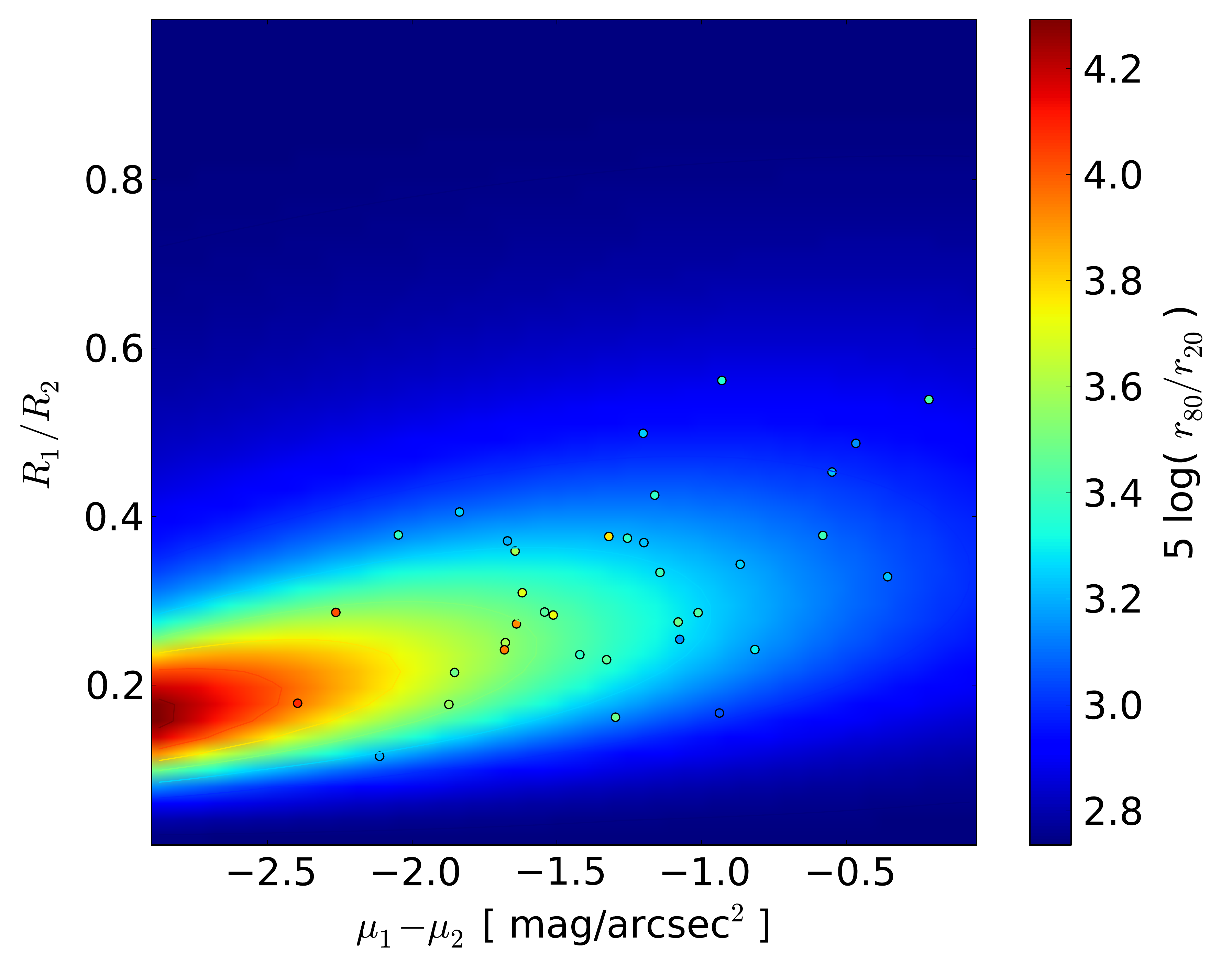} \ \ \ \ \ \ \ \ \ \ 
\includegraphics[height=6.5cm, angle=0]{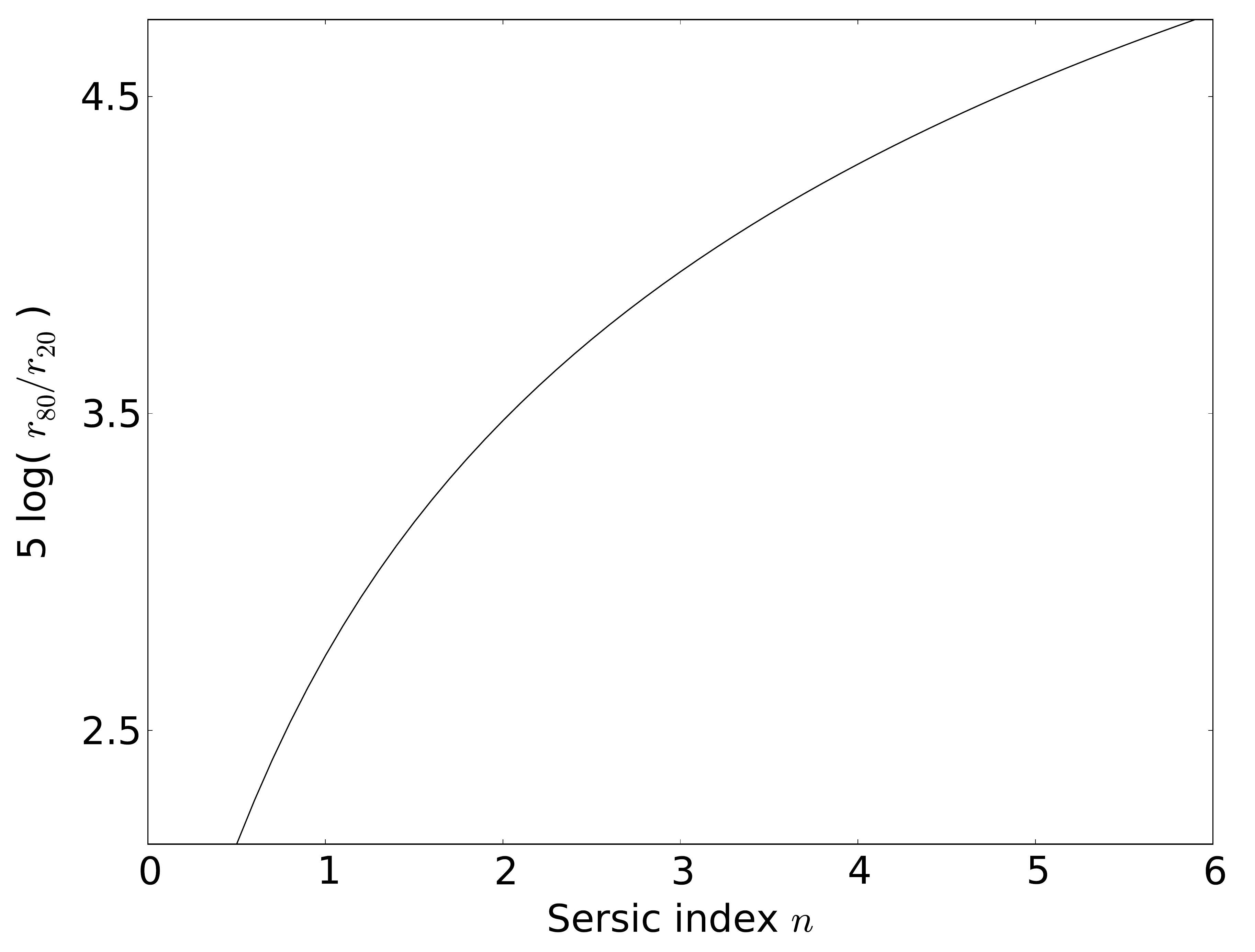}
\caption{The galaxy concentration (color) is shown as a function of the ratio of scale lengths of the two components and their difference of surface brightness at $r_e$.   The points are the observed two-component galaxies.  \textit{Right panel}: in the case of a single S\'ersic function, there is a one-to-one relation of S\'ersic $n$ and the concentration.
 \label{fig:sersicprofiles}}
\end{center}
\end{figure}

\begin{figure}
\begin{center}
\centering
\includegraphics[height=8.2cm, angle=-90]{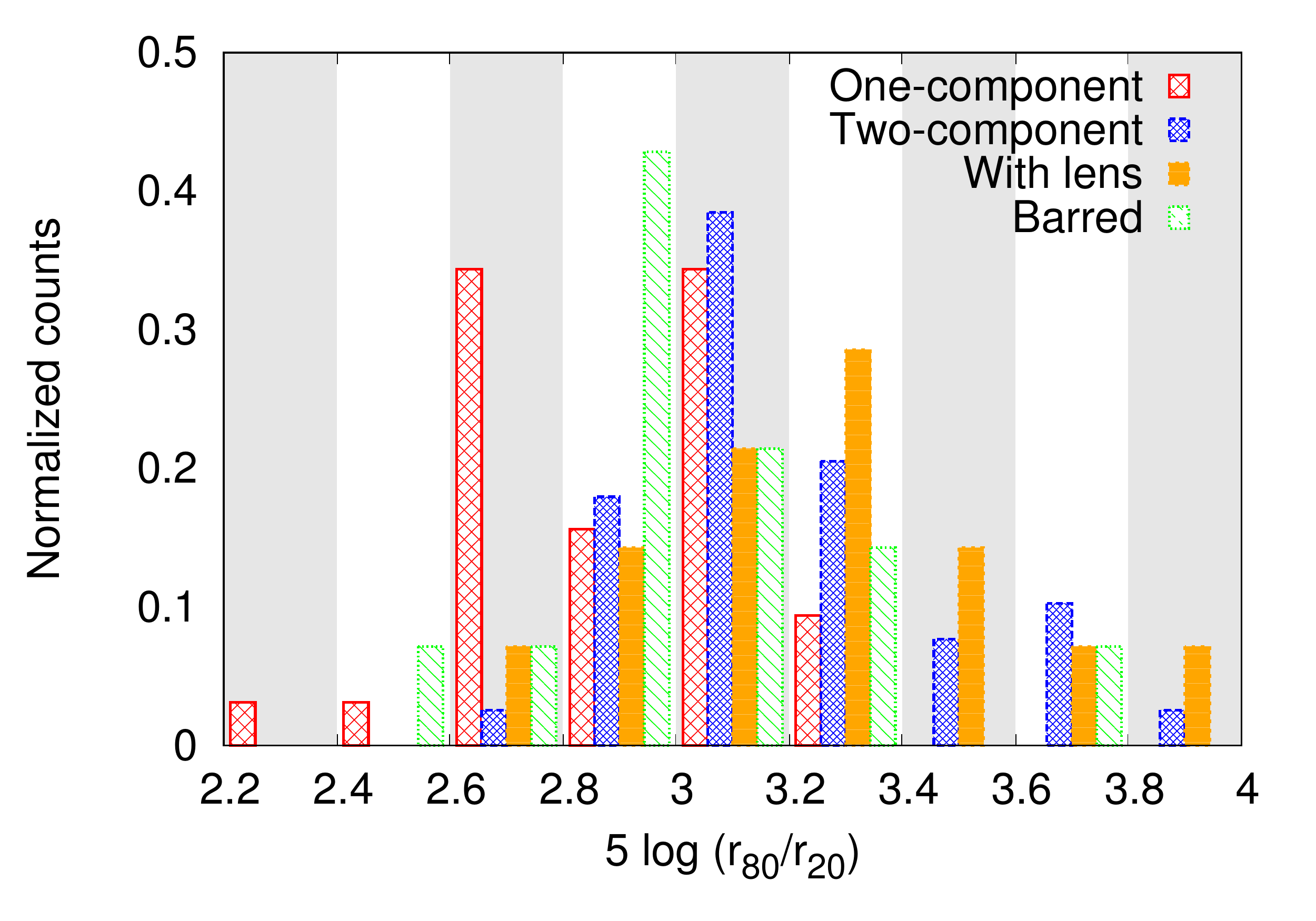}\ \ \ \ \
\includegraphics[height=8.2cm, angle=-90]{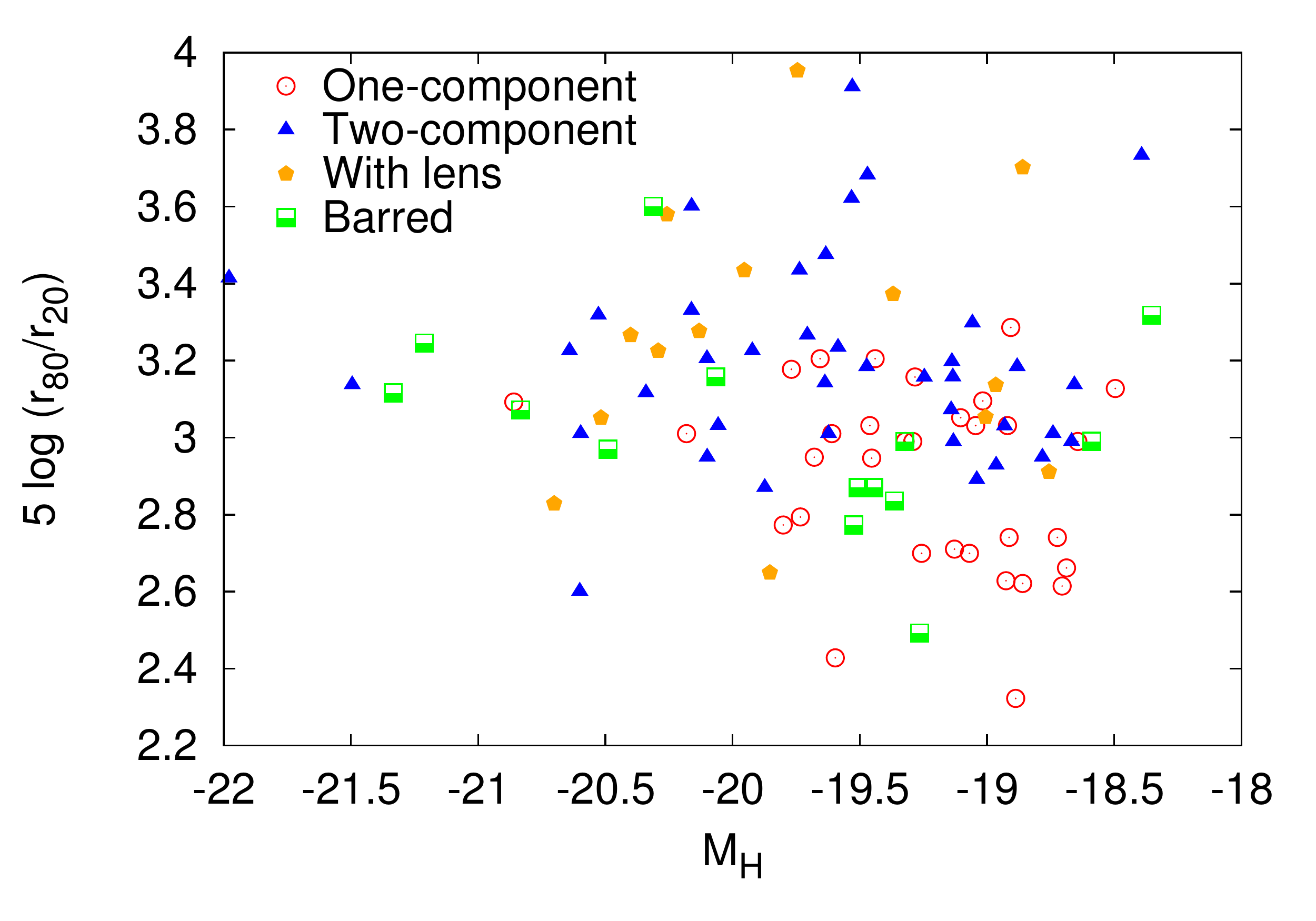}
\caption{\textit{Left panel:} histogram of concentrations normalized to the numbers of galaxies in the groups.  \textit{Right panel:} galaxy concentration as a function of galaxy brightness. \label{fig:conc}}
\end{center}
\end{figure}

\section{Figures with local projected galaxy density instead of clustercentric distance, and with effective surface brightness instead of brightness}
\label{sec:additional}
In Fig.\ \ref{fig:dens10F} local projected galaxy density is shown as a measure of the local environment, alternatively to the clustercentric distance.
The local projected galaxy number density at the position of a given galaxy is computed from the circular projected area $A$ centered on that position and encompassing 10 neighbor galaxies brighter than $M_r\leq-15.2$ mag (the r-magnitude completeness limit of the VCC, see \citealt{Weinmann:2011p3976}). Known background galaxies as well as likely members of the more distant M- and W-clouds \citep{Gavazzi3d} are excluded. The resulting density is $11/A$ in units of number per square degree.

In Figs.\ \ref{fig:kormendy2MASS}\&\ref{fig:nirs0s_mu} the photometric scaling relations are shown with surface brightness instead of brightness. In Fig.\ \ref{fig:kormendy2MASS} the Kormendy relation is shown for the whole galaxies, the data for the bright galaxies is taken from the 2MASS extended source catalog. Fig.\ \ref{fig:nirs0s_mu} shows the scalings of the components in comparison to those of the galaxies in NIRS0S and OSUBSGS.

\begin{figure}[h]
\begin{center}
\includegraphics[height=8.2cm, angle=-90]{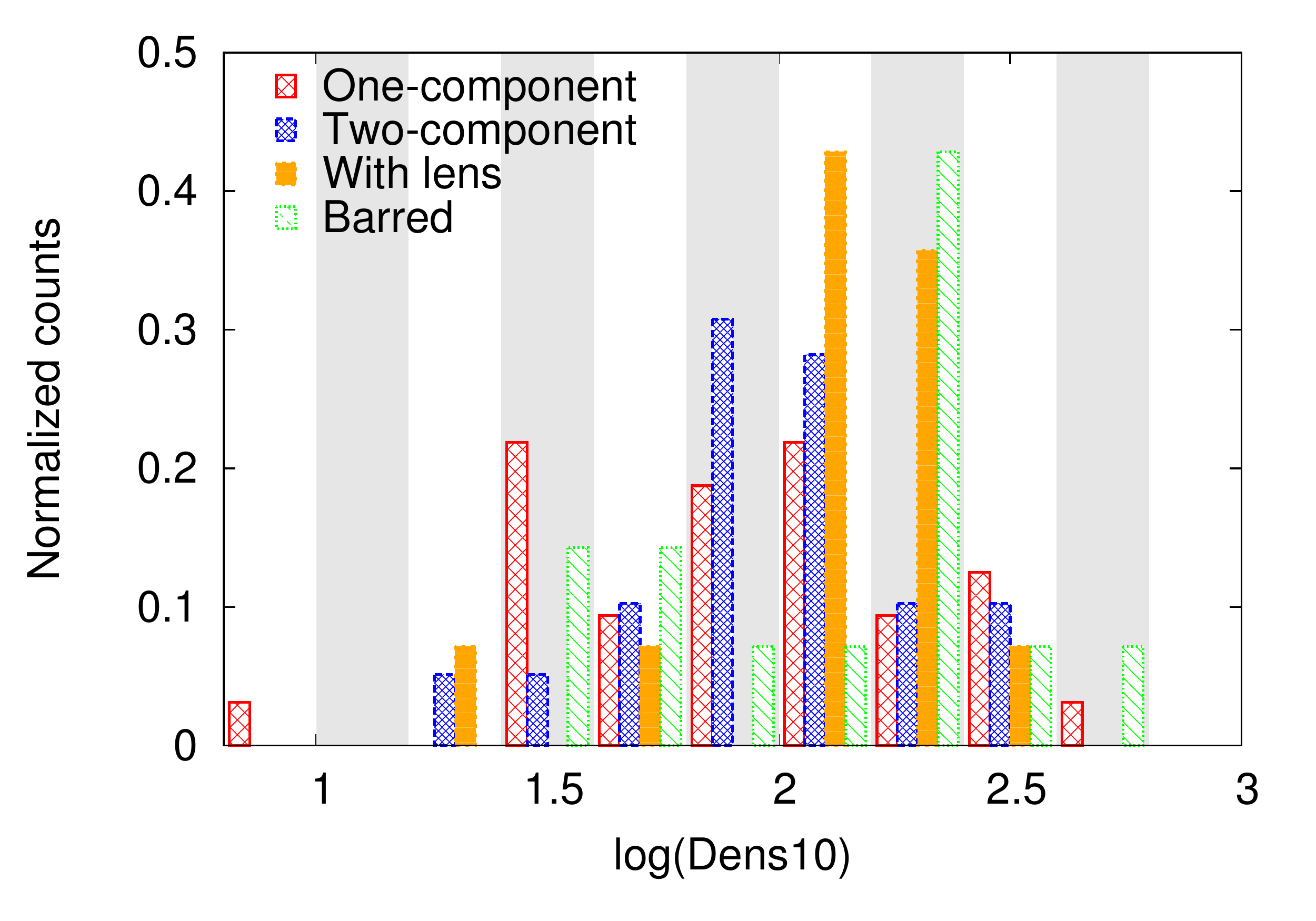}
\includegraphics[height=8.2cm, angle=-90]{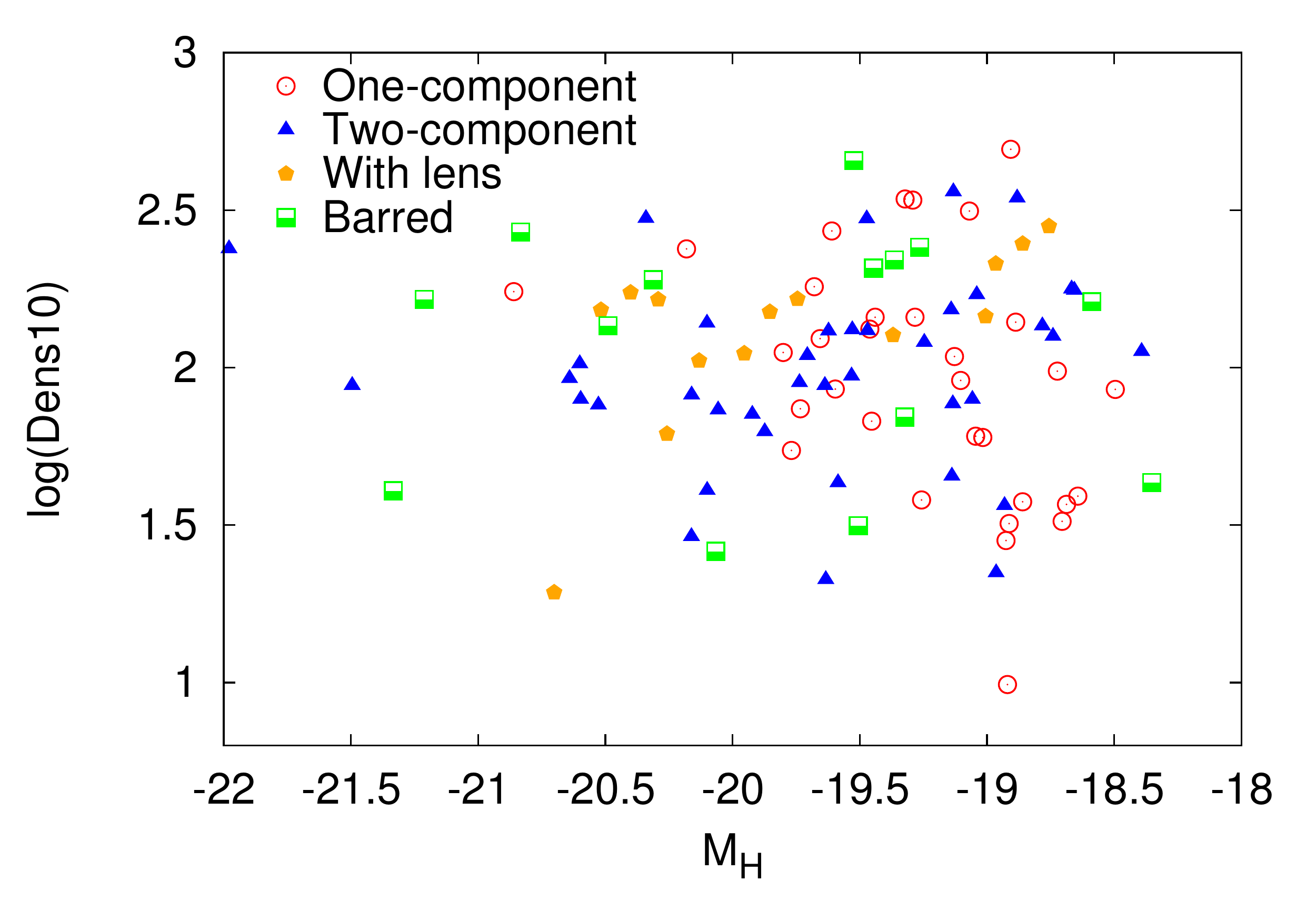}
\caption{\textit{Left panel:} Histogram of local galaxy density normalized by the number of galaxies in the groups. \textit{Right panel:} Local galaxy density versus galaxy brightness.
\label{fig:dens10F}}
\end{center}
\end{figure}

\begin{figure}[h]
\begin{center}
\includegraphics[height=8.2cm, angle=-90]{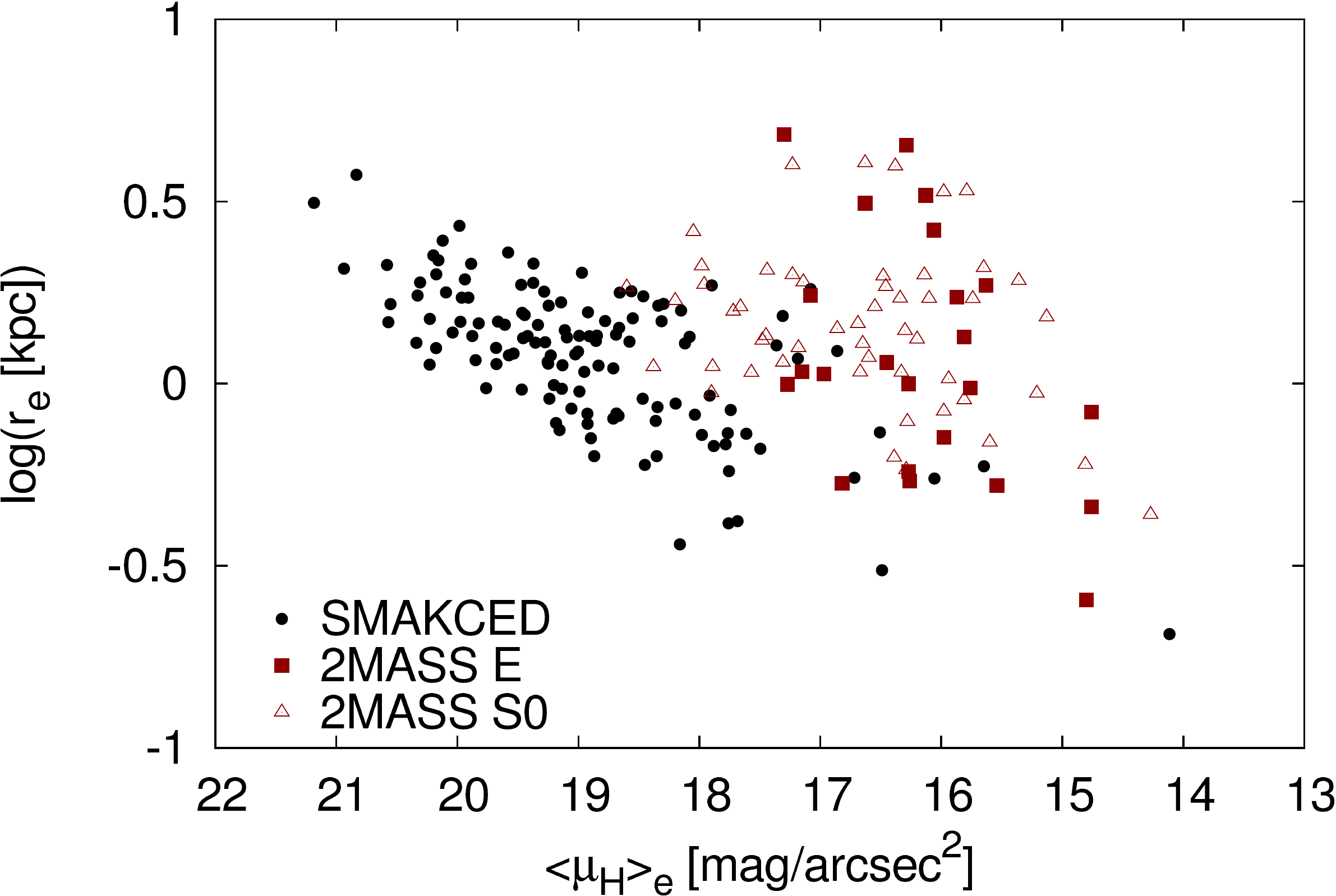}
\includegraphics[height=8.2cm, angle=-90]{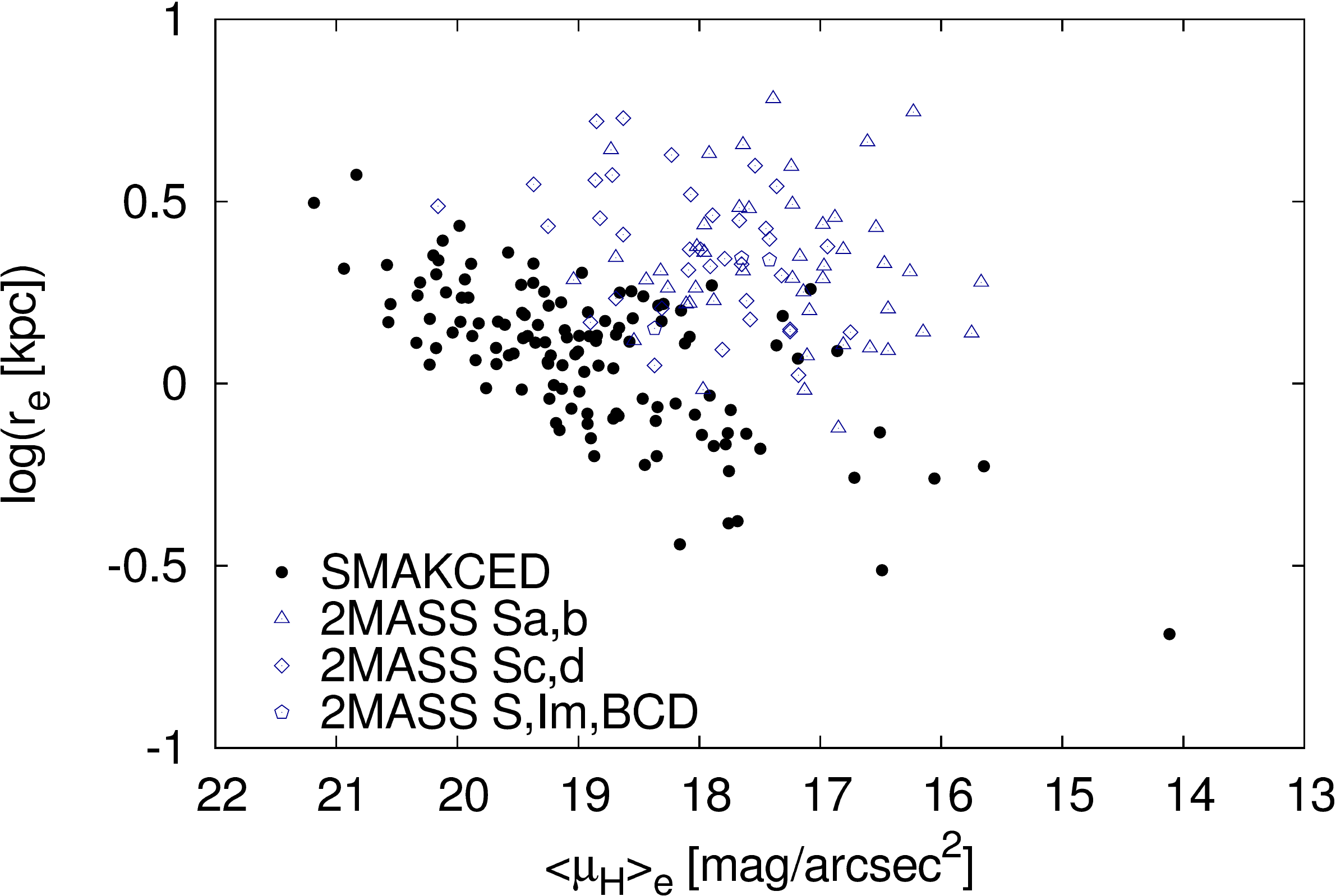}
\caption{Same as Fig.\ \ref{fig:2MASS}, but showing the Kormendy relation instead of size versus brightness, both measured for the whole galaxy.
\label{fig:kormendy2MASS}}
\end{center}
\end{figure}

\begin{figure}[h]
\begin{center}
\includegraphics[height=0.8\textwidth, angle=-90]{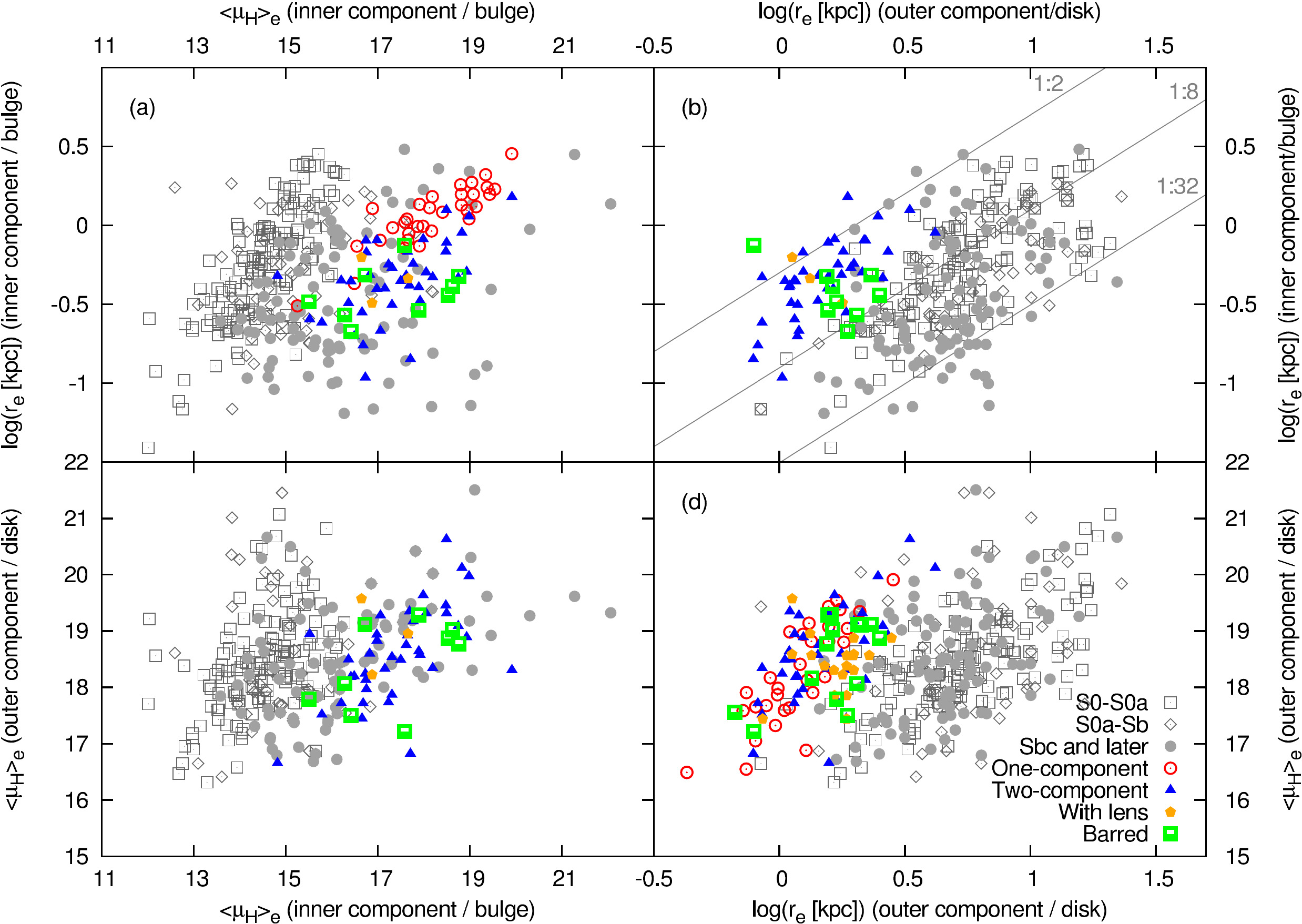}
\caption{Same as Fig.\ \ref{fig:nirs0s}, but showing the comparison to the NIRS0S and OSUBSGS with surface brightness instead of brightness; all quantities are measured for the individual components.
\label{fig:nirs0s_mu}}
\end{center}
\end{figure}

\section{Comparison to McDonald et al.\ (2011) profile fits}
\label{section:mcd}

\citet{Mcdonald:2011p4445} also studied galaxies in the Virgo cluster, and there is a substantial overlap of  88 galaxies with this study, as indicated in Table 3.
They fitted the azimuthally averaged light profiles with one- or two-component models, possibly including an additional nucleus.
We first list the main differences between the two studies, which may be responsible for the 
 differences in the type of model chosen for individual galaxies   \citep{janz+2012}, and subsequently compare
 parameters of models for galaxies that are fitted with similar models in both studies.

\citeauthor{Mcdonald:2011p4445} adopt S\'ersic+exponential models for two-component galaxies, as mostly  done also in this study, but
all their single component galaxies have exactly exponential profiles. This is a major restriction for the available
parameter space for single component galaxies in comparison to this study. 
Furthermore, they employ an automated one-dimensional fitting algorithm, while our visual model selection makes use of all the available information,
and is deliberated by subjective human minds. Also, two of our structural types, galaxies with bars and lenses, are not part
of their model variety.
Finally, the studies are based on different data sets. 

\citeauthor{Mcdonald:2011p4445}'s published  light profiles
 extend to similar radii as in this study for 30 of the 88 galaxies in both samples, when requiring a similar surface brightness error at the outermost radii.
For only two galaxies their data reach larger radii (VCC 1087 and 1549), while for the 56 remaining galaxies our images are deeper. 
 The profiles of 14 of the galaxies with comparable radial extent show good agreement\footnote{VCC 0523, 0856, 0917, 0951, 1036, 1183, 1261, 1431, 1491, 1499, 1545, 1695, 1871, and 1895.}, while 16 galaxies show differences in their profiles between the two studies\footnote{VCC 0437, 0543, 0698, 0751, 1283, 1392, 1422, 1512, 1528, 1567, 1614, 1627, 1743, 1910, 1945, and 2042.}. For all but two of the deviant profiles, the \citeauthor{Mcdonald:2011p4445} surface brightnesses are brighter at larger radii, and for at least six of those \footnote{VCC 0437, 1567, 1614, 1743, 1910, and 1945.}, they show a visible break in the profile that is not present in our data. Several times this is accompanied with discrepancies of the profile in comparison to their optical SDSS profiles that go beyond a normal color gradient. Our profiles are displayed on our dedicated galaxy decomposition webpages, \url{http://www.smakced.net/data.html}, where links to the near-IR as well as the optical profiles of \citet{Mcdonald:2011p4445} are also provided.

In Fig.\ \ref{fig:mcdonald} we compare the decomposition parameters of this study (SMAKCED) to those 
determined by  \citeauthor{Mcdonald:2011p4445}. 
 The comparison is restricted to galaxies that have comparable decompositions, i.e.\ galaxies that are
  fitted with one S\'ersic function or two-component models in both studies. 
    Also, galaxies with a less reliable fit in \citeauthor{Mcdonald:2011p4445} and galaxies, for which the total magnitudes differ
 between the two studies, are omitted. As expected (see \S\ref{section:fitting}) the obtained parameters 
   show a reasonable agreement.

\begin{figure}[h]
\begin{center}
\includegraphics[height=0.45\linewidth, angle=-90]{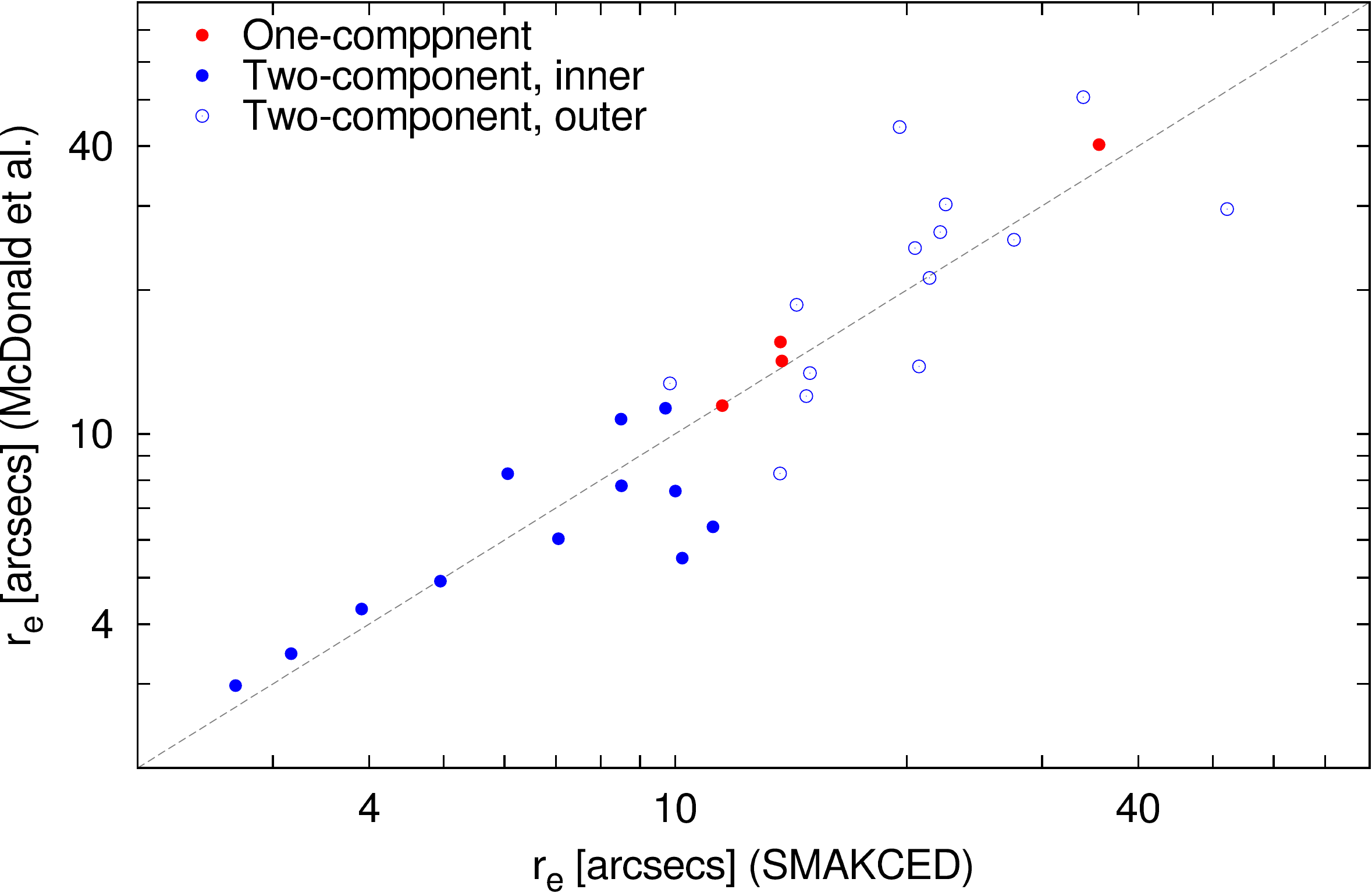}\ \ \ \ \ \  \ \ \ \ \  \ \ \ \ \ 
\includegraphics[height=0.45\linewidth, angle=-90]{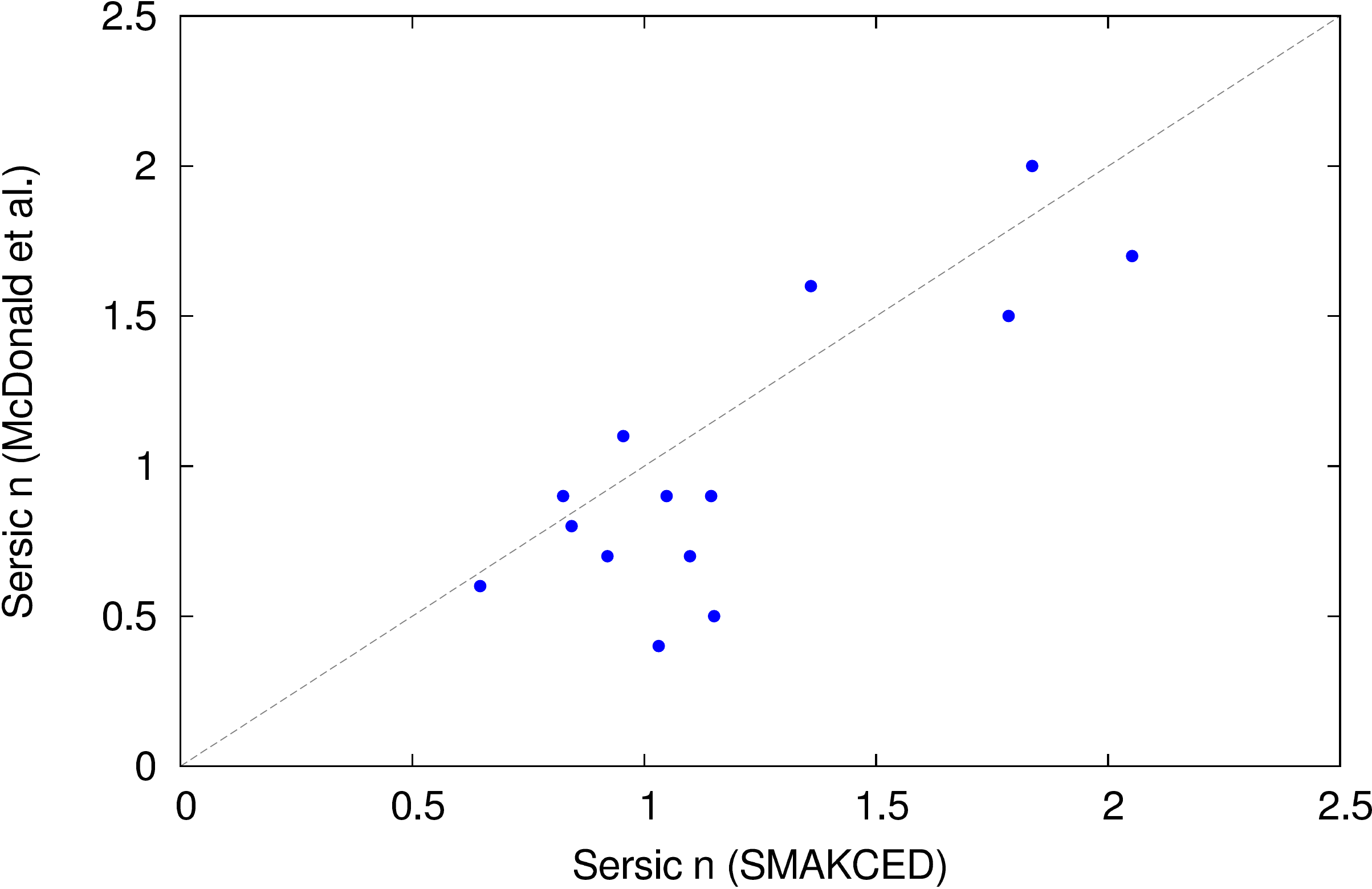}
\caption{Comparison to the profile fits of  \citet{Mcdonald:2011p4445}. Shown are galaxies that were characterized in both studies with a one-component  ({\textit red} points) or  two-component model  ({\textit blue} symbols, {\textit filled and open} symbols for inner and outer components, respectively). Galaxies with a poorer quality flag for the fit in  \citeauthor{Mcdonald:2011p4445} are omitted, as well as galaxies with profile discrepancies (see text). The {\it left} and {\it right} panel show the radii and S\'ersic indices, respectively. A generally good agreement is found.
\label{fig:mcdonald}}
\end{center}
\end{figure}

\section{GALFIT vs. bdbar for NIRS0S galaxies and SMAKCED}
\label{section:bdbar}
The original NIRS0S and OSUBSGS decompositions  were done with \textsc{BDBar} \citep{Laurikainen:2004kt}. 
Moreover, instead of using $\sigma$-images as done in \textsc{galfit}, a weighting $w=1/F_\textrm{model}$ was used in \textsc{BDBar} ($F_\textrm{model}=$ flux of the pixel in the model).
For an ideal model the weighting should not make a difference. However, in a real galaxy there are
anomalies, or parts are not perfectly modeled. Therefore, a different weighting can bias the resulting fit parameters, though the
effect is not expected to be large \citep{Laurikainen:2005gm}.
 We rerun  the NIRS0S decompositions with \textsc{galfit}.
Since the automatic decompositions can not  be used blindly for the comparison, we selected some of those galaxies, for which 
the same features
in the data were fitted. 
The parameters derived with \textsc{BDbar} and \textsc{galfit} 
are compared 
 in Fig.\ \ref{fig:bdbar}. 
The  scatter is reasonably small  and no severe systematic bias is introduced.
There are three outliers for the sizes of the components. This can happen, if there are only a few pixels for fitting the bulge and the resulting fit has an unreasonably high S\'ersic index. 
In a manual decomposition one tries to avoid such pathological behavior.

\begin{figure}[h]
\begin{center}
\includegraphics[height=8cm, angle=-90]{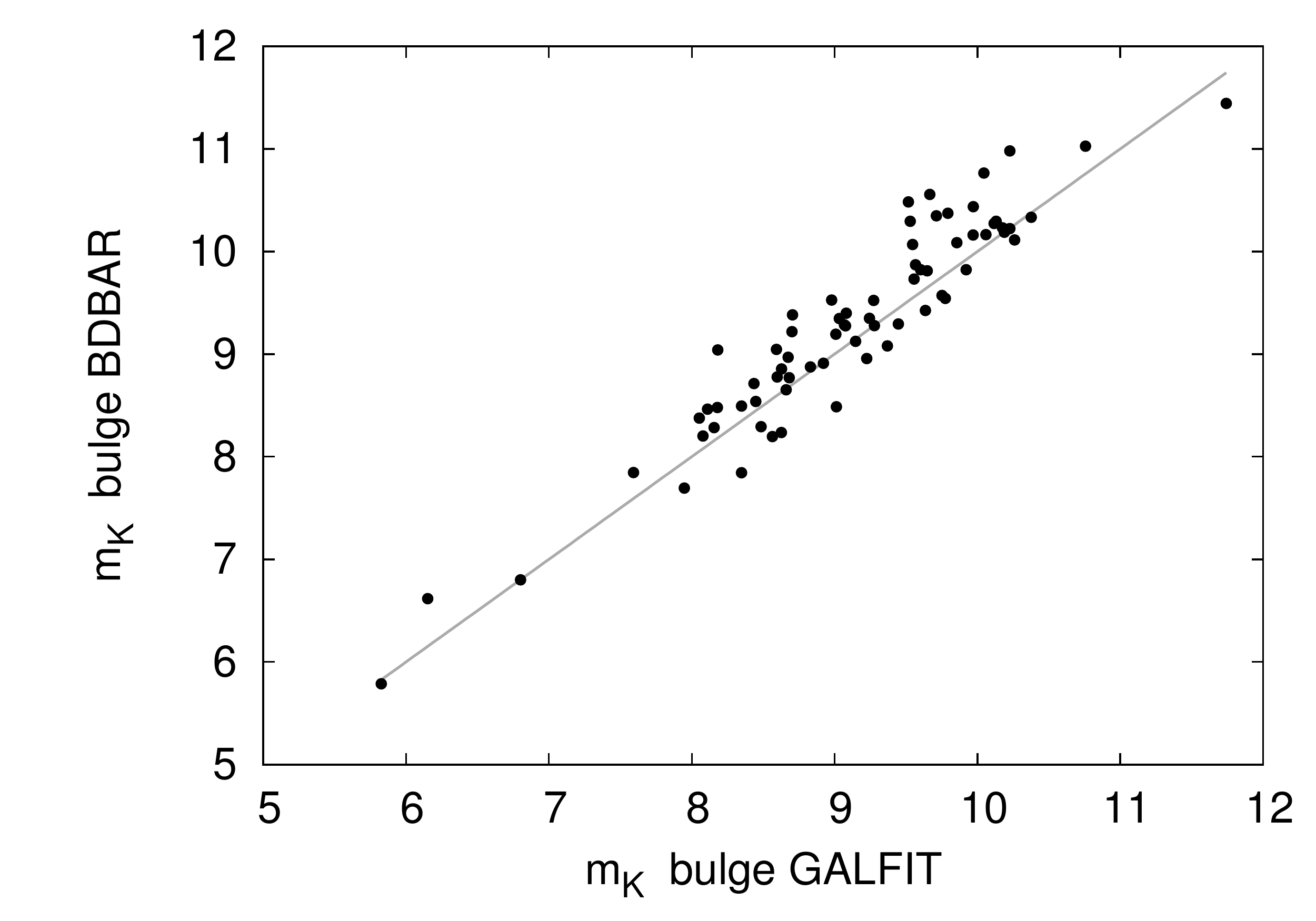}
\includegraphics[height=8cm, angle=-90]{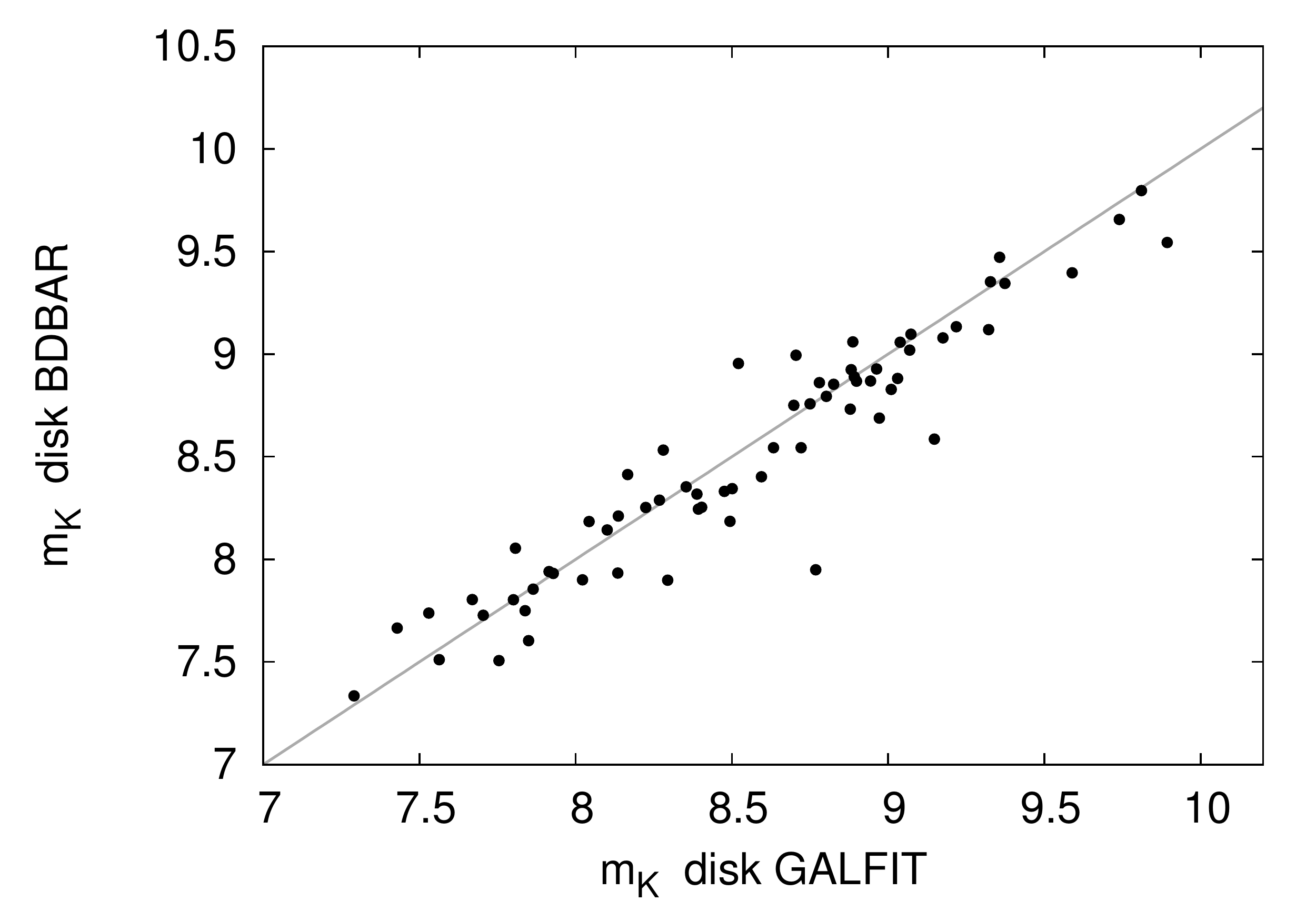}\\
\includegraphics[height=8cm, angle=-90]{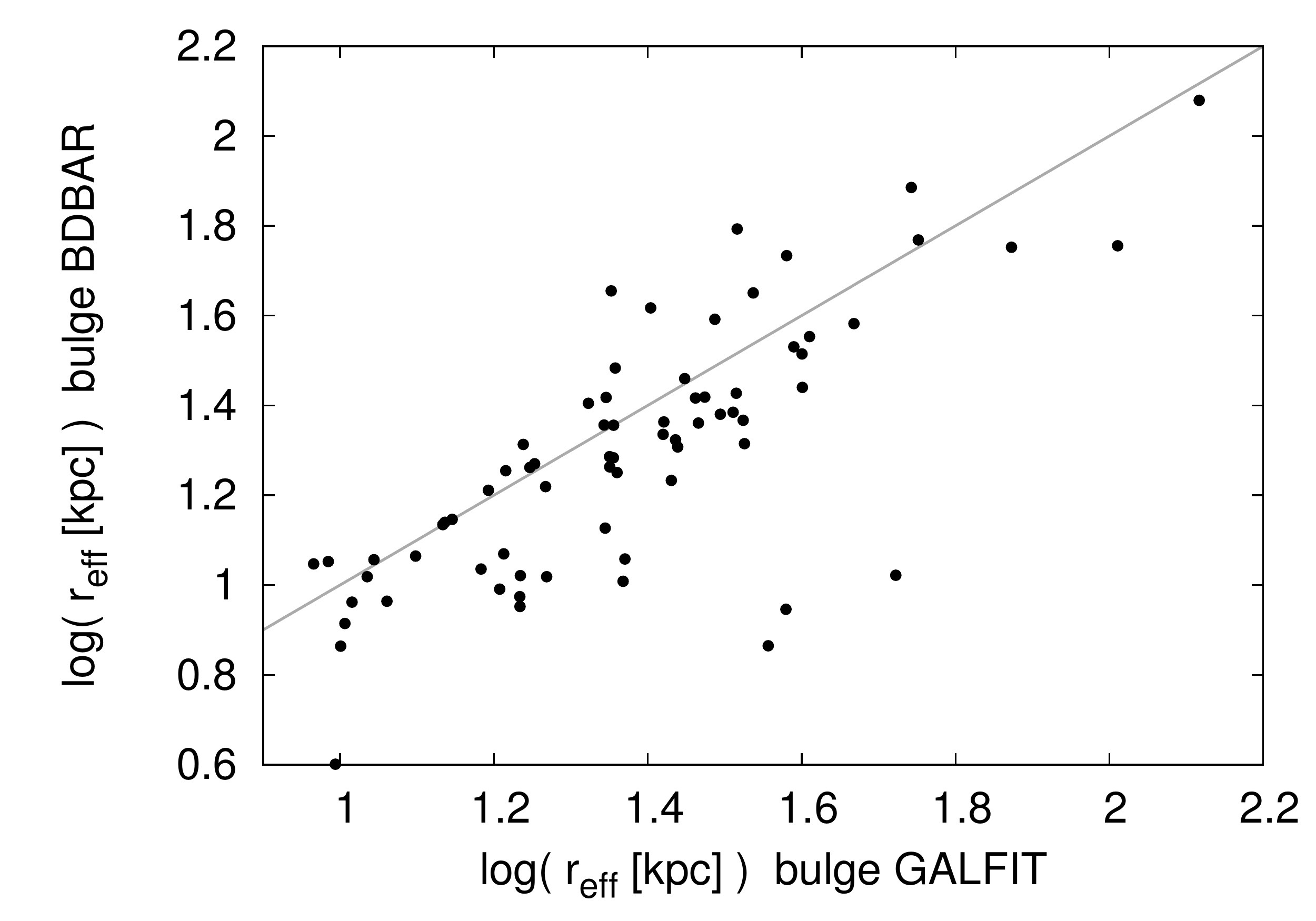}
\includegraphics[height=8cm, angle=-90]{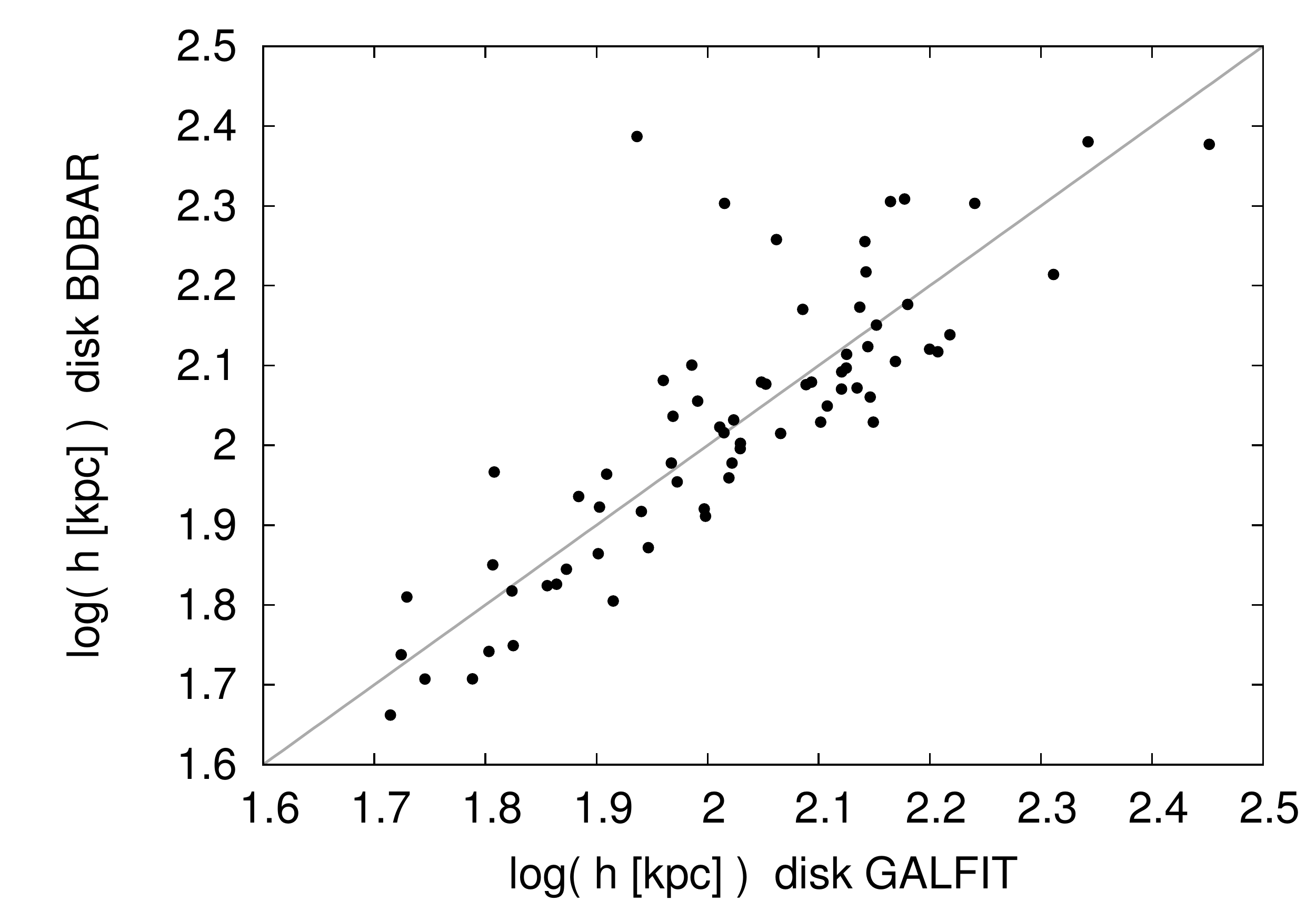}
\caption{This figure compares  the \textsc{BDBar} and \textsc{galfit} decomposition parameters for NIRS0S:  the brightnesses \textit{(upper panels)} and the scale parameters \textit{(lower panels)} for the bulge \textit{(left)} and the disk \textit{(right)} are shown. The agreement in this \textsc{galfit}/\textsc{BDBar} comparison
would be even better if barlenses, that form part of the
bar, would have been fitted in our \textsc{galfit}
decompositions, in a similar manner as in \textsc{Laurikainen:2010p4602}
using \textsc{BDBar}.\label{fig:bdbar}}
\end{center}
\end{figure}

}

\acknowledgments
The observations of this publication are based on proposals of the SMAKCED team (\url{http://www.smakced.net}).

We thank the referee for valuable comments that led to improvements of the paper. Furthermore, we thank A.\ Pasquali for useful discussions concerning the boxiness; as well as J.\ Laine and E.\ Busekool for help with the observations.

J.J.\ acknowledges the financial support of the Gottfried Daimler- and Karl Benz Foundation and the University of Oulu.
J.J., T.L., K.S.A.H., H.T.M., and S.P.\ were supported within the framework of the Excellence Initiative by the German Research Foundation (DFG) through the Heidelberg Graduate School of Fundamental Physics (grant number GSC 129/1), by the DFG though grants LI 1801/2-1 (H.T.M.) and LI 1801/7-1 (K.S.A.H.), and by the IMPRS for Astronomy and Cosmic Physics at the University of Heidelberg (K.S.A.H.\ and S.P.). 
 J.J.\ and H.S.\ acknowledge the support by the Academy of Finland. 
E.T.\ thanks the financial support of the Fulbright Program jointly with the Spanish Ministry of Education.
J.F.-B. acknowledges support from the Ram\'on y Cajal Program and from grant AYA2010-21322-C03-02 by the Spanish Ministry of Science and Innovation, as well as financial support to the DAGAL network from the People Programme (Marie Curie Actions) of the European Union's Seventh Framework Programme FP7/2007-2013/ under REA grant agreement number PITN-GA-2011-289313.
K.S.A.H.\ acknowledges funding from the University of Heidelberg through a Landesgraduiertenf\"orderung (LGFG) fellowship for doctoral training.

Based on observations made with the Italian Telescopio Nazionale Galileo (TNG) operated on the island of La Palma by the Fundaci\'on Galileo Galilei of the INAF (Istituto Nazionale di Astrofisica) at the Spanish Observatorio del Roque de los Muchachos of the Instituto de Astrofisica de Canarias.

Based on observations made with the Nordic Optical Telescope, operated
on the island of La Palma jointly by Denmark, Finland, Iceland,
Norway, and Sweden, in the Spanish Observatorio del Roque de los
Muchachos of the Instituto de Astrofisica de Canarias.

This publication makes use of data products from the Two Micron All Sky Survey, which is a joint project of the University of Massachusetts and the Infrared Processing and Analysis Center/California Institute of Technology, funded by the National Aeronautics and Space Administration and the National Science Foundation.

This work is based in part on data obtained as part of the UKIRT Infrared Deep Sky Survey.

Funding for SDSS-III has been provided by the Alfred P.\ Sloan Foundation, the Participating Institutions, the National Science Foundation, and the U.S.\  Department of Energy Office of Science. The SDSS-III web site is \url{http://www.sdss3.org/}.
SDSS-III is managed by the Astrophysical Research Consortium for the Participating Institutions of the SDSS-III Collaboration including the University of Arizona, the Brazilian Participation Group, Brookhaven National Laboratory, University of Cambridge, Carnegie Mellon University, University of Florida, the French Participation Group, the German Participation Group, Harvard University, the Instituto de Astrofisica de Canarias, the Michigan State/Notre Dame/JINA Participation Group, Johns Hopkins University, Lawrence Berkeley National Laboratory, Max Planck Institute for Astrophysics, Max Planck Institute for Extraterrestrial Physics, New Mexico State University, New York University, Ohio State University, Pennsylvania State University, University of Portsmouth, Princeton University, the Spanish Participation Group, University of Tokyo, University of Utah, Vanderbilt University, University of Virginia, University of Washington, and Yale University.

This research has made use of the NASA/IPAC Extragalactic Database (NED) which is operated by the Jet Propulsion Laboratory, California Institute of Technology, under contract with the National Aeronautics and Space Administration.


\setcounter{section}{0}

\LongTables
\begin{deluxetable*}{lccccrrrrrrrrl}
\tablenum{3}
\tablecaption{Observations and global galaxy parameters\label{table:observations}}
\tablecolumns{14}
\tablehead{\colhead{Galaxy} & \colhead{Right ascension} & \colhead{Declination} & \colhead{Telescope} & \colhead{Campaign} & \colhead{$t_{\textrm{exp}}$} & \colhead{Seeing}  &  \colhead{$S/N$} & \colhead{$m_H$} & \colhead{$r_e [\arcsec]$} &  \colhead{$b/a$} &  \colhead{$C$} &  \colhead{$\log \rho_{10}$} & \colhead{Add. information} }
\startdata
VCC0009 &  12:09:22.3  &  +13:59:22  & TNG  &  2010/Mar  &  222  &  0.88  &  1.3  &  10.59  &  33.9  &  0.83  &  2.83  &  1.29  &  F06  \\
VCC0021 &  12:10:23.1  &  +10:11:23  & NTT & Archive &  12  & 0.74  & - &  12.14  &  14.2  &  0.58  &  3.03  &  0.99  &  F06 \\
VCC0033 &  12:11:07.7  &  +14:16:08  & NTT & Archive &  12  & 0.81  & - &  12.17  &  9.7  &  0.94  &  2.74  &  1.5  &  F06 \\
VCC0140 &  12:15:12.5  &  +14:25:13  & TNG  &  2010/Apr  &  30  &  0.81  &  4.7  &  11.0  &  10.6  &  0.71  &  3.03  &  1.86  &  F06  \\
VCC0165 &  12:15:53.3  &  +13:12:53  & TNG  &  2010/Apr  &  30  &  0.94  &  3.4  &  11.34  &  8.5  &  0.89  &  3.95  &  2.22  &    \\
VCC0170 &  12:15:56.4  &  +14:25:56  & NTT  &  2010/Mar  &  172  &  1.03  &  1.5  &  11.45  &  27.3  &  0.65  &  3.14  &  1.94  &    \\
VCC0173 &  12:16:00.4  &  +08:12:00  & NTT  &  2010/Mar  &  70  &  1.05  &  0.9  &  12.14  &  14.5  &  0.92  &  2.93  &  1.35  &    \\
VCC0178 &  12:16:09.1  &  +12:41:09  & TNG  &  2012/Mar  &  84  &  1.45  &  1.6  &  12.52  &  10.3  &  0.57  &  2.74  &  1.99  &    \\
VCC0200 &  12:16:33.7  &  +13:01:34  & NOT  &  2010/Apr  &  59  &  0.75  &  1.5  &  11.81  &  12.4  &  0.97  &  3.37  &  2.1  &  F06  \\
VCC0209 &  12:16:52.3  &  +14:30:52  & NOT  &  2011/Mar  &  76  &  0.97  &  1.5  &  12.1  &  16.9  &  0.31  &  2.83  &  1.87  &    \\
VCC0216 &  12:17:01.1  &  +09:24:01  & NTT & Archive &  15  & 0.97  & - &  12.17  &  12.1  &  0.65  &  2.63  &  1.45  &  sp, {\it b} \\
VCC0218 &  12:17:05.4  &  +12:17:05  & TNG  &  2010/Apr  &  58  &  0.92  &  1.1  &  11.78  &  22.4  &  0.31  &  3.12  &  1.79  &    \\
VCC0227 &  12:17:14.5  &  +08:56:14  & NTT & Archive &  18  & 1.12  & - &  12.37  &  17.3  &  0.61  &  2.61  &  1.51  &   \\
VCC0230 &  12:17:19.7  &  +11:56:20  & NOT  &  2010/Apr  &  78  &  0.8  &  1.4  &  12.51  &  9.7  &  0.81  &  2.99  &  1.59  &  F06  \\
VCC0308 &  12:18:51.0  &  +07:51:51  & NTT & Archive &  12  & 0.72  & - &  10.99  &  17.5  &  0.94  &  2.95  &  1.61  &  sp \\
VCC0319 &  12:19:02.0  &  +13:58:02  & TNG  &  2012/Mar  &  180  &  1.99  &  0.3  &  12.27  &  25.8  &  0.85  &  2.95  &  1.83  &    \\
VCC0389 &  12:20:03.1  &  +14:57:03  & TNG  &  2010/Apr  &  30  &  0.79  &  2.5  &  10.84  &  16.3  &  0.71  &  3.58  &  1.79  &  M11  \\
VCC0407 &  12:20:18.8  &  +09:32:19  & NOT  &  Archive  &  12  &  0.95  &  0.6  &  11.56  &  18.1  &  0.72  &  2.87  &  1.5  &    \\
VCC0437 &  12:20:48.8  &  +17:29:49  & NTT & Archive &  18  & 0.82  & - &  10.94  &  26.7  &  0.57  &  3.33  &  1.46  &  F06, M11 \\
VCC0490 &  12:21:38.8  &  +15:44:39  & NOT  &  Archive  &  42  &  0.93  &  0.7  &  10.88  &  23.6  &  0.76  &  2.6  &  2.01  &  M11, sp  \\
VCC0510 &  12:21:53.7  &  +15:38:54  & NOT  &  2011/Mar  &  179  &  0.95  &  1.1  &  11.85  &  18.5  &  0.8  &  3.03  &  2.12  &  M11  \\
VCC0523 &  12:22:04.1  &  +12:47:04  & NOT  &  2010/Jan  &  18  &  0.9  &  1.0  &  10.57  &  18.9  &  0.72  &  3.01  &  1.9  &  G00, M11, b, {\it sp}  \\
VCC0543 &  12:22:19.5  &  +14:45:20  & TNG  &  2010/Apr  &  63  &  0.9  &  1.7  &  11.24  &  20.9  &  0.53  &  3.23  &  1.85  &  F06, M11  \\
VCC0571 &  12:22:41.1  &  +07:57:41  & NOT  &  2011/Apr  &  78  &  0.81  &  1.4  &  12.03  &  13.5  &  0.51  &  3.3  &  1.9  &  F06  \\
VCC0575 &  12:22:43.4  &  +08:11:43  & TNG  &  2010/Apr  &  30  &  0.79  &  18.6  &  10.46  &  6.9  &  0.61  &  3.6  &  1.79  &  G00, F06  \\
VCC0608 &  12:23:01.7  &  +15:54:02  & NOT  &  2010/Jan  &  36  &  1.01  &  0.7  &  11.72  &  16.2  &  0.57  &  2.87  &  2.32  &  M11  \\
VCC0634 &  12:23:20.0  &  +15:49:20  & NOT  &  2011/Mar  &  180  &  0.76  &  2.4  &  10.94  &  20.5  &  0.84  &  3.23  &  2.22  &  M11  \\
VCC0698 &  12:24:05.1  &  +11:13:05  & NOT  &  2011/Mar  &  77  &  0.91  &  7.9  &  9.87  &  19.2  &  0.43  &  3.75  &  1.94  &  G00, F06, M11  \\
VCC0745 &  12:24:47.0  &  +07:21:47  & TNG  &  2010/Mar  &  26  &  0.86  &  1.0  &  11.77  &  10.8  &  0.62  &  2.99  &  2.54  &    \\
VCC0750 &  12:24:49.6  &  +06:45:50  & NTT & Archive &  18  & 0.86  & - &  12.09  &  15.0  &  0.72  &  3.05  &  2.16  &   \\
VCC0751 &  12:24:48.3  &  +18:11:48  & NTT  &  2010/Mar  &  12  &  0.86  &  1.3  &  11.46  &  11.0  &  0.62  &  3.48  &  1.33  &  F06, M11  \\
VCC0758 &  12:24:54.9  &  +07:26:55  & TNG  &  2012/Mar  &  84  &  1.43  &  6.1  &  9.57  &  15.4  &  0.56  &  3.41  &  2.38  &  G00  \\
VCC0762 &  12:25:03.0  &  +07:30:03  & TNG  &  2012/Mar  &  79  &  1.38  &  1.0  &  12.57  &  9.3  &  0.85  &  2.99  &  2.25  &    \\
VCC0781 &  12:25:15.2  &  +12:42:15  & NTT  &  2010/Mar  &  6  &  1.87  &  0.7  &  11.69  &  13.8  &  0.57  &  3.18  &  2.47  &  G00, M11  \\
VCC0786 &  12:25:14.5  &  +11:50:15  & NOT  &  2011/Mar  &  79  &  0.78  &  1.2  &  11.89  &  26.7  &  0.42  &  2.97  &  2.19  &  M11  \\
VCC0794 &  12:25:21.6  &  +16:25:22  & TNG  &  2010/Mar  &  183  &  1.11  &  1.4  &  12.17  &  28.1  &  0.36  &  3.16  &  1.97  &  M11  \\
VCC0816 &  12:25:36.1  &  +15:50:36  & NOT  &  2012/Mar  &  163  &  1.6  &  1.1  &  12.71  &  20.7  &  0.59  &  2.32  &  2.14  &  M11  \\
VCC0817 &  12:25:36.4  &  +15:49:36  & NOT  &  2012/Mar  &  163  &  1.6  &  0.7  &  11.45  &  21.5  &  0.86  &  3.2  &  2.14  &  M11  \\
VCC0856 &  12:25:57.9  &  +10:03:58  & NTT & Archive &  12  & 0.8  & - &  11.25  &  14.0  &  0.89  &  2.77  &  2.05  &  F06, M11, sp \\
VCC0870 &  12:26:05.3  &  +11:48:05  & NOT  &  2010/Apr  &  96  &  0.78  &  1.1  &  12.02  &  16.7  &  0.55  &  3.07  &  2.18  &    \\
VCC0916 &  12:26:33.2  &  +12:44:33  & NOT  &  2012/Mar  &  74  &  0.59  &  3.3  &  12.18  &  5.2  &  0.93  &  3.29  &  2.69  &    \\
VCC0917 &  12:26:32.4  &  +13:34:32  & TNG  &  2010/Apr  &  26  &  0.83  &  0.7  &  12.41  &  10.2  &  0.55  &  3.14  &  2.25  &  M11  \\
VCC0929 &  12:26:40.5  &  +08:26:41  & NOT  &  2011/Mar  &  88  &  0.76  &  3.1  &  10.25  &  16.1  &  0.88  &  3.09  &  2.24  &  G00  \\
VCC0940 &  12:26:47.1  &  +12:27:47  & NTT  &  2010/Mar  &  174  &  1.0  &  1.4  &  11.57  &  18.1  &  0.73  &  2.77  &  2.66  &  M11, $^a$  \\
VCC0949 &  12:26:54.6  &  +10:39:55  & TNG  &  2012/Mar  &  190  &  1.54  &  0.6  &  12.24  &  21.8  &  0.63  &  3.05  &  1.96  &  M11  \\
VCC0951 &  12:26:54.3  &  +11:39:54  & NTT & Archive &  6  & 0.81  & - &  11.43  &  19.3  &  0.71  &  2.95  &  2.26  &  M11 \\
VCC0990 &  12:27:17.0  &  +16:01:17  & NOT  &  2010/Jan  &  8  &  0.89  &  1.0  &  11.48  &  10.3  &  0.62  &  3.27  &  2.04  &  M11  \\
VCC1010 &  12:27:27.4  &  +12:17:27  & NOT  &  2010/Jan  &  8  &  1.21  &  0.9  &  10.4  &  20.4  &  0.58  &  3.07  &  2.43  &  G00, M11, b, {\it sp}  \\
VCC1036 &  12:27:41.3  &  +12:18:41  & NTT & Archive &  6  & 0.74  & - &  10.65  &  19.8  &  0.43  &  3.49  &  2.4  &  M11 \\
VCC1049 &  12:27:54.8  &  +08:05:55  & TNG  &  2010/Mar  &  10  &  0.97  &  1.5  &  12.13  &  7.9  &  0.85  &  3.14  &  2.33  &  F06  \\
VCC1073 &  12:28:08.6  &  +12:05:09  & TNG  &  2010/Apr  &  44  &  1.0  &  1.6  &  10.87  &  19.6  &  0.66  &  3.6  &  2.28  &  M11  \\
VCC1075 &  12:28:12.3  &  +10:17:12  & NOT  &  2011/Mar  &  181  &  0.79  &  1.2  &  11.96  &  18.3  &  0.65  &  3.16  &  2.08  &  F06, M11  \\
VCC1087 &  12:28:14.9  &  +11:47:15  & NOT  &  2011/Mar  &  72  &  0.94  &  1.6  &  10.86  &  18.6  &  0.7  &  3.05  &  2.18  &  F06, M11  \\
VCC1104 &  12:28:28.1  &  +12:49:28  & TNG  &  2010/Apr  &  90  &  0.83  &  1.2  &  12.4  &  14.1  &  0.68  &  2.91  &  2.45  &  M11  \\
VCC1122 &  12:28:41.7  &  +12:54:42  & TNG  &  2010/Apr  &  60  &  0.92  &  3.3  &  11.79  &  16.9  &  0.43  &  3.05  &  2.5  &  M11  \\
VCC1148 &  12:28:58.2  &  +12:39:58  & TNG  &  2010/Apr  &  24  &  0.8  &  2.0  &  12.26  &  5.2  &  0.97  &  3.7  &  2.39  &  M11  \\
VCC1167 &  12:29:14.7  &  +07:52:15  & NOT  &  2012/Mar  &  180  &  1.7  &  0.9  &  12.32  &  15.6  &  0.97  &  2.99  &  2.53  &    \\
VCC1178 &  12:29:21.3  &  +08:09:21  & NOT  &  2011/Mar  &  76  &  0.67  &  32.5  &  9.67  &  7.4  &  0.71  &  3.93  &  2.41  &  G00, F06  \\
VCC1183 &    {\it 12:29:22.5} & {\it +11:26:02}  & NTT & Archive &  12  & 0.87  & - &  11.12  &  17.0  &  0.61  &  3.43  &  2.04  &  M11, b \\
VCC1185 &  12:29:23.5  &  +12:27:24  & NOT  &  2012/Mar  &  185  &  1.26  &  0.7  &  12.4  &  16.2  &  0.91  &  3.01  &  2.43  &  F06, M11  \\
VCC1196 &  12:29:30.9  &  +14:02:31  & TNG  &  2010/Mar  &  30  &  0.9  &  5.9  &  10.34  &  16.8  &  0.73  &  3.71  &  2.15  &  G00, M11  \\
VCC1254 &  12:30:05.1  &  +08:04:05  & NOT  &  2011/Mar  &  78  &  1.13  &  2.0  &  12.02  &  11.4  &  0.98  &  2.99  &  2.56  &    \\
VCC1261 &  12:30:10.3  &  +10:46:10  & NOT  &  2011/Apr  &  63  &  0.89  &  2.2  &  10.39  &  21.7  &  0.58  &  3.23  &  1.97  &  G00, F06, M11  \\
VCC1283 &  12:30:18.4  &  +13:34:18  & TNG  &  2010/Mar  &  29  &  1.78  &  7.0  &  9.86  &  15.9  &  0.55  &  3.25  &  2.22  &  G00, F06, M11  \\
VCC1297 &  12:30:31.9  &  +12:29:32  & TNG  &  2010/Mar  &  34  &  0.97  &  90.9  &  10.2  &  2.6  &  0.84  &  3.51  &  2.61  &  F06, M11  \\
VCC1304 &  12:30:39.8  &  +15:07:40  & NOT  &  2010/Apr  &  72  &  0.79  &  1.7  &  12.03  &  16.4  &  0.33  &  2.67  &  2.07  &  M11  \\
VCC1308 &  12:30:45.9  &  +11:20:46  & NOT  &  2012/Mar  &  82  &  0.91  &  0.9  &  12.43  &  10.7  &  0.67  &  3.01  &  2.1  &  M11  \\
VCC1348 &  12:31:15.7  &  +12:19:16  & NOT  &  2011/Mar  &  108  &  0.74  &  2.4  &  12.28  &  8.9  &  0.93  &  3.18  &  2.54  &  M11  \\
VCC1355 &  12:31:20.3  &  +14:06:20  & NOT  &  2011/Mar  &  180  &  0.9  &  1.1  &  11.55  &  22.3  &  0.76  &  3.01  &  2.12  &  F06  \\
VCC1386 &  12:31:51.4  &  +12:39:51  & NOT  &  2011/Apr  &  43  &  0.89  &  0.9  &  11.78  &  25.0  &  0.63  &  2.83  &  2.34  &  M11  \\
VCC1392 &  12:31:55.9  &  +12:10:56  & NTT & Archive &  18  & 0.68  & - &  11.81  &  16.9  &  0.95  &  2.49  &  2.38  &  M11 \\
VCC1407 &  12:32:02.7  &  +11:53:03  & NOT  &  2010/Apr  &  52  &  0.86  &  0.9  &  11.85  &  11.9  &  0.82  &  3.16  &  2.16  &  F06, M11  \\
VCC1422 &  12:32:14.2  &  +10:15:14  & NOT  &  2010/Jan  &  23  &  1.0  &  1.3  &  10.61  &  17.8  &  0.86  &  3.32  &  1.88  &  F06, M11, {\it b}  \\
VCC1431 &  12:32:23.5  &  +11:15:23  & NOT  &  2012/Mar  &  75  &  0.91  &  3.8  &  11.02  &  9.1  &  0.97  &  3.01  &  2.38  &  F06, M11  \\
VCC1440 &  12:32:33.5  &  +15:24:34  & TNG  &  2010/Apr  &  30  &  0.82  &  3.8  &  11.52  &  7.2  &  0.97  &  3.91  &  2.12  &  F06, M11  \\
VCC1453 &  12:32:44.2  &  +14:11:44  & NOT  &  2010/Apr  &  47  &  0.69  &  1.5  &  11.0  &  16.9  &  0.77  &  3.6  &  1.91  &  M11  \\
VCC1475 &  12:33:05.0  &  +16:15:05  & NOT  &  2011/Mar  &  80  &  0.92  &  11.8  &  9.91  &  9.2  &  0.81  &  3.93  &  1.87  &  G00, F06, M11  \\
VCC1479 &  12:33:07.5  &  +14:34:08  & TNG  &  2010/Mar  &  60  &  1.23  &  4.7  &  10.05  &  23.3  &  0.4  &  3.22  &  2.03  &  G00, M11  \\
VCC1488 &  12:33:13.4  &  +09:23:13  & NOT  &  2011/Mar  &  75  &  0.78  &  1.4  &  11.96  &  14.0  &  0.6  &  3.16  &  1.89  &  F06, M11  \\
VCC1491 &  12:33:14.0  &  +12:51:14  & NTT & Archive &  8  & 0.8  & - &  12.04  &  10.0  &  0.79  &  2.89  &  2.23  &  M11 \\
VCC1499 &  12:33:19.8  &  +12:51:20  & NTT & Archive &  8  & 0.8  & - &  12.49  &  7.5  &  0.63  &  2.99  &  2.21  &  F06, M11 \\
VCC1501 &  12:33:24.7  &  +08:41:25  & TNG  &  2012/Mar  &  94  &  1.54  &  0.6  &  12.86  &  12.0  &  0.45  &  3.13  &  1.93  &  M11  \\
VCC1512 &  12:33:34.6  &  +11:15:35  & NTT  &  2010/Mar  &  57  &  0.86  &  0.9  &  12.69  &  12.1  &  0.74  &  3.73  &  2.05  &  F06, M11  \\
VCC1514 &  12:33:37.7  &  +07:52:38  & NTT  &  2010/Mar  &  180  &  1.01  &  1.3  &  12.19  &  24.1  &  0.32  &  2.89  &  1.88  &  M11  \\
VCC1521 &  12:33:45.0  &  +10:59:45  & TNG  &  2010/Apr  &  26  &  1.19  &  5.6  &  10.64  &  14.6  &  0.32  &  3.46  &  2.17  &  G00, M11  \\
VCC1528 &  12:33:51.6  &  +13:19:52  & NTT & Archive &  6  & 0.81  & - &  11.37  &  8.4  &  0.9  &  3.43  &  1.95  &  F06, M11 \\
VCC1545 &  12:34:11.5  &  +12:02:12  & NOT  &  2010/Apr  &  33  &  0.69  &  0.7  &  11.72  &  9.9  &  0.77  &  3.68  &  2.12  &  F06, M11  \\
VCC1549 &  12:34:14.8  &  +11:04:15  & NTT & Archive &  12  & 0.6  & - &  11.4  &  11.4  &  0.81  &  3.2  &  2.09  &  M11 \\
VCC1567 &  12:34:31.2  &  +09:37:31  & NTT  &  2010/Mar  &  104  &  1.11  &  0.7  &  11.26  &  28.6  &  0.34  &  2.89  &  1.79  &  M11  \\
VCC1614 &  12:35:27.3  &  +12:45:27  & NOT  &  2011/Mar  &  93  &  0.78  &  4.9  &  11.36  &  9.0  &  0.86  &  2.65  &  2.18  &  M11  \\
VCC1627 &  12:35:37.3  &  +12:22:37  & TNG  &  2010/Apr  &  27  &  1.03  &  2.0  &  11.72  &  3.8  &  0.89  &  3.2  &  2.16  &  F06, M11  \\
VCC1684 &   {\it 12:36:39.4} & {\it+11:06:07}   & NTT  &  2010/Mar  &  167  &  0.97  &  1.1  &  12.48  &  18.5  &  0.29  &  2.97  &  1.89  &  M11  \\
VCC1695 &  12:36:54.8  &  +12:31:55  & NTT & Archive &  12  & 0.63  & - &  11.57  &  16.2  &  0.79  &  3.62  &  1.97  &  F06, M11, sp, {\it b} \\
VCC1743 &  12:38:06.8  &  +10:04:07  & NOT  &  2012/Mar  &  150  &  0.9  &  1.1  &  12.53  &  15.1  &  0.39  &  2.95  &  2.13  &  F06, M11  \\
VCC1779 &  12:39:04.7  &  +14:43:05  & NTT  &  2010/Mar  &  67  &  0.95  &  1.1  &  11.99  &  19.6  &  0.41  &  3.2  &  1.66  &  F06, M11  \\
VCC1827 &  12:40:11.9  &  +08:23:12  & NOT  &  2012/Mar  &  75  &  1.65  &  1.6  &  10.15  &  20.7  &  0.6  &  3.12  &  1.61  &  G00, M11  \\
VCC1828 &  12:40:13.5  &  +12:52:13  & NOT  &  2011/Mar  &  151  &  0.81  &  1.4  &  12.16  &  15.7  &  0.7  &  3.1  &  1.78  &  F06, M11  \\
VCC1833 &  12:40:19.8  &  +15:56:20  & NOT  &  2011/Mar  &  65  &  0.89  &  4.0  &  11.3  &  8.3  &  0.68  &  3.31  &  1.35  &  F06, M11  \\
VCC1836 &  12:40:19.6  &  +14:42:20  & NTT  &  2010/Mar  &  150  &  0.98  &  0.6  &  11.51  &  46.8  &  0.47  &  3.23  &  1.63  &  M11, sp  \\
VCC1857 &  12:40:53.0  &  +10:28:53  & NOT  &  2012/Mar  &  167  &  1.7  &  0.2  &  12.19  &  26.5  &  0.52  &  2.43  &  1.93  &  F06, M11  \\
VCC1861 &  12:40:58.6  &  +11:11:59  & NOT  &  2011/Mar  &  76  &  0.84  &  1.9  &  11.13  &  15.3  &  0.97  &  3.28  &  2.02  &  F06, M11  \\
VCC1871 &  12:41:15.8  &  +11:23:16  & TNG  &  2010/Apr  &  29  &  0.86  &  11.7  &  10.68  &  6.9  &  0.87  &  3.27  &  2.24  &  F06, M11  \\
VCC1876 &  12:41:20.6  &  +14:42:21  & NOT  &  2011/Mar  &  112  &  0.87  &  1.4  &  12.05  &  16.9  &  0.51  &  2.7  &  1.58  &  M11  \\
VCC1890 &  12:41:46.7  &  +11:29:47  & NOT  &  2011/Mar  &  193  &  0.79  &  0.8  &  12.03  &  23.7  &  0.6  &  2.7  &  2.5  &  M11  \\
VCC1895 &  12:41:51.9  &  +09:24:52  & NTT & Archive &  12  & 1.01  & - &  12.04  &  15.0  &  0.48  &  3.03  &  1.78  &  F06, M11 \\
VCC1896 &  12:41:54.6  &  +09:35:55  & NOT  &  Archive  &  14  &  1.17  &  0.5  &  11.8  &  14.9  &  0.55  &  2.99  &  1.84  &  M11, b, sp  \\
VCC1897 &  12:41:54.3  &  +13:46:54  & NOT  &  2011/Apr  &  107  &  0.85  &  0.7  &  11.38  &  30.8  &  0.55  &  2.87  &  1.8  &  M11  \\
VCC1902 &  12:41:59.5  &  +12:56:59  & NOT  &  2012/Mar  &  82  &  1.55  &  1.2  &  9.97  &  22.4  &  0.87  &  3.14  &  1.94  &  G00, M11  \\
VCC1910 &  12:42:08.7  &  +11:45:09  & NOT  &  2010/Jan  &  8  &  1.0  &  1.0  &  10.77  &  11.6  &  0.81  &  3.12  &  2.47  &  F06, M11, {\it b}  \\
VCC1912 &  12:42:09.1  &  +12:35:09  & NOT  &  2010/Jan  &  8  &  1.16  &  0.6  &  11.06  &  22.2  &  0.37  &  3.23  &  1.97  &  M11, b  \\
VCC1945 &  12:42:54.1  &  +11:26:54  & NOT  &  2011/Mar  &  96  &  0.99  &  0.8  &  11.73  &  23.4  &  0.37  &  2.64  &  2.42  &  M11  \\
VCC1947 &  12:42:56.3  &  +03:40:56  & NTT & Archive &  12  & 0.93  & - &  11.06  &  9.1  &  0.76  &  3.16  &  1.42  &  b \\
VCC1949 &  12:42:57.8  &  +12:17:58  & NOT  &  2011/Apr  &  142  &  1.13  &  1.9  &  10.72  &  25.2  &  0.44  &  2.97  &  2.13  &  M11, b  \\
VCC2008 &  12:44:47.4  &  +12:03:47  & TNG  &  2012/Mar  &  185  &  1.61  &  0.3  &  12.24  &  39.2  &  0.41  &  2.71  &  2.04  &  M11  \\
VCC2012 &  12:45:05.7  &  +10:54:06  & NOT  &  2011/Mar  &  180  &  1.04  &  1.2  &  11.5  &  21.5  &  0.83  &  2.79  &  1.87  &  M11  \\
VCC2019 &  12:45:20.4  &  +13:41:20  & NOT  &  Archive  &  20  &  1.05  &  0.7  &  11.33  &  16.7  &  0.69  &  3.18  &  1.74  &  F06, M11, {\it sp}  \\
VCC2042 &  12:46:38.3  &  +09:18:38  & NTT & Archive &  18  & 0.69  & - &  12.42  &  18.4  &  0.84  &  2.66  &  1.57  &  M11, {\it sp} \\
VCC2048 &  12:47:15.3  &  +10:12:15  & NOT  &  2010/Jan  &  8  &  1.22  &  1.0  &  10.79  &  18.5  &  0.51  &  3.43  &  1.7  &  G00, F06, M11, {\it b}  \\
VCC2050 &  12:47:20.6  &  +12:09:21  & NTT  &  2010/Mar  &  44  &  1.13  &  0.9  &  12.19  &  14.3  &  0.48  &  3.03  &  1.56  &  F06, M11  \\
VCC2080 &  12:48:58.4  &  +10:35:58  & NTT  &  2010/Mar  &  96  &  1.1  &  0.8  &  12.82  &  18.8  &  0.42  &  3.32  &  1.64  &    \\
VCC2083 &  12:50:14.5  &  +10:32:14  & NOT  &  2012/Mar  &  168  &  1.54  &  0.3  &  12.66  &  14.1  &  0.84  &  2.62  &  1.57  &  M11  
 \enddata
\tablecomments{The coordinates give the position of the galaxy centers of the galaxy as found by the assymetry centering and by identification of 2MASS stars
(the coordinats printed in italics list the NED values).
The total on target integration times ($t_{\textrm{exp}}$)  are given in minutes,  and the seeing is specified as the full width at half maximum (FWHM) in arcsecs, and the signal-to-noise is given per pixel at 2$r_e$ ($S/N$).
The next columns are total brightness ($m_H$), effective radius ($r_e$), the axis ratio $b/a$ at $2r_e$, the concentration index $C$ (see \S\ref{sec:nonpar}), and the local projected density ($\rho_{10}$, see Appendix \ref{sec:additional}). The references given in the colum with additional information are: G01 -- \citet{gavazzi+profiles}, F06 -- \citet{Ferrarese:2006p586}, and M11 -- \citet{Mcdonald:2011p4445}. Furthermore, b and sp indicate that \citet{Lisker:2006p385} found a bar or spiral arms, or indications for those when printed with italics.\\
$^a$While \citet{Lisker:2006p385} did not find the bar in VCC0940 with SDSS images, it was reported by \citet{barazza} using VLT data.}
\end{deluxetable*}

 \LongTables
\clearpage
\begin{landscape}
\begin{deluxetable*}{lc|rrrr|rrrr|rrrr|rr|rr|cl}
\centering
\tablenum{5}
\tablecaption{Parameters from the decompositions\label{table:params}}
\tablecolumns{20}
\tablehead{&   & \multicolumn{4}{c}{Simple} &  \multicolumn{4}{c}{Inner} &  \multicolumn{4}{c}{Global / outer} & \multicolumn{2}{c}{Simple}  & \multicolumn{2}{c}{Final} & & \colhead{}   \\
\colhead{Galaxy} & \colhead{Group}  & \multicolumn{4}{c}{model} &  \multicolumn{4}{c}{component} &  \multicolumn{4}{c}{component} & \multicolumn{2}{c}{decomposition}  & \multicolumn{2}{c}{decomposition}  & \colhead{Nucleus} & \colhead{Additional}\\
 & & \colhead{$m$} & \colhead{$r_e$} & \colhead{$n$} & \colhead{$b/a$} & \colhead{$m$} & \colhead{$r_e$} & \colhead{$n$} & \colhead{$b/a$} & \colhead{$m$} & \colhead{$r_e$} & \colhead{$n$} & \colhead{$b/a$} &  \colhead{RFF} & \colhead{EVI}  &  \colhead{RFF} & \colhead{EVI}  & & \colhead{information}}
\startdata
VCC0009 & L & - & - & - & - & - & - & - & - & 10.58 & 35.0 & 0.93 & 0.85 & 0.019 & 0.112 & -0.001 & 0.05 & N  \\
VCC0021 & 1 & 12.19 & 12.3 & 1.24 & 0.62 & - & - & - & - & - & - & - & - & - & - & - & - & -  \\
VCC0033 & 1 & 12.2 & 8.8 & 0.98 & 0.89 & - & - & - & - & - & - & - & - & - & - & - & - & -  \\
VCC0140 & 2 & - & - & - & - & 12.12 & 5.6 & 0.96 & 0.69 & 11.52 & 14.3 & 0.72 & 0.73 & 0.02 & 0.26 & -0.012 & -0.069 & N  \\
VCC0165 & L & 11.27 & 9.6 & 3.4 & 0.87 & 11.58 & 7.8 & 3.54 & 0.87 & 13.31 & 14.1 & 0.27 & 0.81 & 0.012 & 0.495 & 0.001 & 0.315 & -  \\
VCC0170 & 2 & - & - & - & - & 12.23 & 15.6 & 1.14 & 0.65 & 12.17 & 41.4 & 0.41 & 0.7 & - & - & 0.025 & 0.142 & N  \\
VCC0173 & {\it 2} & 12.06 & 15.7 & 1.48 & 0.89 & 14.66 & 6.3 & 1.63 & 0.64 & 12.25 & 15.6 & 1.0 & 0.93 & 0.018 & 0.083 & 0.012 & 0.048 & - & sp \\
VCC0178 & 1 & 12.53 & 10.1 & 0.95 & 0.55 & - & - & - & - & - & - & - & - & -0.006 & -0.043 & - & - & N  \\
VCC0200 & L & 11.8 & 12.3 & 1.82 & 0.91 & 13.43 & 5.8 & 1.6 & 0.74 & 12.2 & 16.6 & 1.0 & 0.92 & 0.017 & 0.111 & 0.009 & 0.055 & -  \\
VCC0216 & 1 & 12.18 & 12.5 & 0.87 & 0.67 & - & - & - & - & - & - & - & - & - & - & - & - & N & sp \\
VCC0227 & 1 & 12.37 & 17.5 & 0.87 & 0.62 & - & - & - & - & - & - & - & - & - & - & - & - & N  \\
VCC0230 & 1 & 12.54 & 9.2 & 1.25 & 0.82 & - & - & - & - & - & - & - & - & 0.011 & 0.595 & - & - & N  \\
VCC0308 & 2 & 10.81 & 21.1 & 1.85 & 0.94 & 13.07 & 5.0 & 1.14 & 0.88 & 11.18 & 19.4 & 0.63 & 0.96 & -0.011 & -0.028 & -0.029 & -0.087 & N & sp  \\
VCC0319 & 1 & 12.29 & 21.2 & 1.26 & 0.88 & - & - & - & - & - & - & - & - & -0.057 & -0.147 & - & - & N  \\
VCC0389 & {\it L} & 10.85 & 15.8 & 1.91 & 0.79 & - & - & - & - & 11.16 & 20.6 & 0.89 & 0.85 & 0.025 & 0.652 & 0.012 & 0.584 & N  \\
VCC0407 & B & 11.64 & 14.5 & 0.99 & 0.71 & - & - & - & - & 11.68 & 16.9 & 1.0 & 0.69 & 0.026 & 0.069 & 0.02 & 0.048 & -  \\
VCC0437 & 2 & 10.88 & 24.6 & 1.72 & 0.64 & 12.52 & 8.5 & 1.14 & 0.78 & 11.19 & 33.9 & 1.0 & 0.54 & - & - & - & - & N  \\
VCC0490 & 2 & 10.86 & 23.0 & 0.94 & 0.78 & 13.05 & 18.9 & 3.38 & 0.78 & 11.05 & 23.4 & 0.67 & 0.73 & 0.031 & 0.78 & 0.019 & 0.388 & -  & sp \\
VCC0510 & 1 & 11.8 & 19.7 & 1.51 & 0.83 & - & - & - & - & - & - & - & - & 0.017 & 0.128 & - & - & N  \\
VCC0523 & 2 & 10.53 & 18.5 & 1.5 & 0.77 & 12.75 & 7.1 & 1.15 & 0.63 & 10.73 & 21.4 & 1.0 & 0.69 & 0.015 & 0.051 & 0.0 & -0.013 & -  \\
VCC0543 & 2 & 11.25 & 19.9 & 1.65 & 0.54 & 13.12 & 6.7 & 0.92 & 0.53 & 11.49 & 24.1 & 1.0 & 0.57 & 0.017 & 0.164 & 0.009 & 0.059 & -  \\
VCC0571 & 2 & 11.83 & 17.2 & 2.78 & 0.57 & 14.2 & 2.5 & 1.62 & 0.68 & 12.24 & 14.7 & 1.0 & 0.56 & 0.038 & 0.332 & 0.018 & 0.119 & -  \\
VCC0608 & {\it B} & 11.73 & 17.1 & 1.2 & 0.64 & 14.73 & 5.9 & 1.77 & 0.58 & 11.96 & 19.4 & 0.9 & 0.71 & 0.012 & 0.057 & 0.005 & 0.011 & -  \\
VCC0634 & {\it L} & 10.95 & 19.4 & 1.51 & 0.81 & - & - & - & - & 11.13 & 24.7 & 1.0 & 0.78 & 0.03 & 0.453 & 0.017 & 0.268 & N  \\
VCC0745 & 1 & 11.78 & 10.1 & 1.34 & 0.63 & - & - & - & - & - & - & - & - & -0.001 & 0.007 & - & - & N  \\
VCC0750 & L & 11.98 & 16.1 & 1.51 & 0.77 & - & - & - & - & 12.04 & 16.8 & 1.29 & 0.72 & - & - & - & - & N  \\
VCC0751 & 2 & 11.46 & 12.2 & 2.04 & 0.58 & 12.73 & 5.5 & 1.38 & 0.49 & 11.95 & 14.5 & 1.0 & 0.74 & 0.012 & 0.058 & 0.003 & 0.004 & N  \\
VCC0758 & 2 & 9.56 & 15.4 & 2.04 & 0.54 & 11.01 & 6.0 & 1.04 & 0.47 & 9.93 & 19.7 & 1.0 & 0.63 & 0.032 & 3.045 & 0.023 & 1.573 & -  \\
VCC0762 & 2 & 12.62 & 8.3 & 1.27 & 0.83 & 13.53 & 5.2 & 0.87 & 0.86 & 13.15 & 13.8 & 1.0 & 0.8 & 0.002 & 0.015 & -0.001 & -0.007 & N  \\
VCC0781 & 2 & 11.68 & 12.8 & 1.52 & 0.59 & 13.94 & 3.9 & 0.65 & 0.73 & 11.87 & 14.8 & 1.0 & 0.57 & 0.006 & -0.003 & 0.003 & -0.015 & N  \\
VCC0816 & 1 & 12.61 & 21.8 & 0.73 & 0.54 & - & - & - & - & - & - & - & - & 0.008 & 0.031 & - & - & -  \\
VCC0817 & 2 & 11.44 & 19.6 & 1.52 & 0.91 & 13.38 & 6.3 & 0.7 & 0.93 & 11.59 & 26.8 & 1.0 & 0.86 & 0.023 & 0.132 & 0.004 & 0.017 & -  \\
VCC0856 & 1 & 11.26 & 14.0 & 1.02 & 0.9 & - & - & - & - & - & - & - & - & - & - & - & - & N & sp \\
VCC0870 & {\it 2} & 11.98 & 16.4 & 1.57 & 0.59 & 14.49 & 4.1 & 0.82 & 0.68 & 12.15 & 17.8 & 1.0 & 0.58 & 0.022 & 0.271 & 0.017 & 0.265 & N  \\
VCC0916 & 1 & 12.16 & 5.4 & 1.87 & 0.94 & - & - & - & - & - & - & - & - & 0.007 & 0.073 & - & - & N  \\
VCC0917 & 2 & 12.36 & 9.1 & 1.82 & 0.57 & 14.53 & 2.2 & 1.19 & 0.78 & 12.62 & 10.3 & 1.0 & 0.52 & -0.003 & -0.018 & -0.013 & -0.061 & -  \\
VCC0929 & 1 & 10.26 & 16.0 & 1.5 & 0.87 & - & - & - & - & - & - & - & - & 0.016 & 0.37 & - & - & N  \\
VCC0940 & B & 11.55 & 19.3 & 1.14 & 0.81 & 14.4 & 5.1 & 0.85 & 0.93 & 11.81 & 20.4 & 0.74 & 0.92 & 0.019 & 0.143 & 0.014 & 0.103 & N  \\
VCC0949 & 1 & 12.28 & 19.7 & 1.18 & 0.67 & - & - & - & - & - & - & - & - & -0.046 & -0.148 & - & - & -  \\
VCC0951 & 1 & 11.39 & 19.1 & 1.29 & 0.72 & - & - & - & - & - & - & - & - & - & - & - & - & N  \\
VCC0990 & {\it 2} & 11.5 & 10.0 & 1.66 & 0.67 & 16.76 & 1.8 & 0.95 & 0.38 & 11.52 & 9.8 & 1.59 & 0.69 & 0.001 & 0.011 & -0.001 & -0.029 & -  \\
VCC1010 & B & 10.3 & 23.4 & 1.74 & 0.54 & 13.93 & 2.6 & 0.9 & 0.71 & 10.52 & 23.4 & 1.0 & 0.57 & -0.0 & -0.015 & -0.009 & -0.052 & -  \\
VCC1049 & {\it L} & 12.08 & 7.9 & 1.81 & 0.82  & - & - & - & - & 12.23 & 7.9 & 2.04 & 0.83 & 0.006 & 0.023 & 0.0 & 0.017 & -  \\
VCC1073 & {\it B} & 10.8 & 20.8 & 2.46 & 0.74 & 12.29 & 6.1 & 1.45 & 0.81 & 11.36 & 29.0 & 0.68 & 0.76 & 0.007 & 0.09 & -0.002 & 0.086 & -  \\
VCC1075 & 2 & 11.97 & 18.3 & 1.39 & 0.69 & 13.64 & 8.5 & 0.84 & 0.6 & 12.2 & 22.5 & 1.0 & 0.79 & 0.021 & 0.14 & 0.015 & 0.094 & N  \\
VCC1087 & {\it L} & 10.79 & 18.8 & 1.47 & 0.72 & - & - & - & - & 10.93 & 20.8 & 1.0 & 0.68 & 0.012 & 0.076 & 0.007 & 0.046 & N  \\
VCC1104 & L & 12.46 & 12.8 & 1.08 & 0.69 & - & - & - & - & 12.51 & 14.1 & 1.0 & 0.69 & 0.008 & 0.051 & 0.004 & 0.028 & N  \\
VCC1148 & {\it L} & 12.24 & 5.5 & 2.47 & 0.96 & - & - & - & - & 12.27 & 5.8 & 3.02 & 0.94 & -0.011 & 0.116 & -0.024 & -0.126 & N  \\
VCC1167 & 1 & 12.28 & 15.5 & 1.48 & 0.96 & - & - & - & - & - & - & - & - & 0.004 & 0.02 & - & - & N  \\
VCC1183 & L & 11.15 & 13.6 & 1.72 & 0.8 & 13.49 & 4.0 & 1.02 & 0.7 & 11.39 & 22.3 & 1.0 & 0.54 & - & - & - & - & N  \\
VCC1185 & 1 & 12.37 & 16.4 & 1.48 & 0.94 & - & - & - & - & - & - & - & - & 0.007 & 0.036 & - & - & N  \\
VCC1254 & 2 & 11.93 & 12.7 & 1.4 & 0.98 & 15.4 & 1.4 & 1.22 & 0.93 & 12.02 & 12.8 & 1.0 & 0.93 & 0.032 & 0.495 & 0.013 & 0.093 & N  \\
VCC1261 & 2 & 10.33 & 22.3 & 1.87 & 0.62 & 13.27 & 3.5 & 1.23 & 0.92 & 10.48 & 23.1 & 1.22 & 0.57 & 0.019 & 0.243 & 0.003 & 0.033 & N  \\
VCC1283 & B & 9.86 & 18.9 & 1.64 & 0.59 & 12.05 & 4.1 & 0.98 & 0.71 & 10.45 & 21.2 & 1.0 & 0.96 & 0.067 & 4.557 & 0.015 & 0.754 & -  \\
VCC1308 & {\it 2} & 12.47 & 9.6 & 1.22 & 0.67 & 13.81 & 5.1 & 0.92 & 0.84 & 12.82 & 13.7 & 1.0 & 0.57 & 0.008 & 0.038 & 0.005 & 0.027 & N  \\
VCC1348 & 2 & 12.24 & 9.6 & 1.59 & 0.96 & 13.79 & 5.8 & 2.29 & 0.84 & 12.6 & 10.7 & 1.0 & 0.87 & 0.028 & 1.296 & 0.021 & 1.801 & N  \\
VCC1355 & {\it 2} & 11.46 & 26.3 & 1.48 & 0.81 & 12.89 & 14.2 & 1.17 & 0.68 & 11.81 & 30.9 & 0.8 & 0.96 & 0.013 & 0.094 & 0.012 & 0.089 & N  \\
VCC1386 & {\it B} & 11.77 & 23.3 & 1.06 & 0.64 & - & - & - & - & 11.82 & 25.7 & 1.0 & 0.62 & 0.223 & -0.19 & 0.015 & 0.13 & N  \\
VCC1392 & {\it B} & 11.8 & 21.3 & 0.83 & 0.54 & - & - & - & - & 11.83 & 26.6 & 1.0 & 0.57 & - & - & - & - & -  \\
VCC1407 & 1 & 11.9 & 11.1 & 1.44 & 0.83 & - & - & - & - & - & - & - & - & 0.003 & 0.017 & - & - & N  \\
VCC1422 & 2 & 10.54 & 19.2 & 1.95 & 0.79 & 11.58 & 10.0 & 1.53 & 0.72 & 11.13 & 27.2 & 1.0 & 0.81 & 0.005 & 0.01 & -0.0 & -0.007 & -  \\
VCC1431 & 1 & 11.01 & 9.2 & 1.48 & 0.96 & - & - & - & - & - & - & - & - & 0.006 & 0.092 & - & - & N  \\
VCC1440 & 2 & 11.57 & 6.5 & 3.33 & 0.97 & 12.3 & 3.2 & 1.84 & 0.97 & 12.5 & 14.4 & 1.0 & 0.91 & -0.015 & 0.21 & -0.023 & -0.126 & N  \\
VCC1453 & 2 & 10.93 & 18.3 & 2.22 & 0.77 & 12.48 & 6.1 & 1.36 & 0.74 & 11.34 & 22.1 & 1.0 & 0.82 & 0.012 & 0.158 & 0.01 & 0.106 & N  \\
VCC1488 & 2 & 12.01 & 13.0 & 1.39 & 0.6 & 13.47 & 6.0 & 0.93 & 0.64 & 12.31 & 17.9 & 1.0 & 0.59 & 0.02 & 0.135 & 0.015 & 0.093 & -  \\
VCC1491 & 2 & 12.11 & 8.8 & 1.06 & 0.79 & 13.22 & 5.5 & 0.78 & 0.84 & 12.5 & 13.2 & 1.0 & 0.72 & - & - & - & - & N  \\
VCC1499 & B & 12.54 & 7.8 & 1.15 & 0.63 & - & - & - & - & 12.62 & 8.3 & 1.0 & 0.68 & - & - & - & - & -  \\
VCC1501 & 1 & 12.95 & 11.5 & 1.4 & 0.46 & - & - & - & - & - & - & - & - & -0.006 & -0.025 & - & - & -  \\
VCC1512 & 2 & 12.3 & 20.8 & 4.04 & 0.79 & 14.34 & 2.7 & 2.05 & 0.85 & 13.0 & 15.0 & 1.0 & 0.72 & 0.024 & 0.127 & 0.014 & 0.064 & -  \\
VCC1528 & 2 & 11.35 & 8.8 & 2.06 & 0.83 & 12.94 & 3.0 & 1.22 & 0.75 & 11.71 & 10.7 & 1.0 & 0.92 & - & - & - & - & -  \\
VCC1545 & {\it 2} & 11.67 & 9.6 & 2.57 & 0.84 & 12.72 & 4.0 & 1.9 & 0.89 & 12.36 & 14.0 & 1.0 & 0.73 & 0.002 & 0.007 & -0.001 & -0.004 & N  \\
VCC1549 & 1 & 11.36 & 12.1 & 1.71 & 0.83 & - & - & - & - & - & - & - & - & - & - & - & - & N  \\
VCC1614 & L & 11.43 & 8.6 & 0.83 & 0.88 & - & - & - & - & 11.61 & 10.7 & 1.0 & 0.93 & 0.039 & 1.084 & 0.02 & 0.634 & N  \\
VCC1627 & 1 & 11.72 & 3.9 & 1.95 & 0.87 & - & - & - & - & - & - & - & - & -0.032 & -0.064 & - & - & N  \\
VCC1695 & 2 & - & - & - & - & 13.25 & 5.0 & 1.05 & 0.63 & 11.85 & 20.5 & 1.0 & 0.81 & - & - & - & - & N  \\
VCC1743 & {\it 2} & 12.55 & 15.7 & 1.17 & 0.44 & 13.94 & 9.7 & 0.82 & 0.35 & 12.83 & 19.6 & 1.0 & 0.51 & 0.008 & 0.041 & 0.004 & 0.021 & -  \\
VCC1779 & 2 & 12.04 & 16.9 & 1.52 & 0.48 & 13.49 & 7.1 & 0.94 & 0.51 & 12.32 & 24.8 & 1.0 & 0.45 & 0.041 & 0.262 & 0.029 & 0.175 & -  \\
VCC1827 & B & 10.03 & 24.7 & 1.91 & 0.71 & 13.07 & 3.4 & 0.86 & 0.85 & 10.53 & 25.5 & 0.52 & 0.8 & 0.081 & 1.165 & 0.013 & 0.135 & N  \\
VCC1828 & 1 & 12.13 & 15.1 & 1.46 & 0.71 & - & - & - & - & - & - & - & - & 0.009 & 0.052 & - & - & N  \\
VCC1836 & 2 & 11.46 & 35.5 & 1.65 & 0.55 & 13.18 & 11.2 & 1.03 & 0.72 & 11.69 & 52.2 & 1.0 & 0.43 & -0.034 & -0.087 & -0.047 & -0.121 & -  \\
VCC1857 & 1 & 12.19 & 26.1 & 0.72 & 0.53 & - & - & - & - & - & - & - & - & 0.01 & 0.04 & - & - & -  \\
VCC1861 & L & 11.15 & 14.5 & 1.48 & 0.97 & - & - & - & - & 11.32 & 18.9 & 1.0 & 0.93 & 0.031 & 0.423 & 0.015 & 0.148 & N  \\
VCC1871 & L & 10.64 & 7.4 & 1.97 & 0.87 & - & - & - & - & 10.75 & 7.2 & 2.21 & 0.86 & 0.013 & 5.85 & -0.001 & 2.587 & -  \\
VCC1876 & 1 & 12.07 & 16.1 & 0.92 & 0.51 & - & - & - & - & - & - & - & - & 0.006 & 0.052 & - & - & N  \\
VCC1890 & 1 & 12.01 & 23.4 & 0.97 & 0.6 & - & - & - & - & - & - & - & - & 0.021 & 0.108 & - & - & -  \\
VCC1895 & 1 & 12.08 & 13.1 & 1.32 & 0.47 & - & - & - & - & - & - & - & - & - & - & - & - & -  \\
VCC1896 & B & 11.77 & 16.6 & 1.28 & 0.71 & 14.61 & 3.6 & 0.73 & 0.79 & 12.26 & 19.6 & 0.47 & 0.85 & 0.004 & 0.002 & -0.006 & -0.029 & N  \\
VCC1897 & 2 & 11.31 & 30.5 & 1.5 & 0.59 & 14.26 & 5.8 & 1.12 & 0.93 & 11.45 & 32.3 & 1.0 & 0.54 & 0.018 & 0.072 & 0.009 & 0.034 & -  \\
VCC1902 & 2 & 9.84 & 27.3 & 1.99 & 0.8 & 11.38 & 10.0 & 1.79 & 0.69 & 10.34 & 27.6 & 0.64 & 0.86 & 0.043 & 0.612 & 0.011 & 0.093 & N  \\
VCC1910 & {\it 2} & 10.73 & 12.5 & 1.58 & 0.84 & 11.71 & 8.4 & 1.48 & 0.74 & 11.35 & 15.5 & 1.0 & 0.92 & -0.005 & -0.034 & -0.006 & -0.039 & -  \\
VCC1947 & {\it B} & 11.08 & 9.6 & 1.52 & 0.79 & 12.39 & 9.3 & 3.12 & 0.69 & 11.56 & 9.9 & 1.0 & 0.94 & 0.004 & 0.103 & 0.002 & 0.002 & N  \\
VCC1949 & B & 10.66 & 29.0 & 1.47 & 0.5 & 14.96 & 4.5 & 1.4 & 0.65 & 11.12 & 31.4 & 0.52 & 0.64 & 0.069 & 0.651 & 0.021 & 0.166 & - & sp \\
VCC2008 & 1 & 12.3 & 35.6 & 0.86 & 0.44 & - & - & - & - & - & - & - & - & -0.078 & -0.174 & - & - & -  \\
VCC2012 & 1 & 11.45 & 22.6 & 1.18 & 0.85 & - & - & - & - & - & - & - & - & 0.015 & 0.092 & - & - & -  \\
VCC2019 & 1 & 11.37 & 17.0 & 1.36 & 0.71 & - & - & - & - & - & - & - & - & 0.012 & 0.033 & - & - & N & sp  \\
VCC2042 & 1 & 12.38 & 18.5 & 1.04 & 0.96 & - & - & - & - & - & - & - & - & - & - & - & - & N  \\
VCC2050 & 2 & 12.2 & 15.0 & 1.38 & 0.5 & 13.08 & 10.2 & 1.1 & 0.44 & 12.8 & 20.8 & 1.0 & 0.63 & 0.012 & 0.048 & 0.011 & 0.043 & N  \\
VCC2080 & B & - & - & - & - & - & - & - & - & 12.87 & 20.1 & 1.61 & 0.46 & - & - & 0.016 & 0.071 & N  \\
VCC2083 & 1 & 12.71 & 13.8 & 0.86 & 0.85 & - & - & - & - & - & - & - & - & 0.005 & 0.019 & - & - & N  \\
\enddata
\tablecomments{{ The table lists the details of the decompositions for the 99 galaxies, for which the analysis was done. The columns list }
 the structural group (with less certain group assignments printed with italics), the parameters (integrated brightness $m$, half-light radius $r_e$, S\'ersic index $n$, { and the axis ratio $b/a$}) of the simple, as well as the inner and { global or} outer components of the final model, where applicable. { `Global component' refers to the main component in galaxies with a lens in addition to one other component. } Furthermore, we give the RFF and EVI indices for both models (see \S\ref{EVIRFF}). Values are omitted for archive data and in case the one-component models with nucleus failed for galaxies, for which the final model includes a nucleus { (`N' in the second last column). The parameters of nuclei, lenses, and bars that resulted from our best fits are not used in our analysis, but can be provided on request for specific objects.} The final column indicates where there are hints of spiral arms in the residual images. }
\end{deluxetable*}
\clearpage
\end{landscape}

\end{document}